\documentclass [] {aa} 

\usepackage{txfonts,graphicx,lscape,longtable}

\begin{document}

\title{Oxygen abundances in planet-harbouring stars.
\thanks{Based on data from the {\footnotesize FEROS} spectrograph at the 2.2-m ESO/MPI 
    telescope (observing run ID 074.C-0135), at the La Silla Observatory, ESO (Chile), and the UVES 
    spectrograph at VLT/UT2 Kueyen telescope (observing run ID 074.C-0134), at the Paranal Observatory, 
    ESO (Chile), and on observations made with the SARG spectrograph at 
    3.5-m TNG, operated by the Fundaci\'on Galileo Galilei of the INAF, 
    and with the UES spectrograph at the 4-m William Hershel Telescope 
    (WHT), operated by the Isaac Newton Group, both at the Spanish 
    Observatorio del Roque de los Muchachos 
    of the Instituto de Astrofisica de Canarias.}}
    
\subtitle{Comparison of different abundance indicators.}

\author{A.~Ecuvillon\inst{1}, G.~Israelian\inst{1}, N. C.~Santos\inst{2,3}, 
N. G.~Shchukina\inst{4}, M.~Mayor\inst{3} \and R.~Rebolo\inst{1,5}}

\offprints{ \email{aecuvill@ll.iac.es}}

\institute{Instituto de Astrof\'{\i}sica de Canarias, E-38200 La Laguna, Tenerife, Spain \and Centro de 
Astronomia e Astrofisica de Universidade de Lisboa, Observatorio Astronomico de Lisboa, Tapada de Ajuda, 1349-018 Lisboa, 
Portugal \and Observatoire de Gen\`eve, 51 ch.  des  Maillettes, CH--1290 Sauverny, Switzerland \and Main
Astronomical Observatory, National Academy of Sciences, 27 Zabolotnogo Street, 03680 Kyiv-127, Ukraine \and Consejo Superior de
Investigaciones Cient\'{\i}ficas, Spain}
\date{Received 19 May 2005 / Accepted 16 August 2005} 

\titlerunning{Oxygen abundances in Planet-harbouring stars} 
\authorrunning{A. Ecuvillon et al.}

\abstract{We present a detailed and uniform study of oxygen abundances in 155
solar type stars,
96 of which are planet hosts and 59 of which form part of a volume--limited
comparison sample with no known planets.
EW measurements were carried out for the [O\,I] 6300 \AA\ line and the O\,I triplet, and spectral synthesis 
was performed for several OH lines. NLTE corrections were calculated and applied to the LTE abundance 
results derived from the O\,I 7771--5 \AA\ triplet. Abundances from [O\,I], the O\,I triplet and near-UV OH 
were obtained in 103, 87 and 77 dwarfs, respectively. 
We present the first detailed and uniform comparison of these three oxygen 
indicators in a large sample of solar-type stars. There is good agreement between the [O/H] ratios from forbidden and 
OH lines, while the NLTE triplet shows a systematically lower abundance. We found that discrepancies between OH, [O\,I] and the O\,I triplet do not exceed 0.2\,dex in most cases. We have studied abundance trends in 
planet host and comparison sample stars, and no obvious 
anomalies related to the presence of planets have been detected. All three indicators show that, 
on average, [O/Fe] decreases with [Fe/H] in the metallicity range $-0.8<$ [Fe/H] $< 0.5$. The planet host stars 
present an average oxygen overabundance of 0.1--0.2\,dex with respect to the comparison sample. 
 
	  }
\maketitle

\section{Introduction}

The discoveries of more than 120 planetary-mass companions orbiting around solar-type stars have 
provided important opportunities to understand the formation and evolution of planetary systems. Several studies
have shown that planet-harbouring stars are on average more metal-rich than dwarfs of the same spectral
type with no known planets (Gonzalez \cite{Gon97}; Gonzalez et al.\ \cite{Gon01}; Laws et al.\ \cite{Law03};
Santos et al.\ \cite{San01}, \cite{San03b}, \cite{San04a}, \cite{San05}; for a review see Santos et al.\ \cite{San03a}). 
Two possible explanations have been suggested to link the metalllicity excess to the presence of planets. 
The first has been proposed by Gonzalez (\cite{Gon97}), who has suggested that the iron enhancement observed 
in stars with planets is  due mainly to the accretion of large amounts of protoplanetary material onto the 
star. The other hypothesis by Santos et al.~(\cite{San00}, \cite{San01}) attributes the metallicity excess of 
planet host stars to the high metal content of the primordial cloud out of which the planetary system formed.

Detailed chemical analysis of  planet-harbouring stars can provide useful information in the understanding 
of how the systems with giant planets have formed. Searching for chemical anomalies related to the presence 
of planets, in addition to the observed iron excess, is thus of high interest in discriminating between possible planetary
formation hypotheses. For instance, light elements can give important evidence of pollution events
(Israelian et al.\ \cite{Isr01b}, \cite{Isr03a}, \cite{Isr04a}; Sandquist et al.\ \cite{Sand02}; Santos et
al.\ \cite{San02}, \cite{San04b}). 

Abundance trends of volatile and refractory elements are also of
 interest in investigating planetary system formation. If the accretion processes were  mainly 
responsible for the metallicity excess found in planet host stars, a relative overabundance of refractory 
elements would be observed, since volatiles (with low condensation temperatures) are expected to be deficient  in accreted materials (Smith et al.\ \cite{Smi01}). Likewise, if planet host stars had undergone 
significant pollution, their volatile abundances should show clear differences with respect to those of 
field stars. In this framework, it is very important to achieve abundance trends for as many planet host 
stars and as many elements as possible, and to carry out a homogeneous comparison with field stars with no 
known planetary-mass companion. 

Several studies on abundances of metals other than iron have been carried 
out in planet host stars, but most of them have included only a reduced number of targets with planets and
 their results have been compared inhomogeneously  with abundance trends of field stars from other authors (Gonzalez \& 
Laws \cite{Gon00}; Gonzalez et al.\ \cite{Gon01}; Santos et al.\ \cite{San00}; Takeda et al.\ \cite{Tak01}; 
Sadakane et al.\ \cite{Sad02}; for a review see Israelian et al.\ \cite{Isr03b}). 
Recently, some refractories (e.g.\ Ca, Ti, Si, etc.) and volatiles (N, C, S and Zn) have been analysed 
homogeneously in a large number of planet host targets, as well as in a comparison set of stars with no known 
planets (Bodaghee et al.\ \cite{Bod03}; Ecuvillon et al.\ \cite{Ecu04a}, \cite{Ecu04b}; Beirao et al.\ \cite{Bei05}; Gilli et al.\ \cite{Gil05}). Takeda \& Honda
(\cite{Tak05}) presented a study of CNO abundances in 27 planet host stars included in a large 
sample of 160 F, G and K dwarfs and subgiants.  
   
Oxygen is the third most abundant element in the Universe, after hydrogen and helium. 
By analysing elemental abundances in the atmospheres of F and G dwarfs stars it is possible to determine the
chemical composition of the gas out of which the stars were born and to understand the chemical evolution of the 
Galaxy and its formation history (e.g.\ McWilliam \cite{McW97}).
Oxygen is essentially primary. It is formed by $\alpha$-processing in massive stars and released in the 
interstellar medium (ISM) during Type II SN explosions (e.g.\ Arnett \cite{Arn78}; Tinsley \cite{Tin79};
Woosley \& Weaver \cite{Woo95}).

Several indicators have been used in the determination of oxygen abundances in disc and halo stars: the
near-IR O\,I triplet at 7771--5 \AA\  (e.g.\ Abia \& Rebolo \cite{Abi89}; Tomkin et al.\ \cite{Tom92}; King \&
Boesgaard \cite{Kin95}; Cavallo, Pilachowski, \& Rebolo \cite{Cav97}; Mishenina et al.\ \cite{Mis00};
 Israelian et al.\ \cite{Isr01a};
Fulbright \& Johnson \cite{Ful03}; Takeda \cite{Tak03}; Bensby, Feltzing, \& Lundstr\"om \cite{Ben04}; Schukina et al.\ \cite{Shc05}),
the forbidden lines of [O\,I] at 6300 and 6363 \AA\  (e.g.\ King \& Boesgaard \cite{Kin95}; Fulbright \& 
Johnson \cite{Ful03}; Takeda \cite{Tak03}; Bensby, Feltzing, \& Lundstr\"om \cite{Ben04}; Schukina et al.\ \cite{Shc05}), and the near-UV 
OH lines at 3100 \AA\  (e.g.\ Bessell, Sutherland, \& Ruan \cite{Bes91}; Nissen et al.\ \cite{Nis94}; 
Israelian, Garc\'{\i}a-L\'opez, \& Rebolo \cite{Isr98}; Boesgaard et al.\ \cite{Boe99}; Israelian et al.\ 
\cite{Isr01a}). 
Unfortunately, results from different indicators show discrepancies. Israelian et al.\
(\cite{Isr04b}) have reported the largest conflict between the O\,I triplet at 7771--5 \AA\ and the forbidden 
line at 6300 \AA\, with discrepancies in [O/H] ratios of up to 1\,dex.
For stars with [Fe/H] $< -1.0$, many
studies obtained disagreement in the [O/Fe] vs.\ [Fe/H] relationship (e.g.\ Israelian et al.\ \cite{Isr01a}; Nissen et al.\ \cite{Nis02}). 
For solar-metallicity stars as well the situation is unclear (e.g.\ Nissen \& Edvardsson \cite{Nis92}; Feltzing \& Gustafsson \cite{Fel98}; Prochaska et al.\
\cite{Pro00}; Bensby, Feltzing, \& Lundstr\"om \cite{Ben04}). 

The analyses of these lines all have their difficulties. The triplet lines are strongly affected by
deviations from LTE (e.g.\ Shchukina \cite{Shc87}; Kiselman \cite{Kis91}), and by convective 
inhomogeneities (e.g.\ Kiselman \cite{Kis93}). However newer 3D calculations by Asplund et al.\ (\cite{Asp04}) showed that the NLTE effects are very similar in 1D and 3D. The forbidden lines are both very weak and blended by lines 
from other species (e.g.\ Lambert \cite{Lam78}; Bensby, Feltzing, \& Lundstr\"om  \cite{Ben04}).
The OH lines are very sensitive to surface inhomogeneities like granulation (e.g. Kiselman \& Nordlund
\cite{Kis95}).
  
Our work presents a complete and uniform study of the oxygen abundances in two large samples, a set of planet-harbouring stars and a 
volume-limited comparison sample of stars with no known planetary-mass companions, using three different 
indicators in order to check the consistency of the results. We  carry out a detailed comparison among
the abundances provided by different lines and discuss possible discrepancies. We  investigate eventual 
anomalies related to the presence of planets and locate our results within the framework of Galactic chemical
evolution.

\section{Observations}
Most of the spectra for the [O\,I] line and O\,I triplet analysis were collected during several observational 
campaigns with different spectrographs: CORALIE on the 1.2 m Euler Swiss telescope, FEROS on the 2.2 m
ESO/MPI telescope (both at La Silla, Chile), UVES at the VLT/UT2 Kueyen telescope (Paranal, ESO, Chile), SARG
on the 3.5 m TNG and UES on the 4.2 m WHT (both at Roque de los Muchachos, La Palma, Spain). Previous works
have already used them to derive stellar parameters (Santos et al.\ \cite{San04a}) and abundances of 
different species (Bodaghee et al.\ \cite{Bod03}; Ecuvillon et al.\ \cite{Ecu04a}, \cite{Ecu04b}; Beirao et al.\ \cite{Bei05}; Gilli et al.\ \cite{Gil05}). 

New optical spectra were collected with the UVES spectrograph at the Kueyen telescope, 
the FEROS spectrograph at the
 ESO/MPI telescope and the SARG spectrograph at TNG (Roque de los Muchachos, La 
Palma, Spain). A detailed description of the FEROS data is available in the work of Santos et al. 
(\cite{San05}). The new optical SARG and UVES spectra have a resolution $R \sim 57\,000$ and $R\sim85\,000$, as 
well as S/N ratios above 150 and 600, respectively, at $\sim$6000 \AA. 

For the synthesis of the OH lines, we used the near-UV spectra from the UVES spectrograph at the 
 Kueyen telescope, from which Santos et al.\ (\cite{San02}) and 
Ecuvillon et al.\ (\cite{Ecu04a}) have derived beryllium and nitrogen abundances. We refer the reader to 
these papers for a detailed description of the data. New near-UV spectra obtained from the UVES spectrograph at the 
Kueyen telescope were used. These spectra
have a resolution $R \sim 75\,000$ and S/N ratios above 100 in most cases at Be region.

The data reduction for the new SARG spectra was done using IRAF\footnote{IRAF is distributed by the National
Optical Astronomy  Observatories, operated by the Association of Universities for Research in Astronomy,
Inc., under cooperative agrement with the National Science Foundation, USA.} tools in the {\tt echelle} package. 
Standard background correction, flatfield and extraction procedures were used. The wavelength calibration was
performed using a ThAr lamp spectrum taken during the same night. The FEROS and UVES spectra were reduced 
using the corresponding pipeline softwares. 

\begin{table}[!]
\caption[]{Atomic parameters adopted for [O\,I] 6300 \AA\ line, O\,I 7771--5 \AA\ triplet and the near-UV OH
lines.}
\begin{center}
\begin{tabular}{cccr}
\hline
\noalign{\smallskip}
Species & $\lambda$ (\AA) & $\chi_l$ (eV) & $\log{gf}$ \\
\hline
\noalign{\smallskip}
[O\,I] & 6300.230 & 0.00 & $-9.689$ \\
Ni\,I  & 6300.399 & 4.27 & $-2.310$ \\
O\,I   & 7771.960 & 9.11 & $ 0.452$ \\
O\,I   & 7774.180 & 9.11 & $ 0.314$ \\
O\,I   & 7775.400 & 9.11 & $ 0.099$ \\
OH     & 3167.169 & 1.11 & $-1.623$ \\
OH     & 3189.312 & 1.03 & $-1.990$ \\ 
OH     & 3255.490 & 1.30 & $-1.940$ \\ 
OH     & 3172.997 & 1.20 & $-1.692$ \\
OH     & 3173.200 & 1.83 & $-1.100$ \\
\noalign{\smallskip}
\hline	      
\end{tabular} 
\end{center}
\label{tab1}  
\end{table}

\begin{figure*}
\centering 
\includegraphics[]{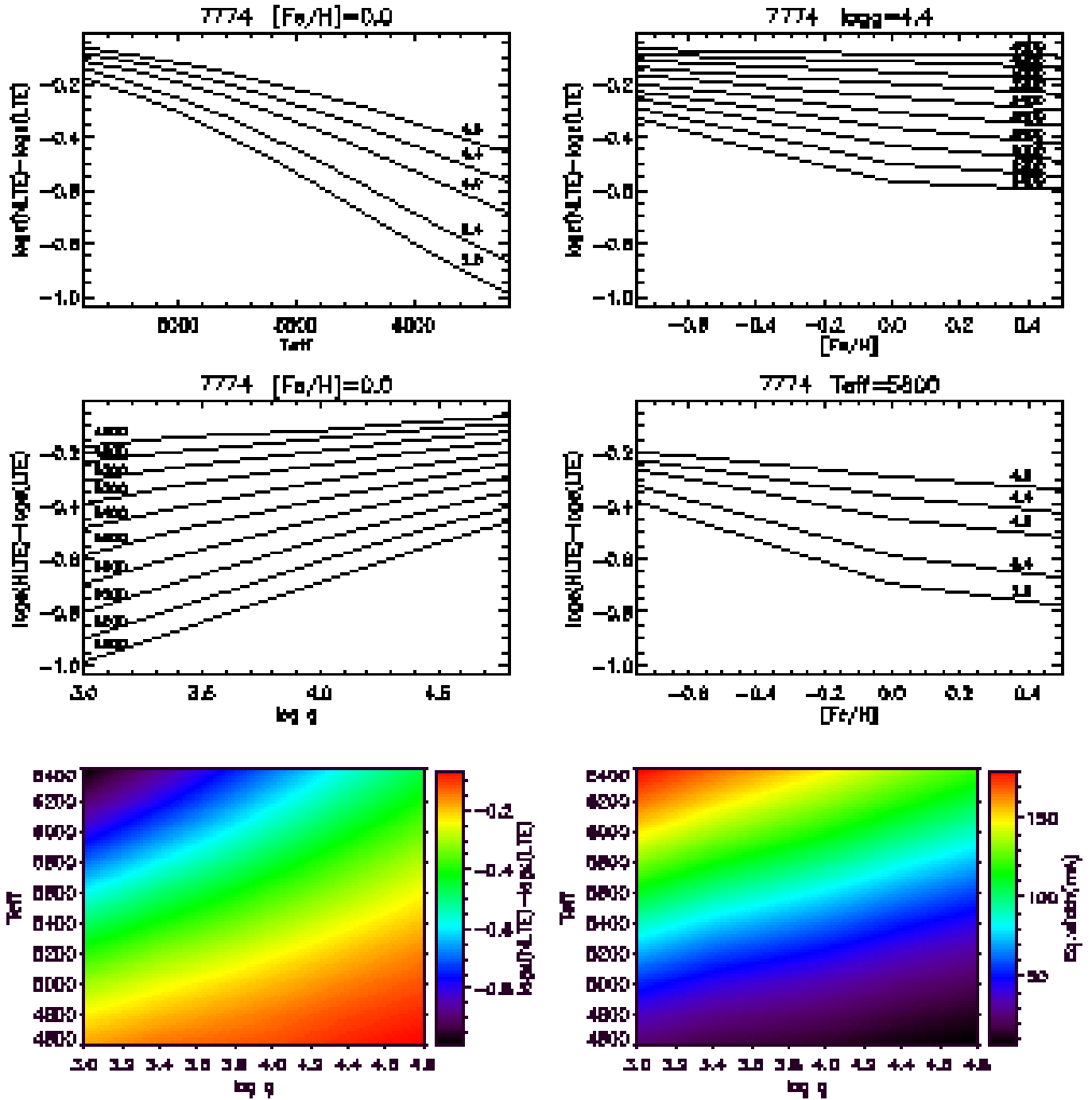}
\caption{NLTE abundance corrections $\log{\epsilon}(NLTE)-\log{\epsilon}(LTE)$ and NLTE equivalent width of
the O\,I IR triplet $\lambda$7774 \AA\ line in a grid of one-dimensional model atmospheres of Kurucz spanning
the range 4600\,K$<T_\mathrm{eff}<$6400\,K, 3.0$<\log{g}<4.8$, and -0.75$<$[Fe/H]$<$0.5. The {\it right-hand 
top} and the {\it right-hand middle} panels show the results for stars with the solar-like gravity 
$\log{g}$=4.4 and the solar-like effective temperature $T_\mathrm{eff}$=5800\,K, while the {\it right-hand 
bottom} and all the {\it left-hand} panels show the results for stars with the solar metallicity 
[Fe/H]=0.0. The {\it bottom} panels show variations of the NLTE abundance corrections ({\it left}) and the NLTE equivalent width
with $T_\mathrm{eff}$ and $\log{g}$ for [Fe/H]=0. The numbers above curves indicate the gravities and
effective temperatures.}
\label{fig1}
\end{figure*}

\begin{figure*}
\centering 
\includegraphics[width=6.7cm]{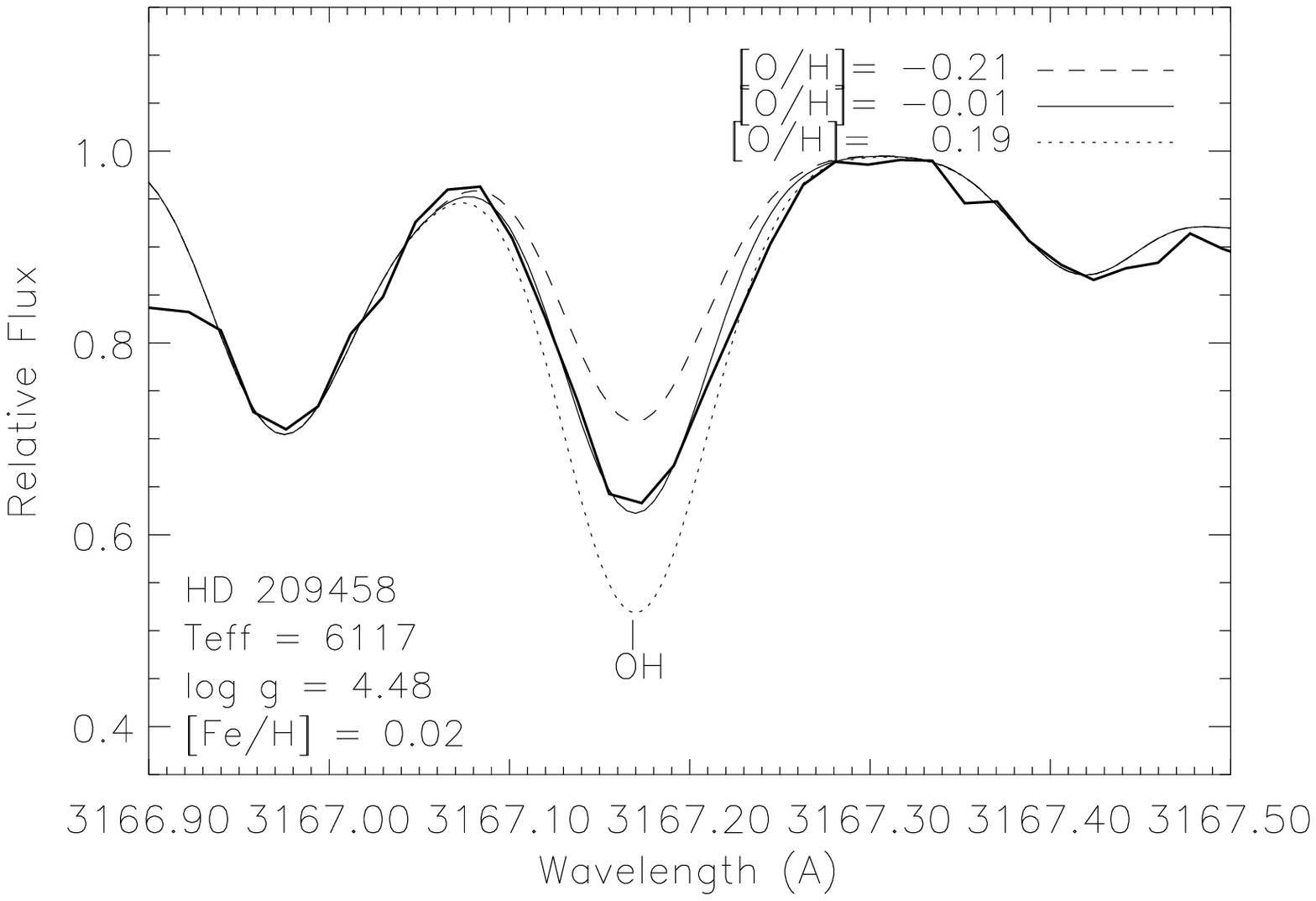}
\includegraphics[width=6.7cm]{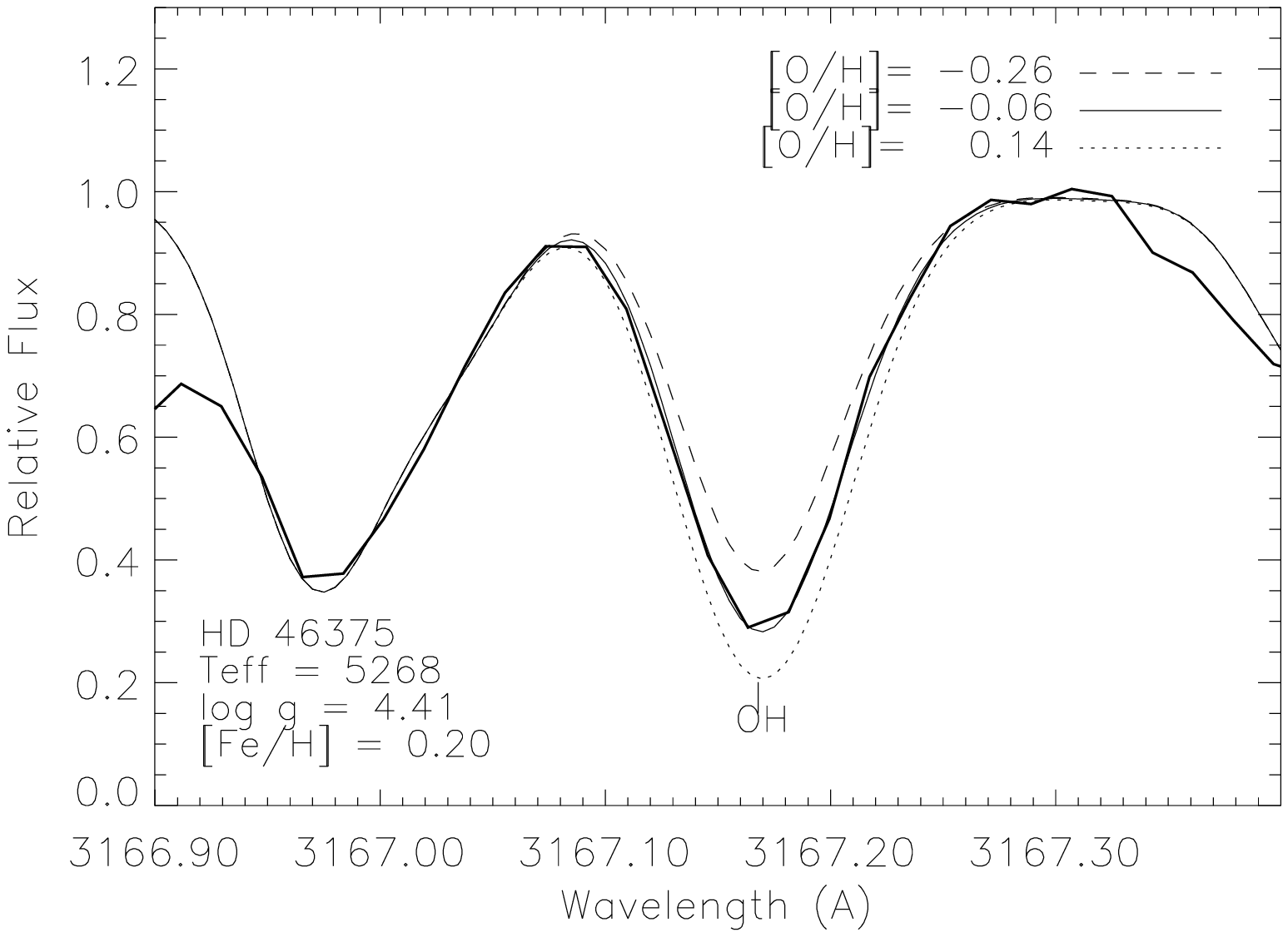}
\includegraphics[width=6.7cm]{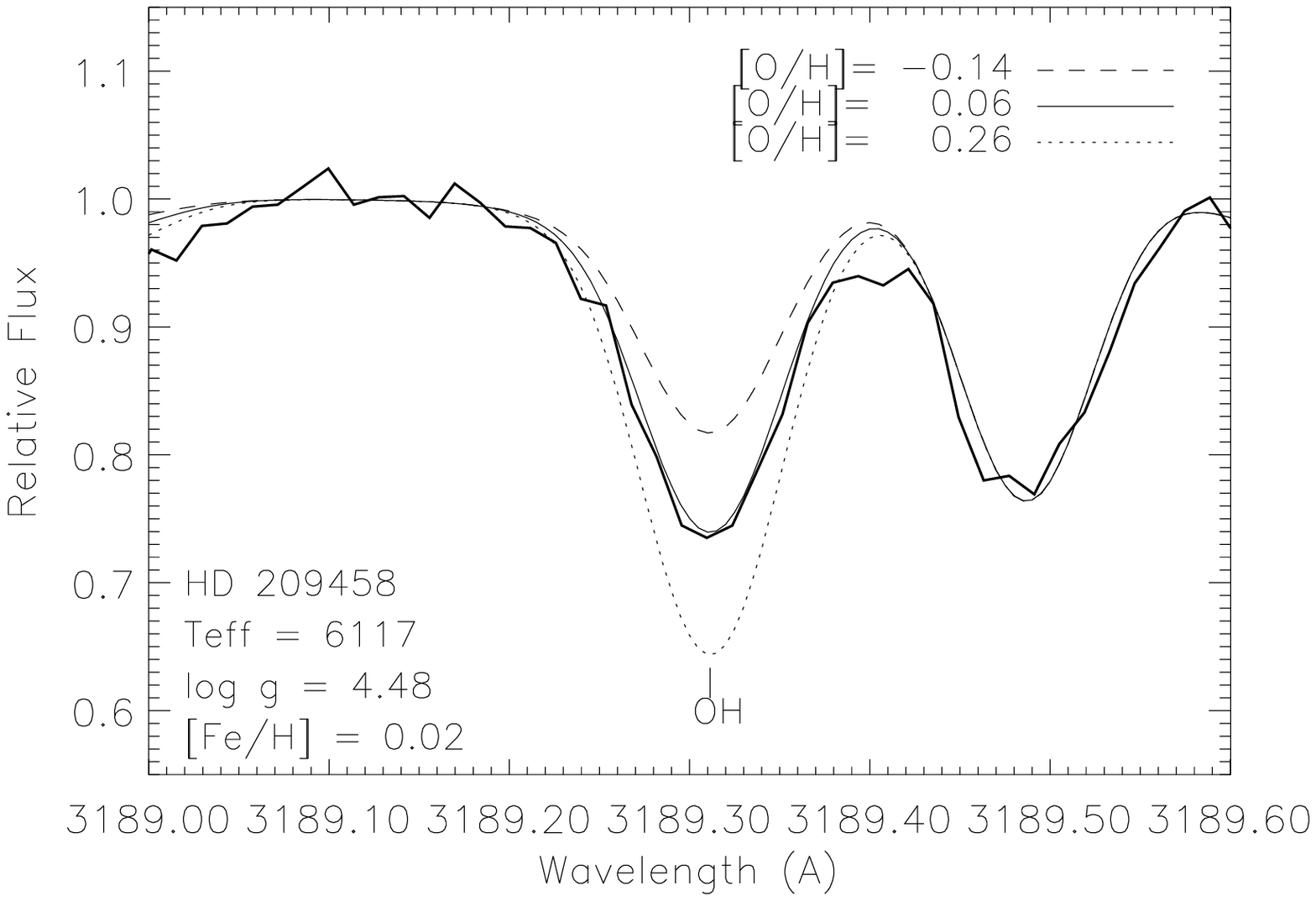}
\includegraphics[width=6.7cm]{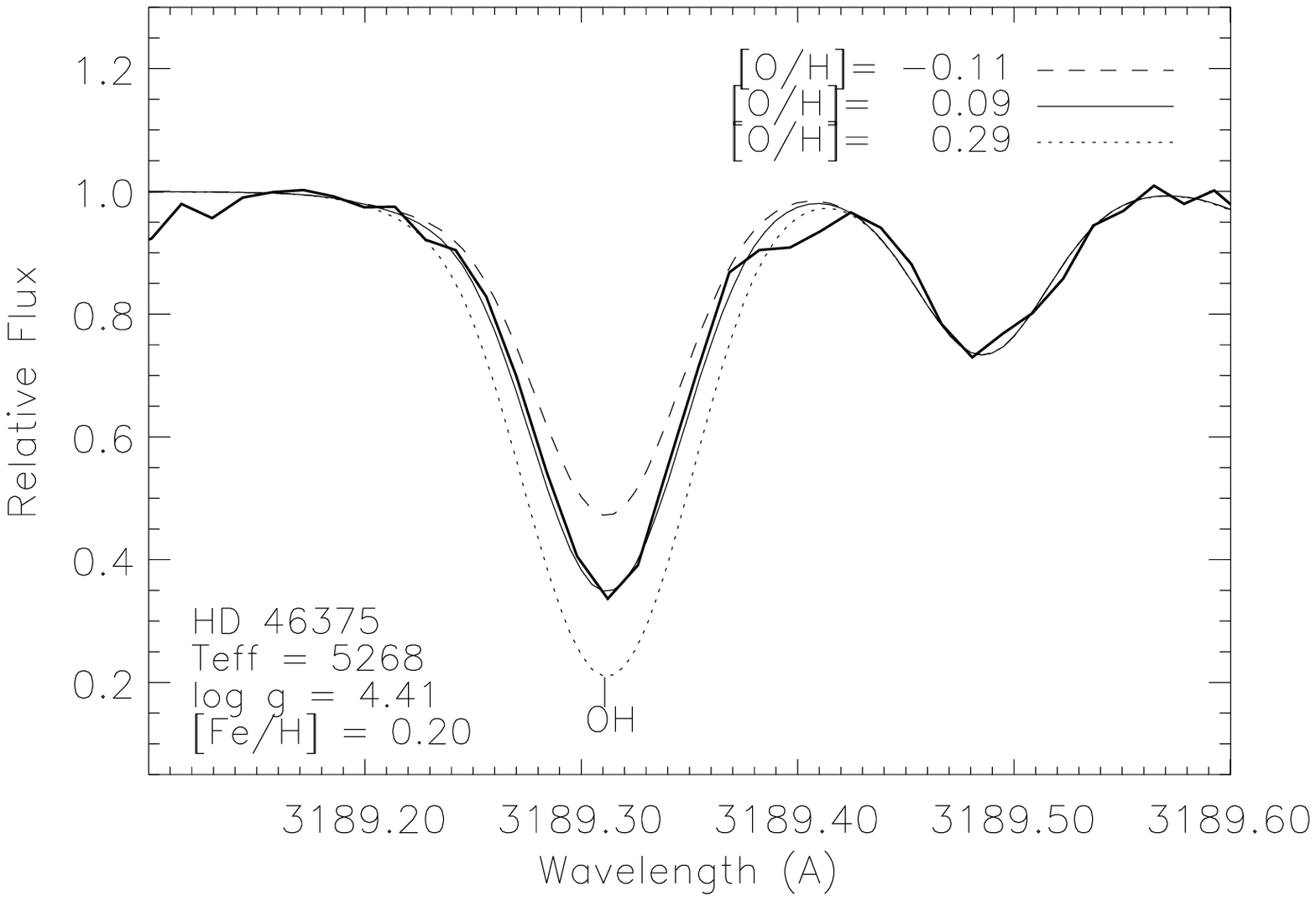}
\includegraphics[width=6.7cm]{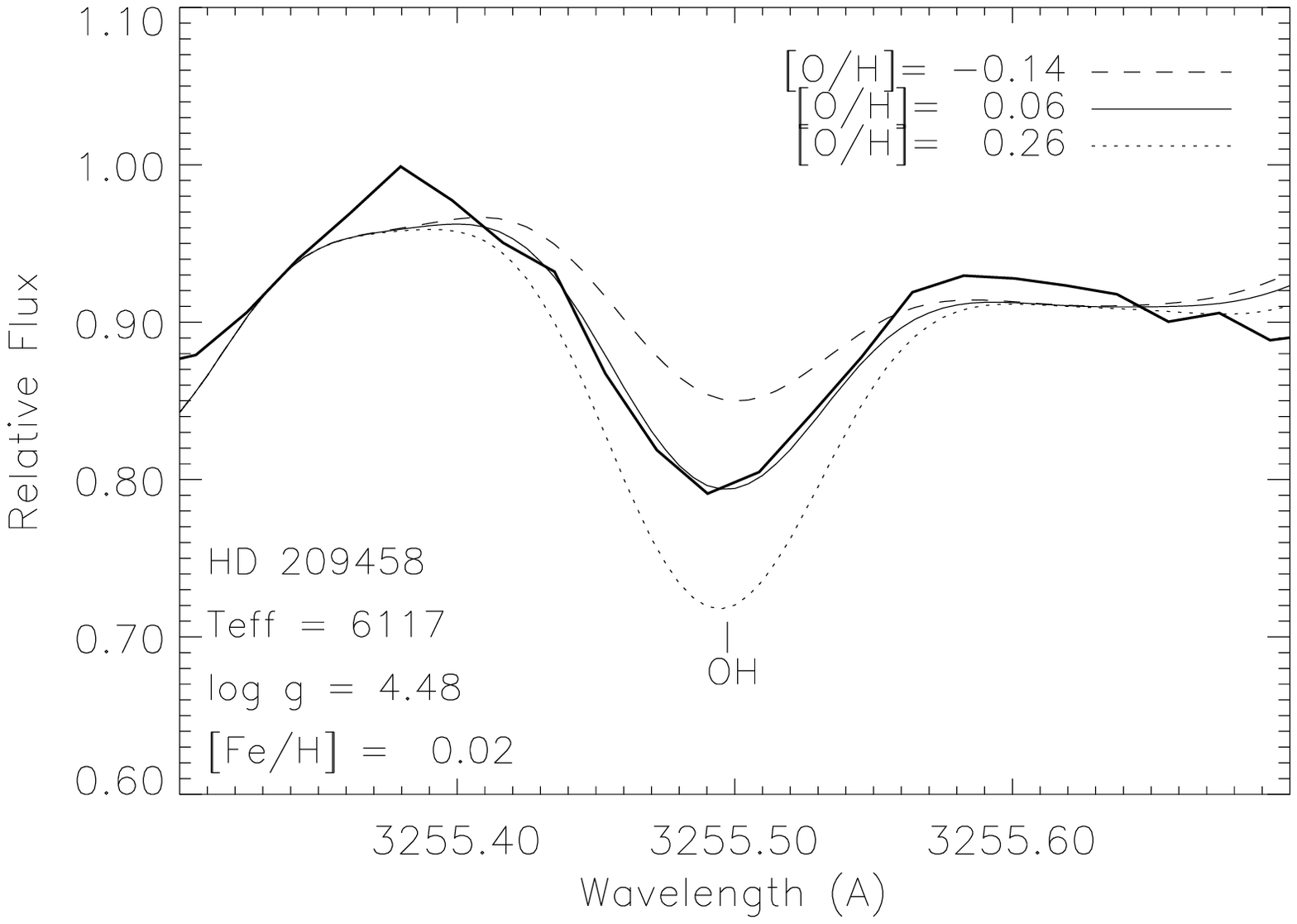}
\includegraphics[width=6.7cm]{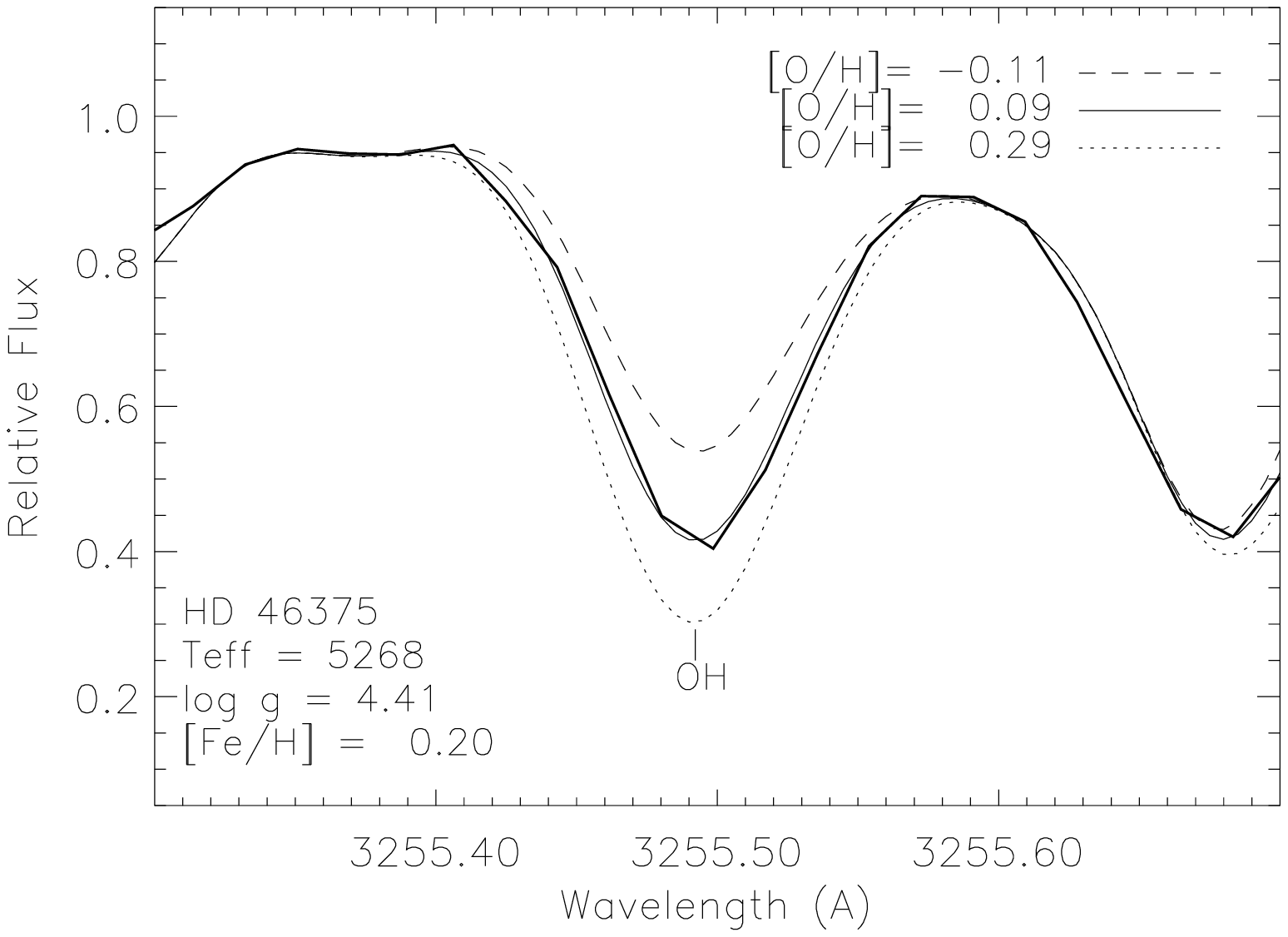}
\includegraphics[width=6.7cm]{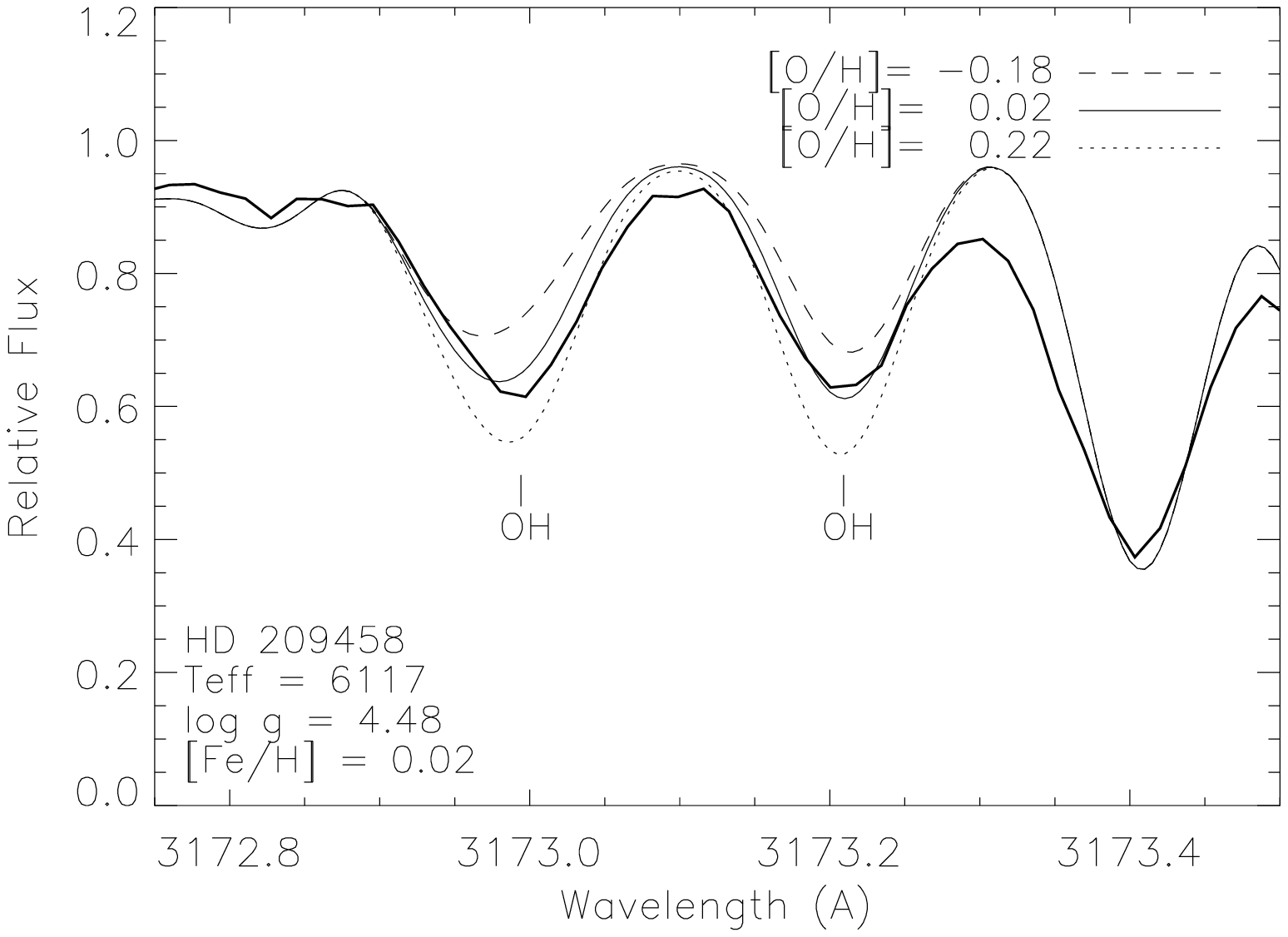}
\includegraphics[width=6.7cm]{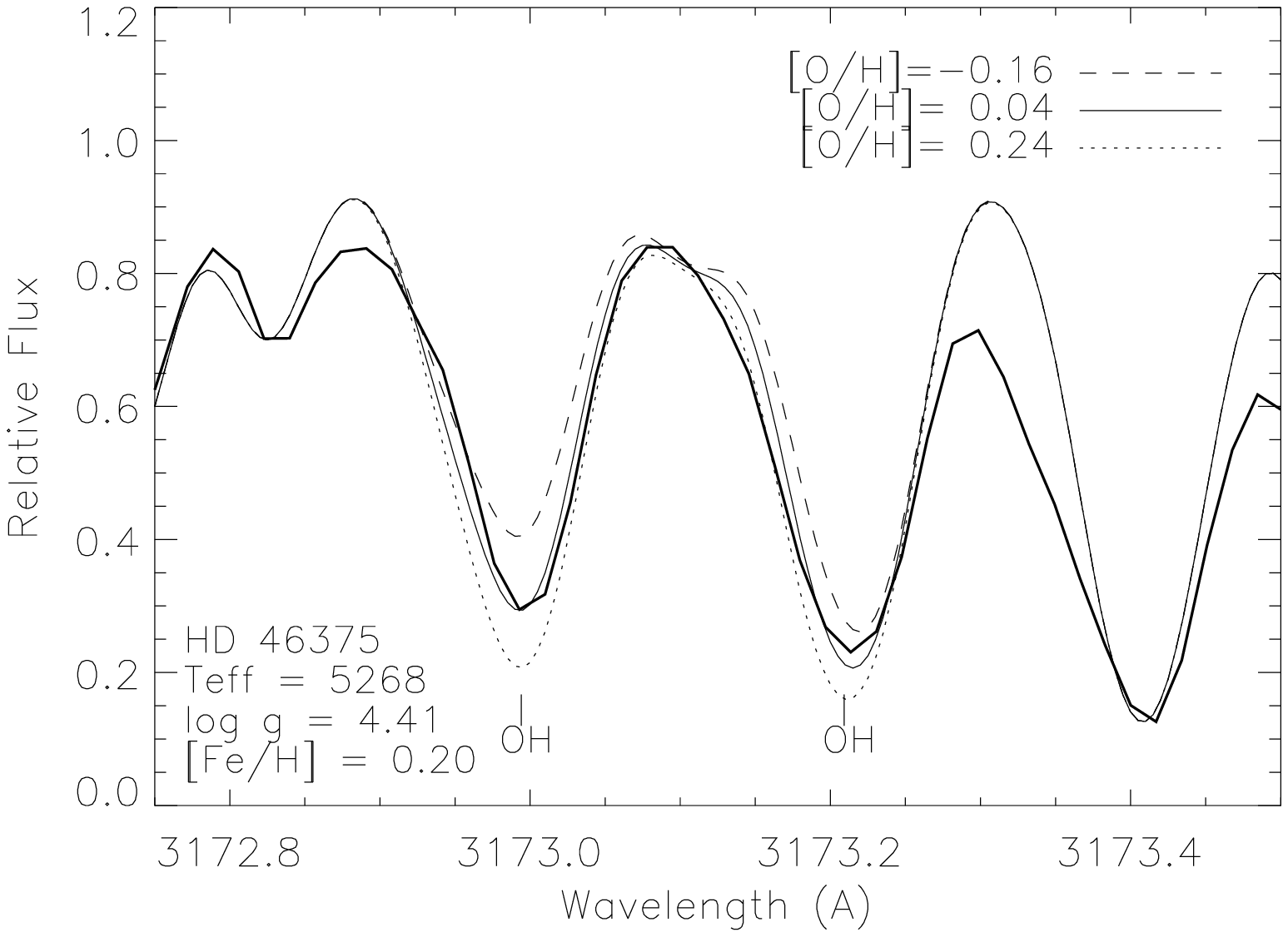}
\caption{The observed spectrum (thick solid line) and three synthetic spectra (dotted, dashed and solid lines) 
for different values of [O/Fe] in the four OH features analysed, for two targets.}
\label{fig2}
\end{figure*}

\begin{table*}[!]
\caption[]{Sensitivity of the three indicators [O\,I], O\,I and OH to changes of 100\,K in effective 
temperature, 0.3\,dex in gravity and metallicity, and 0.5\,km\,s$^{-1}$ in microturbulence}
\begin{center}
\begin{tabular}{|l|c|c|c|c|c|}
\hline
 & & $\Delta T_\mathrm{eff}=\pm100$\,K & $\Delta \log{g}=\pm0.3$\,dex & $\Delta$[Fe/H] =$\pm0.3$\,dex & $\Delta\xi_t=\pm0.5$\,km\,s$^{-1}$\\ \hline \hline
[O\,I] & Star & $\Delta$[O/H] & $\Delta$[O/H]  & $\Delta$[O/H]  & $\Delta$[O/H]  \\ \hline
 & \object{HD\,22049}  & $\pm0.01 $  & $\pm0.14$ & $\pm0.09$ & $\pm0.00^1$\\
 & \object{HD\,37124} & $\pm0.01$   & $\pm0.14$ & $\pm0.08$ & $\pm0.00^1$\\
 & \object{HD\,9826}  & $\pm0.01$   &  $\pm0.14$ & $\pm0.09$ & $\pm0.01$\\ \hline
O\,I & Star & $\Delta$[O/H] & $\Delta$[O/H]  & $\Delta$[O/H] & $\Delta$[O/H] \\ \hline
 & \object{HD\,22049} & $\mp0.14 $  & $\pm0.09$ & $\pm0.02$ & $\mp0.02$\\
 & \object{HD\,37124} & $\mp0.10$   & $\pm0.05$ & $\pm0.02$ & $\mp0.02$\\
 & \object{HD\,9826}  & $\mp0.07$   &  $\pm0.05$ & $\pm0.02$ & $\mp0.03$\\ \hline\hline
OH & Star & $\Delta$[O/H] & $\Delta$[O/H]  & $\Delta$[O/H] & $\Delta$[O/H] \\ \hline
 & \object{HD\,22049} & $\pm0.05 $  & $\mp0.03$ & $\pm0.18$ & $\pm0.00^1$\\
 & \object{HD\,37124} & $\pm0.10$   & $\mp0.05$ & $\pm0.20$ & $\pm0.00^1$\\
 & \object{HD\,9826}  & $\pm0.12$   &  $\mp0.05$ & $\pm0.20$ & $\pm0.00^1$\\ \hline\hline 
\noalign{\smallskip}
\end{tabular}
\end{center}
\footnotesize{$^1$ These sensitivities are based on the method described in Section~\ref{AnForb}. The values 
can be slightly larger if more explicit calculations are carried out.}
\label{tab2}
\end{table*}

\section{Analysis}
Abundance ratios were derived from three different indicators: the forbidden line at 6300 \AA, the 
O\,I  7771--5 \AA\ triplet, and a set of five near-UV OH lines at 3100 \AA. $EW$ measurements were carried 
out for the [O\,I] line and the triplet, while spectral synthesis was performed for the OH lines.

LTE abundances for all the indicators were determined according to a standard analysis with the revised
version of the spectral synthesis code MOOG\footnote{The source code of MOOG2002 can be downloaded from
http://verdi.as.utexas.edu/moog.html.} (Sneden \cite{Sne73}) and a grid of Kurucz (\cite{Kur93}) ATLAS9
atmospheres with overshooting (as well as all other papers in these series). All the atmospheric parameters, $T_\mathrm{eff}$,$\log {g}$, [Fe/H] and $\xi_t$, and the
corresponding uncertainties, were taken from Santos et al.\ (\cite{San04a}, \cite{San05}). The adopted solar 
abundances for iron, oxygen and nickel were $\log{\epsilon}\,({\rm Fe})_{\odot}$ = 7.47\,dex (as used in 
Santos et al.\ \cite{San04a}, \cite{San05}), $\log{\epsilon}\,({\rm O})_{\odot}$ = 8.74\,dex (Nissen et al.\ 
\cite{Nis02}), and $\log{\epsilon}\,({\rm Ni})_{\odot}$ = 6.25\,dex (Anders \& Grevesse \cite{And89}), 
respectively.

\begin{figure*}
\centering 
\includegraphics[width=6.7cm]{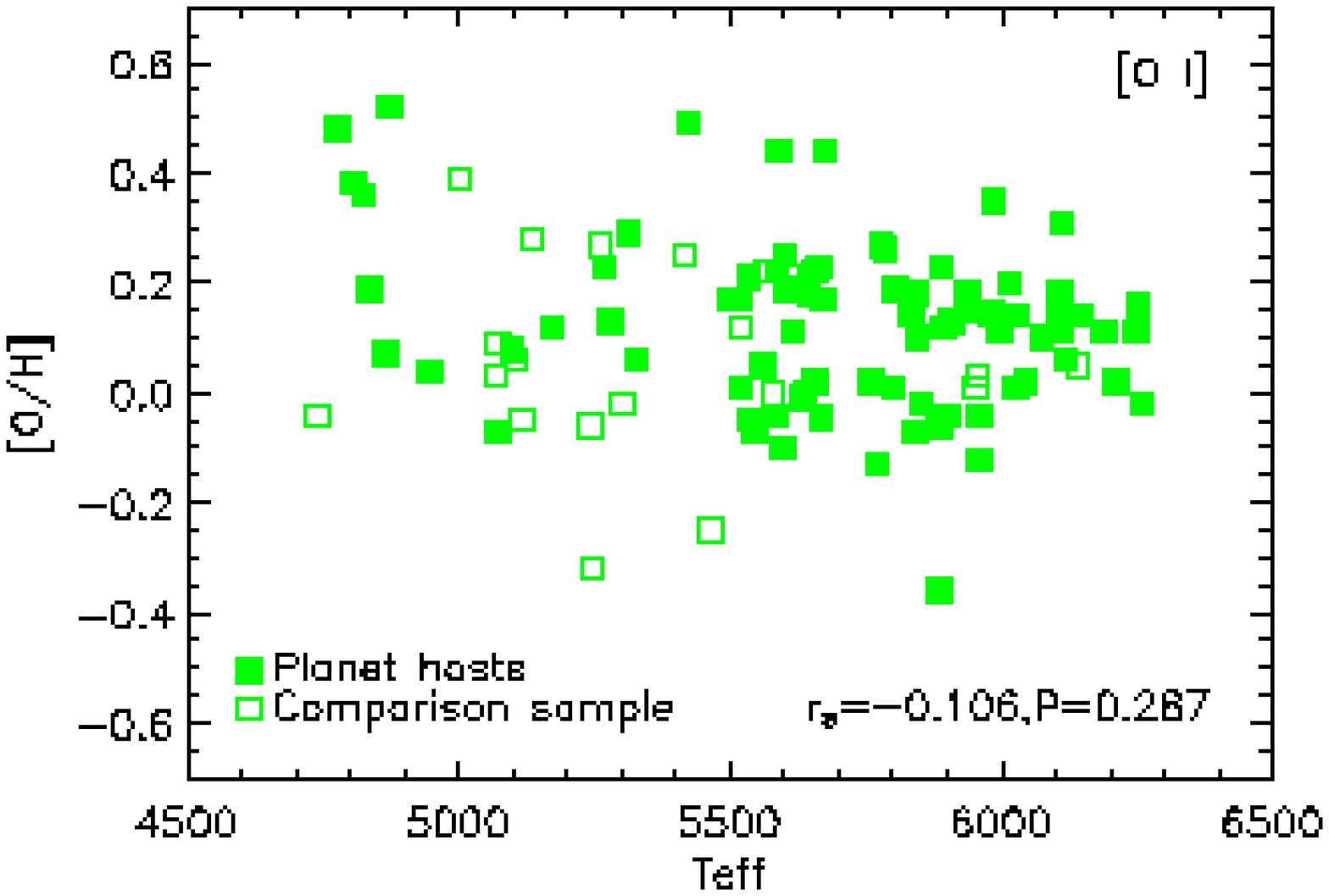}
\includegraphics[width=6.7cm]{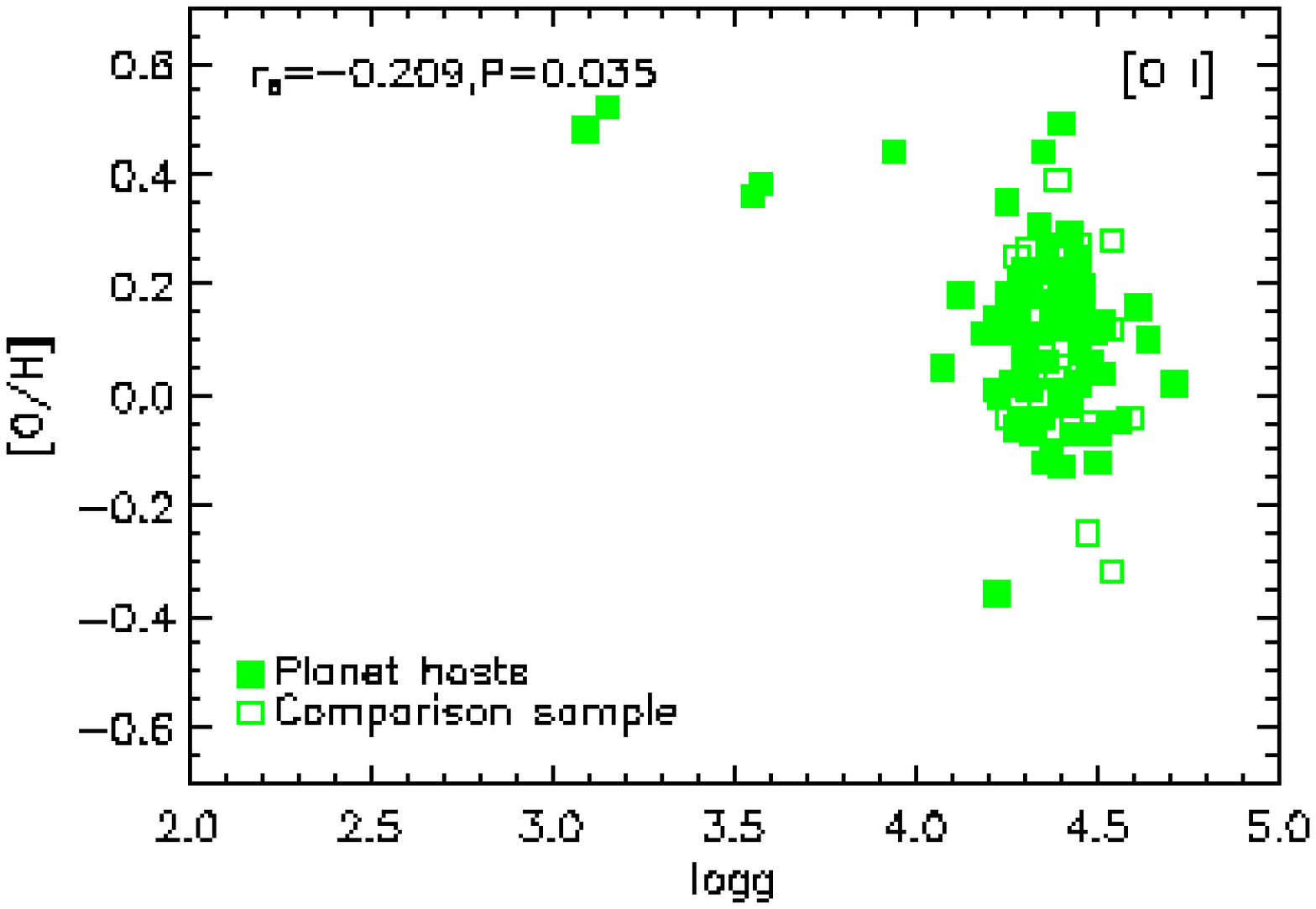}
\includegraphics[width=6.7cm]{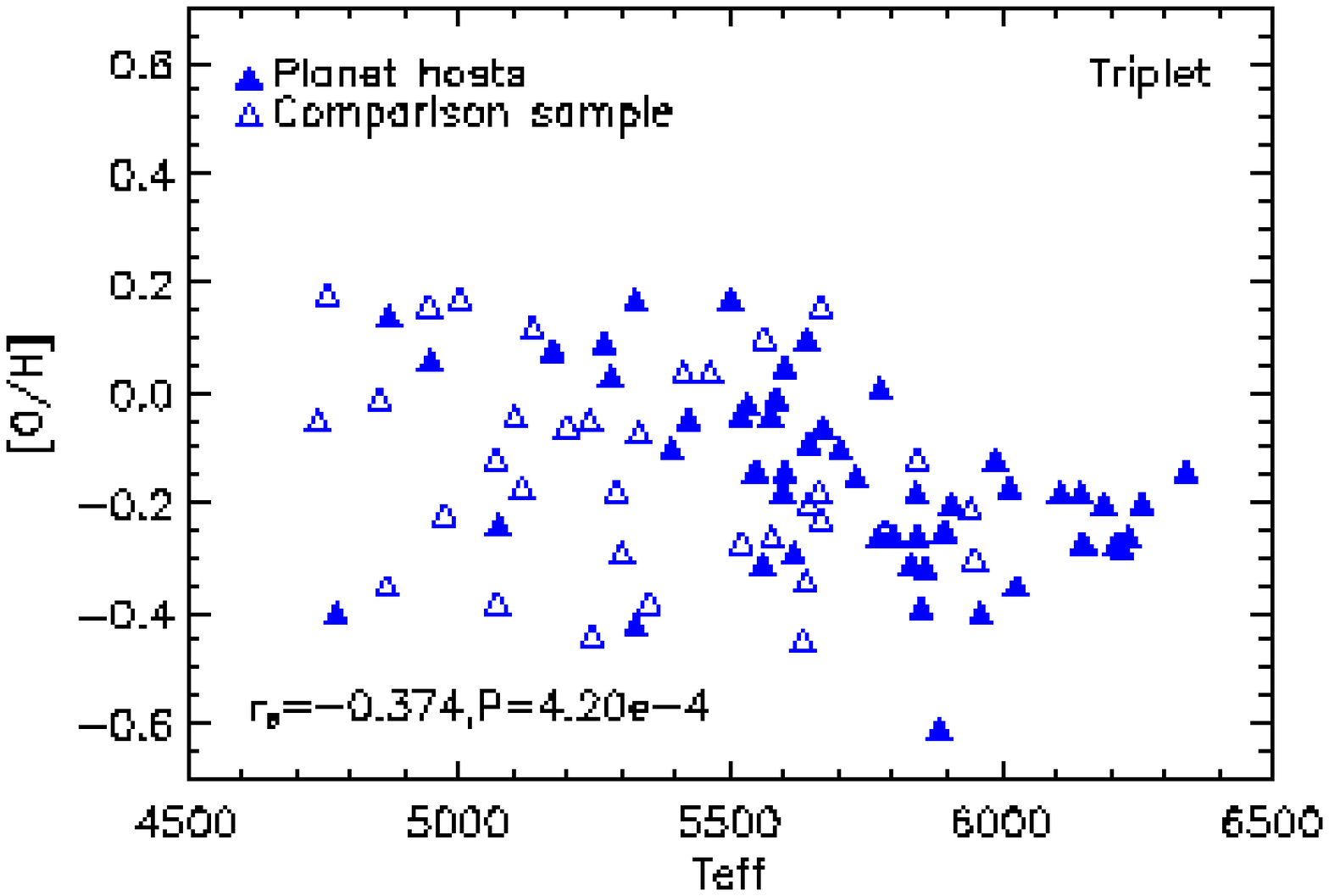}
\includegraphics[width=6.7cm]{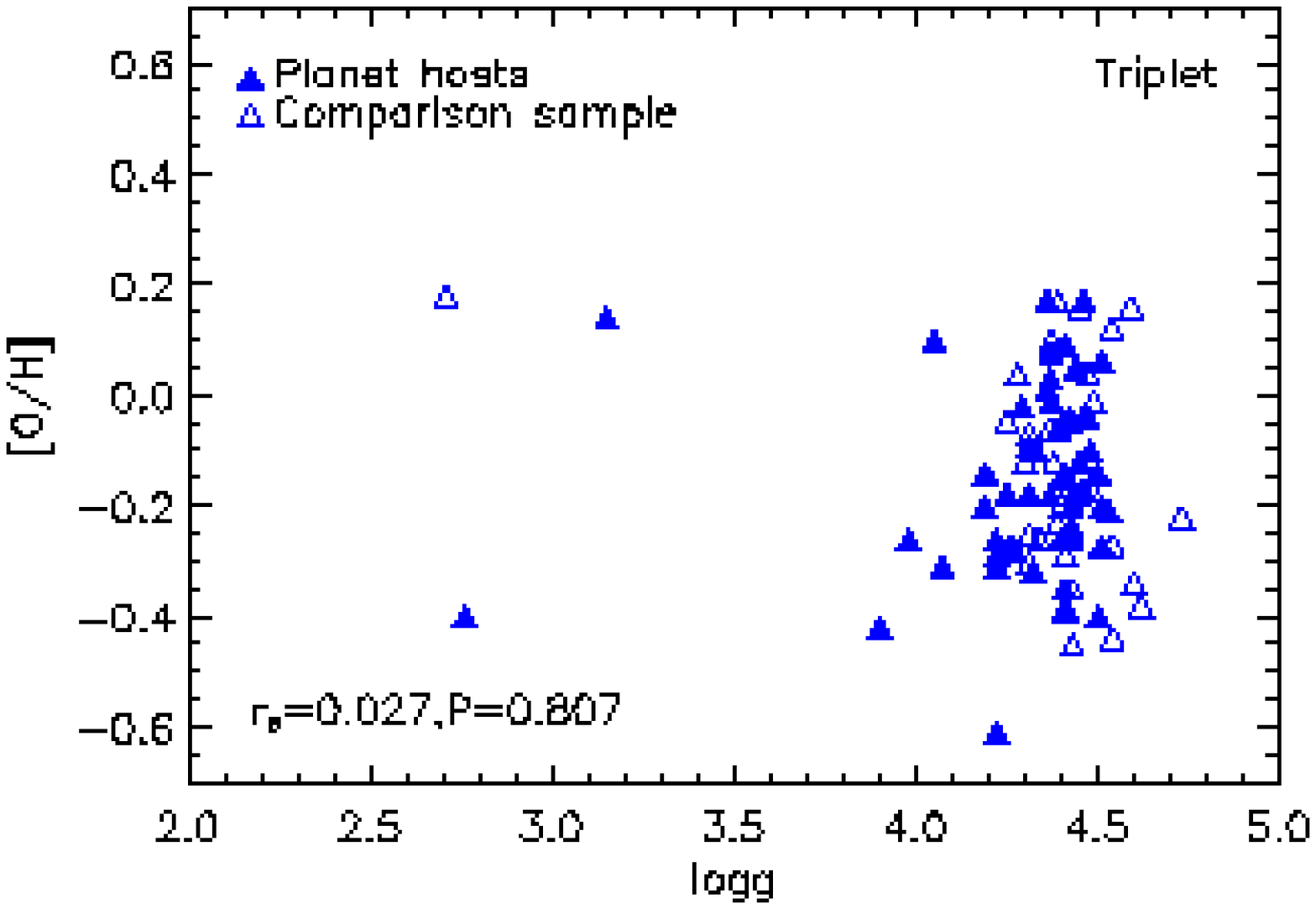}
\includegraphics[width=6.7cm]{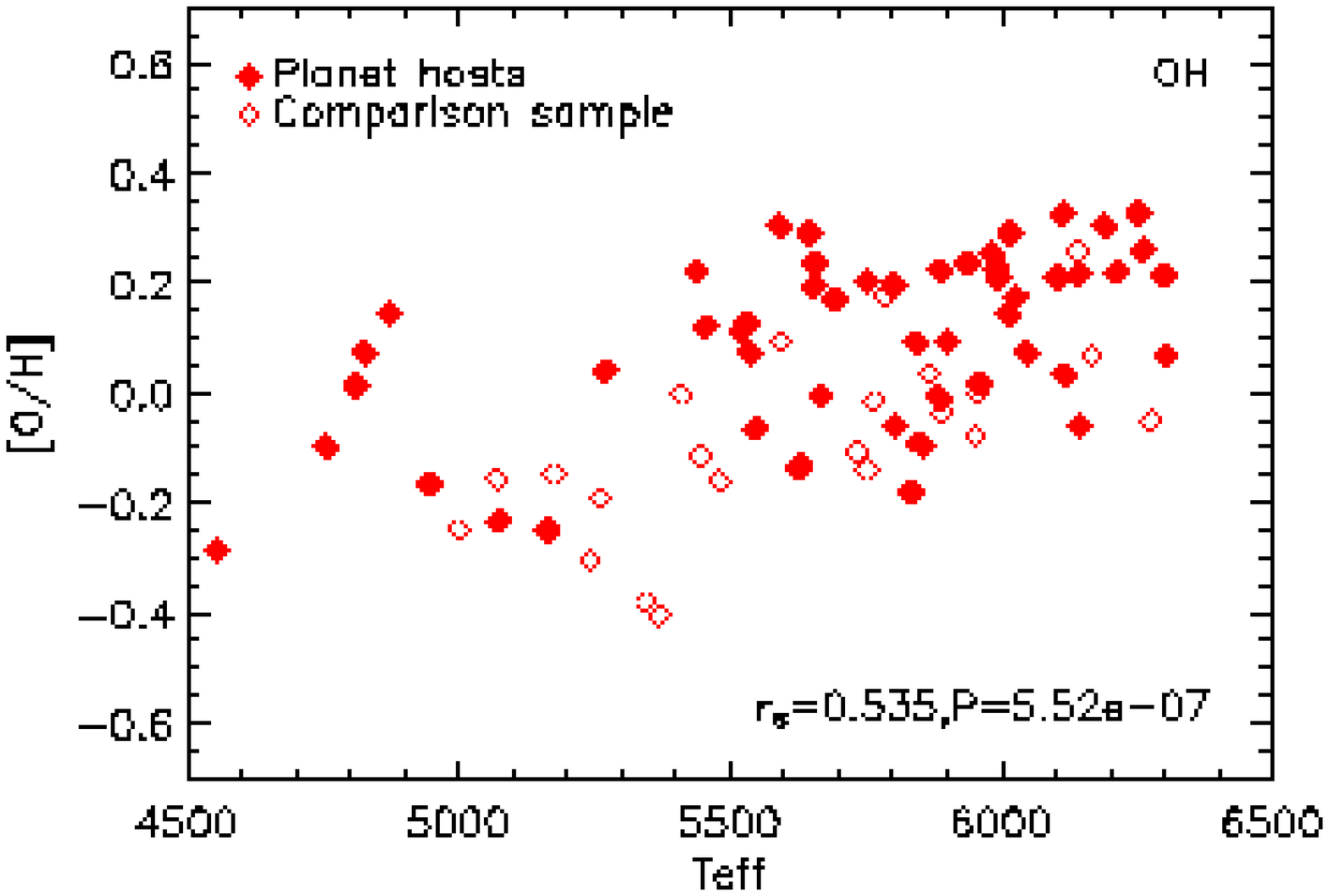}
\includegraphics[width=6.7cm]{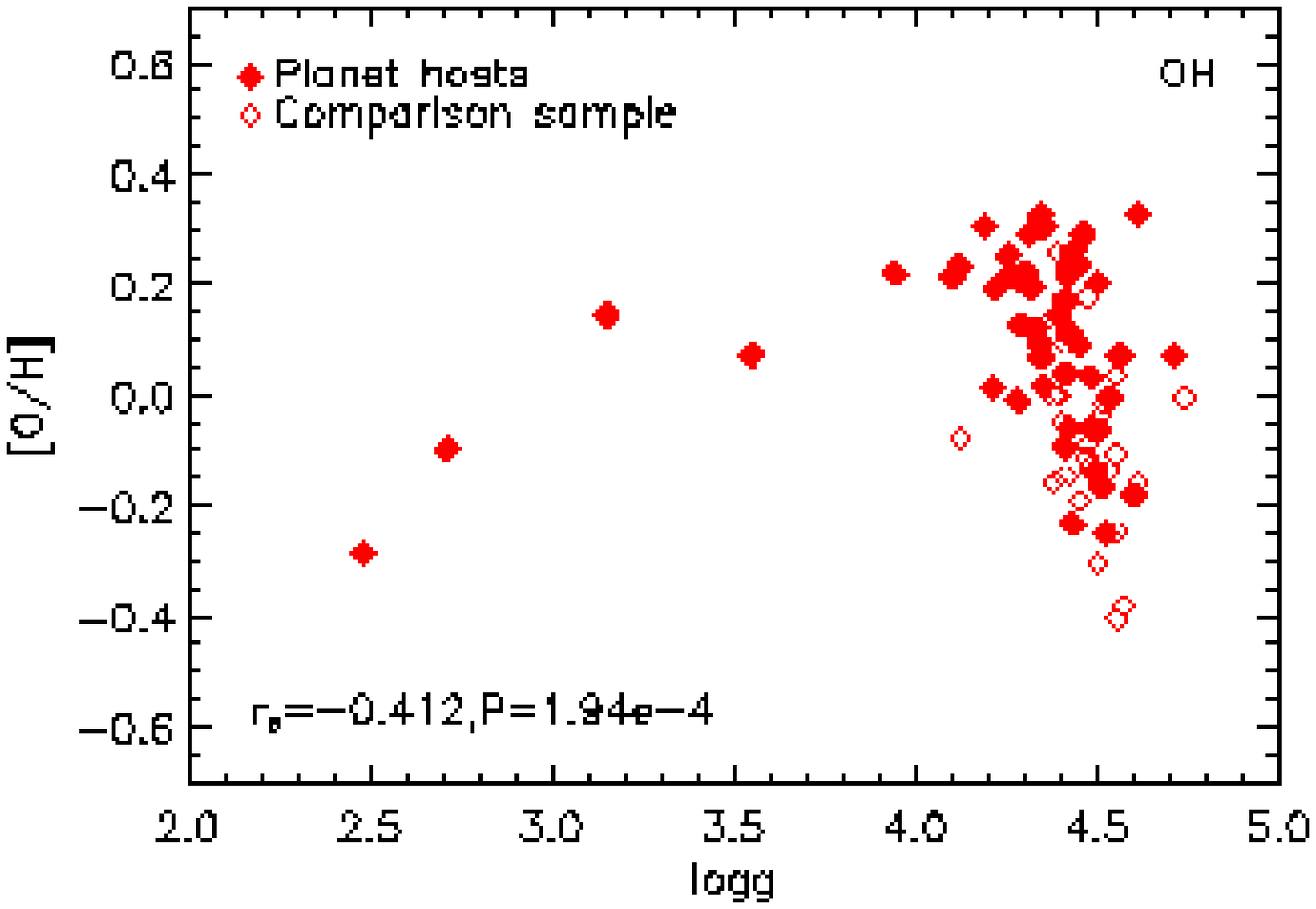}
\caption{[O/Fe] vs.\ $T_\mathrm{eff}$ and $\log {g}$ for the different indicators. Filled and open symbols
represent planet host and comparison sample stars, respectively. The Spearman rank-order correlation coefficient and its significance value are written at the bottom of each plot.}
\label{fig3}
\end{figure*}
 
\subsection{The [O\,I] absorption at 6300 \AA}
\label{AnForb}
A standard LTE analysis was carried out to derive oxygen abundances from the [O\,I] absorption at 6300.3 \AA,
since it is well known that this indicator is not significantly affected by deviations from LTE (e.g.\ Kiselman
\cite{Kis01}).
However, this line is considerably blended by an Ni\,I line at 6300.399 \AA\ (e.g.\ Lambert \cite{Lam78}; 
Allende Prieto, Lambert \& Asplund \cite{All01}).
We estimated the $EW$ of the Ni\,I line, using the {\tt ewfind} driver of MOOG (Sneden \cite{Sne73}) and Ni 
abundances computed by Jorge et al.\ (private communication, see Fig.~11). The oxygen contribution has been obtained by 
subtracting the Ni $EW$ from the whole measured $EW$ of the 6300.3 \AA\ feature.  
 The wavelengths, excitation energies of the lower levels and oscillator strengths of the Ni\,I 
absorption were taken from Allende Prieto, Lambert \& Asplund \cite{All01}), while the adopted atomic data fro [O\,I] are from Lambert (\cite{Lam78}). The $\log{gf}$ value of the [O\,I] line was slightly modified in order to obtain $\log{\epsilon}(O)_{\odot}$=8.74. All these values are listed in Table~\ref{tab1}.
Equivalent widths were determined by Gaussian fitting using the {\tt splot} task of IRAF, and abundances 
were computed with the {\tt abfind} driver of MOOG (Sneden \cite{Sne73}).

Uncertainties in the atmospheric parameters are of the order of 50\,K in $T_\mathrm{eff}$, 0.12\,dex in 
$\log {g}$, 0.08\,km\,s$^{-1}$ in the microturbulence and 0.05\,dex in the metallicity (see Santos et al.\
\cite{San04a}, \cite{San05}).
The sensitivity of the [O\,I] line to variations in atmospheric parameters has been estimated as follows.
We selected a set of three stars having different temperatures (\object{HD\,22049}, \object{HD\,37124} and  \object{HD\,9826}), and we then tested abundance sensitivity to
changes in each atmospheric parameter ($\pm$100\,K for $T_\mathrm{eff}$, $\pm$0.3\,dex for $\log {g}$ and [Fe/H],
$\pm$0.05\,dex for $\xi_t$). The results are shown in Table~\ref{tab2}. 
To take into account the uncertainties caused by the continuum determination, $EW$s for the highest and the
lowest continuum level were measured, and the corresponding abundance errors were added quadratically to the
abundance uncertainties derived from the sensitivity to changes in the atmospheric parameters.
 
\subsection{The O\,I triplet at 7771--5 \AA}
\label{AnTrip}
LTE abundances were derived from $EW$ measurements of the triplet lines. The wavelengths and excitation energies 
of the lower levels for the three triplet lines listed in Table~\ref{tab1} were taken from Kurucz \& Bell (\cite{Kur95}), while oscillator strenght values consistent with $\log{\epsilon}(O)_{\odot}$=8.74 were obtained following an inverted solar analysis in LTE.
Equivalent widths were determined by Gaussian fitting using the {\tt splot} task of IRAF, and abundances 
were computed with the {\tt abfind} driver of MOOG (Sneden \cite{Sne73}).

NLTE corrections were calculated and applied to the LTE results. The NTE computations for the oxygen 
atom were carried out using the atomic model with 23 levels of O\,I and one level of O\,II. Our atomic model is based on the data of Carlsson \& Judge (\cite{Car93}). In our computations only 31 bound--bound and 23 bound--free radiative 
transitions were considered; nevertheless, the consideration of additional levels and transitions does not 
affect our results (Shchukina \cite{Shc87}; Takeda \cite{Tak03}). It is well known that inelastic collisions 
with hydrogen atoms tend to offset the NLTE effects. However, it is often stated that Drawin's formalism 
(Drawin \cite{Dra68}) gives very uncertain results for hydrogen collision rates (e.g.\ Belyaev et al.\ 
\cite{Bel99}). Thus, collisions with H atoms were not taken into account in our computations.
Figure~\ref{fig1} shows the dependence of the NLTE corrections on $T_\mathrm{eff}$, $\log {g}$ and [Fe/H]. Our results are similar to those reported recently by Takeda (\cite{Tak03}). 

The sensitivity of the oxygen abundances derived from triplet lines to variations in atmospheric parameters
has been estimated in the same way as in the [O\,I] case (see Sect.~\ref{AnForb}). The results are shown in
Table~\ref{tab2}. Uncertainties in the final oxygen abundances were determined adding in quadrature the
abundance uncertainty resulting from the continuum determination (0.05\,dex), the standard deviation of each 
mean abundance and the errors due to the abundance sensivities to changes in the atmospheric parameters.

\subsection{Synthesis of the near-UV OH lines}
We determined oxygen abundances by fitting synthetic spectra to the data. Four OH features were analysed at 
3167.2 \AA, at 3189.3 \AA, at 3255.5 \AA, and at 3173 \AA, composed of two OH lines, at 3172.9 \AA\ and at
3173.2 \AA. The atomic line lists for each spectral region were taken from VALD (Kupka et al.\ \cite{Kup99}),
while the molecular data of the OH lines were extracted from Kurucz database.\footnote{Molecular line data
can be downloaded at http://kurucz.harvard.edu/LINELISTS/LINESMOL.} We assumed the 
dissociation potential of the OH molecule $D_0(\rm OH)$= 4.39\,eV (Huber \& Herzberg \cite{Hub79}). Oscillator strength values were slightly modified in order to achieve a good 
fit to the Kurucz Solar Atlas (Kurucz et al.\ \cite{Kur84}), with a solar model having $T_\mathrm{eff}$= 
5777\,K, $\log {g}$= 4.44\,dex, and  $\xi_t$= 1.0\,km\,s$^{-1}$. The adopted data are listed in 
Table~\ref{tab1}. 

The continuum was normalized with 5th order polynomials using the CONT task of IRAF. We then made
further improvements in the placement of the continuum using the DIPSO task of the STARLINK software: 
some points of reference of the continuum level were selected in the Kurucz Solar Flux Atlas (Kurucz et al.\
\cite{Kur84}) and used in the determination of the stellar continuum of our observed spectra.
For the instrumental broadening we used a Gaussian function with $FWHM$ of 0.05 \AA\ and a rotational 
broadening function with $v\sin{i}$ values from CORALIE database. All our targets are slow rotators, with 
$v\sin{i}$ values between 1 and 5 km\,s$^{-1}$ in almost all cases. No macroturbolence broadening was used. 
Two examples of the fitting of the four features are shown in Figure~\ref{fig2}.

The sensitivity of the oxygen abundances from OH lines to changes in the atmospheric parameters was
estimated in the same way as for the [O\,I] and O\,I indicators (see Sect.~\ref{AnForb} and 
Sect.~\ref{AnTrip}). Uncertainties derived from inaccuracies in atmospheric parameters were added in 
quadrature to the abundance uncertainty resulting from the continuum determination (0.05\,dex) and to the 
standard deviation of each mean abundance.\\

The dependence on $T_\mathrm{eff}$ and on $\log {g}$ of the [O/H] results from all the indicators is 
represented in Figure~\ref{fig3}. We note that no significant trends appear for [O I] and OH. This means
that our results are almost free from systematic errors. Only in the case of triplet, does a trend of
decreasing [O/H] with increasing $T_\mathrm{eff}$ exist. This is probably due to the high dependence on 
$T_\mathrm{eff}$ of the NLTE corrections applied to the LTE results of the triplet.
  
\section{Comparison between different indicators}
\begin{figure}[!]
\centering 
\includegraphics[width=6.2cm]{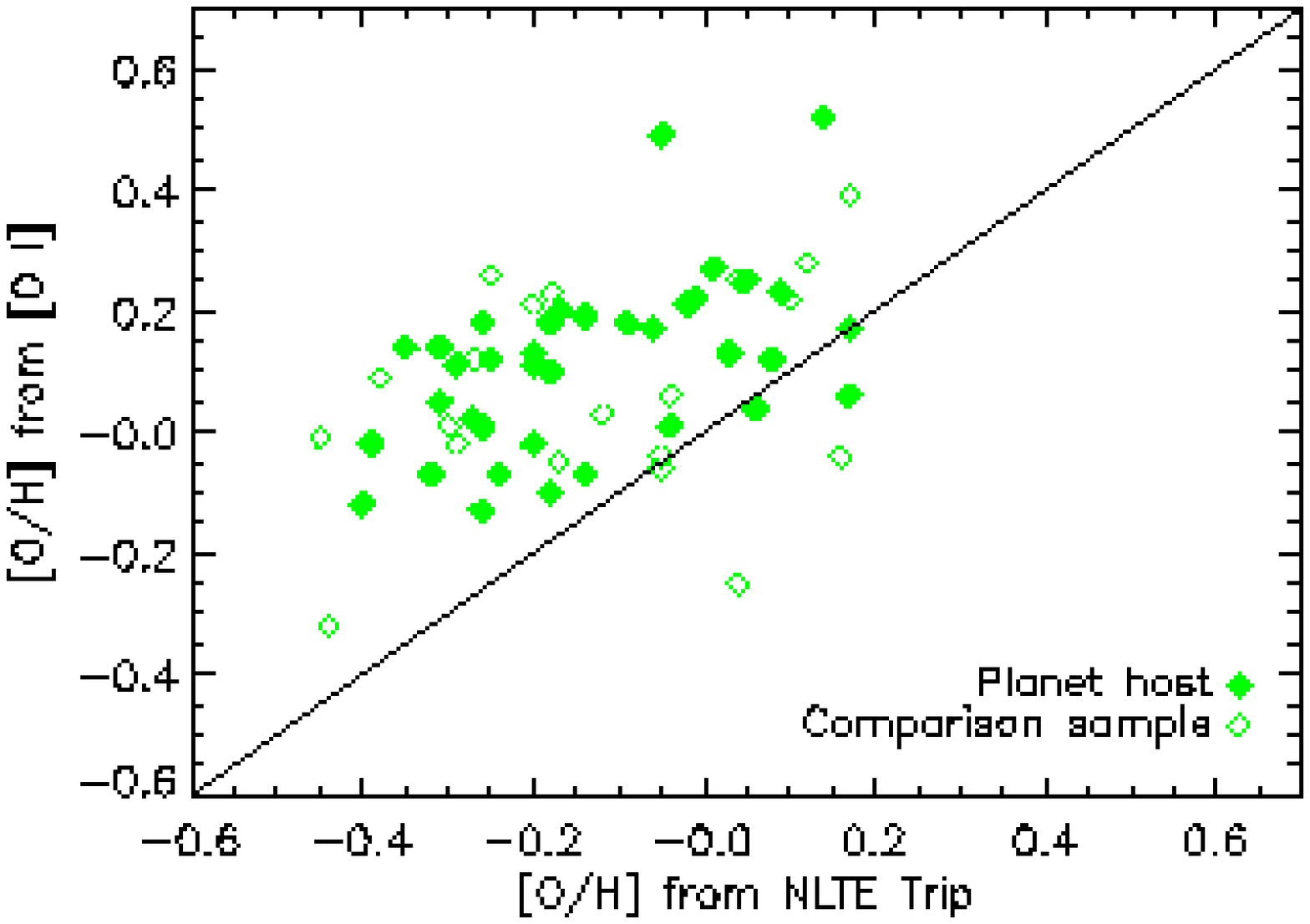}
\includegraphics[width=6.2cm]{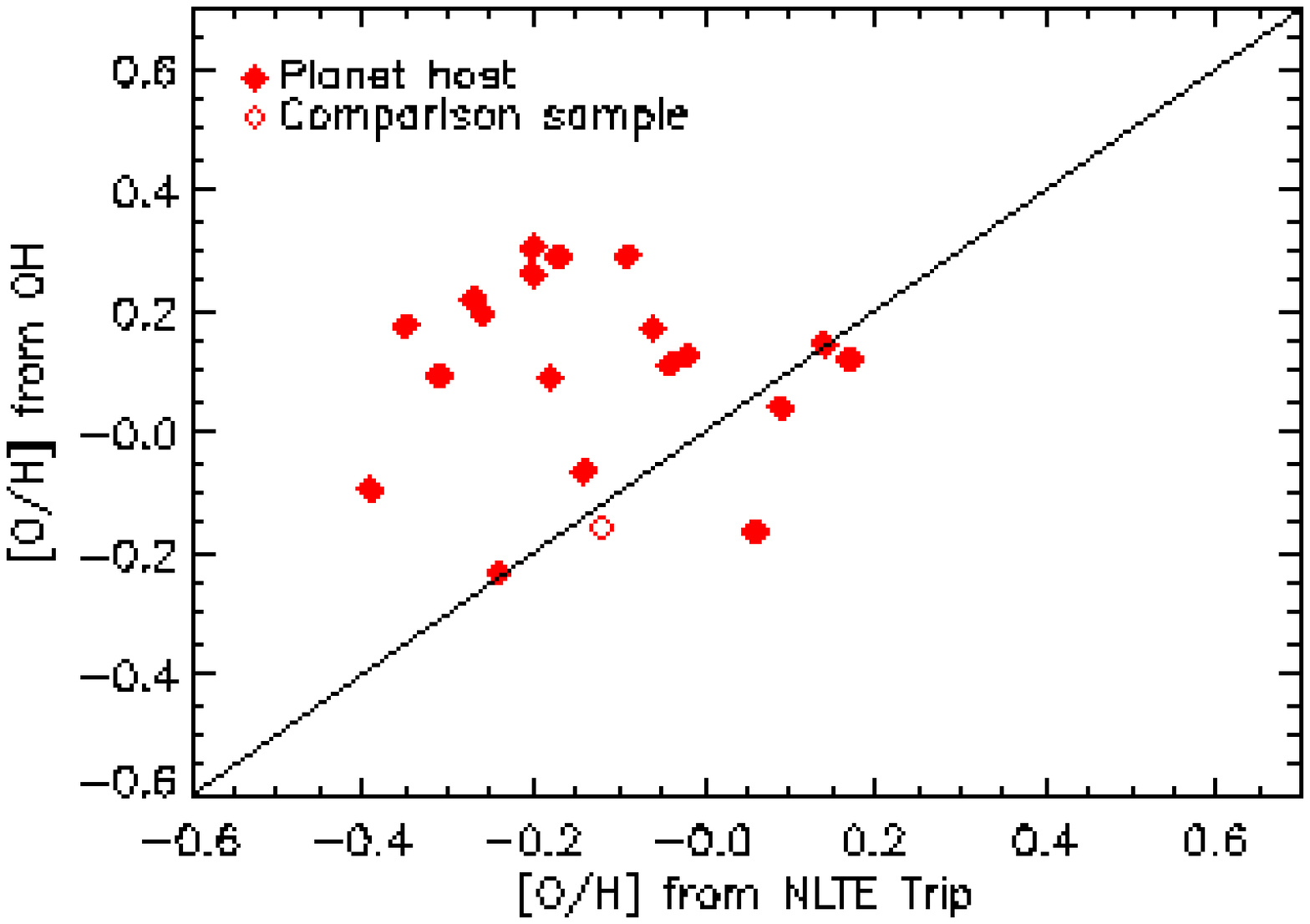}
\includegraphics[width=6.2cm]{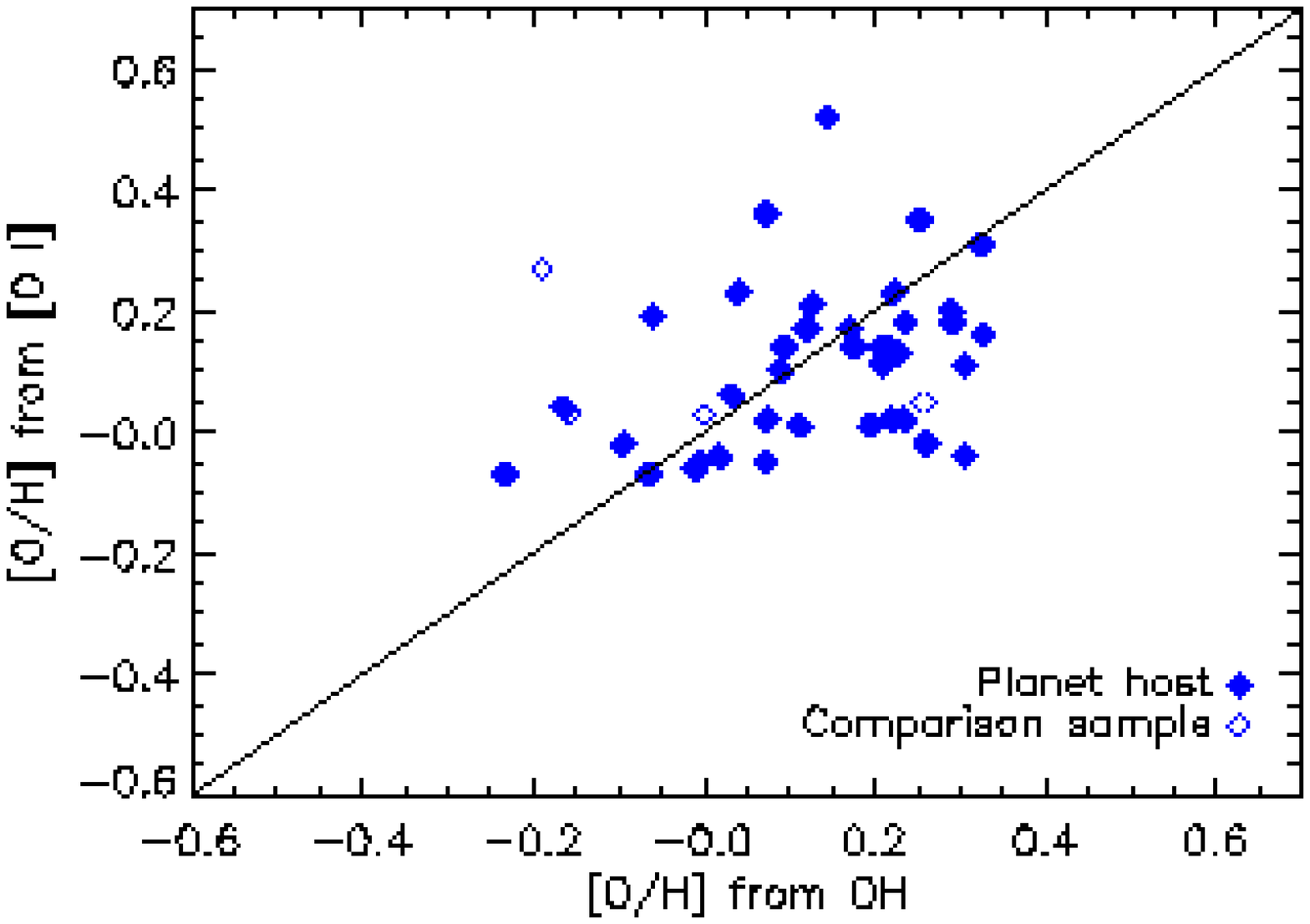}
\caption{Comparisons of the results from different indicators: [O\,I] line, OH lines and triplet lines in
NLTE.}
\label{fig4}
\end{figure}
\begin{figure}[!]
\centering 
\includegraphics[width=6.2cm]{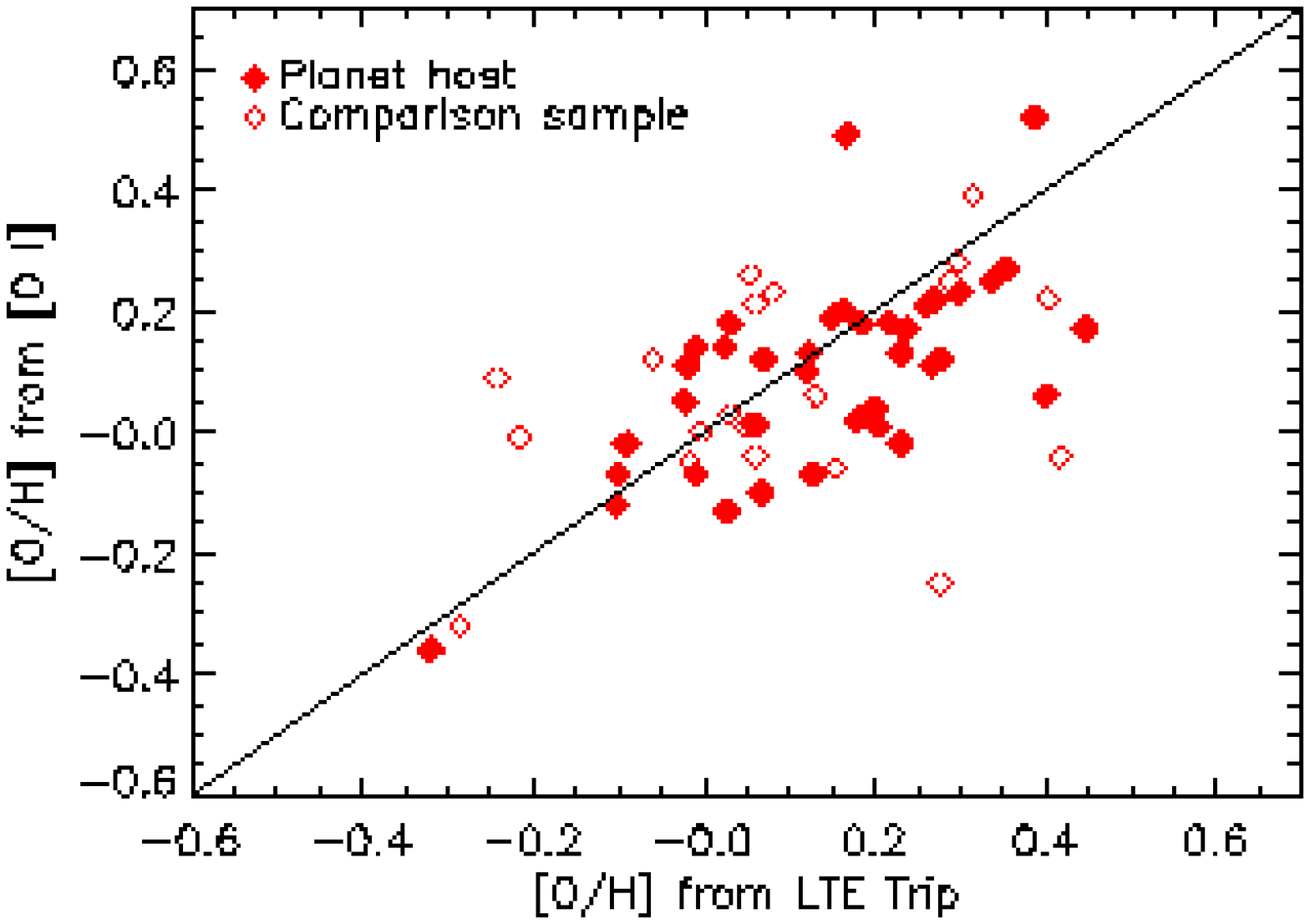}
\includegraphics[width=6.2cm]{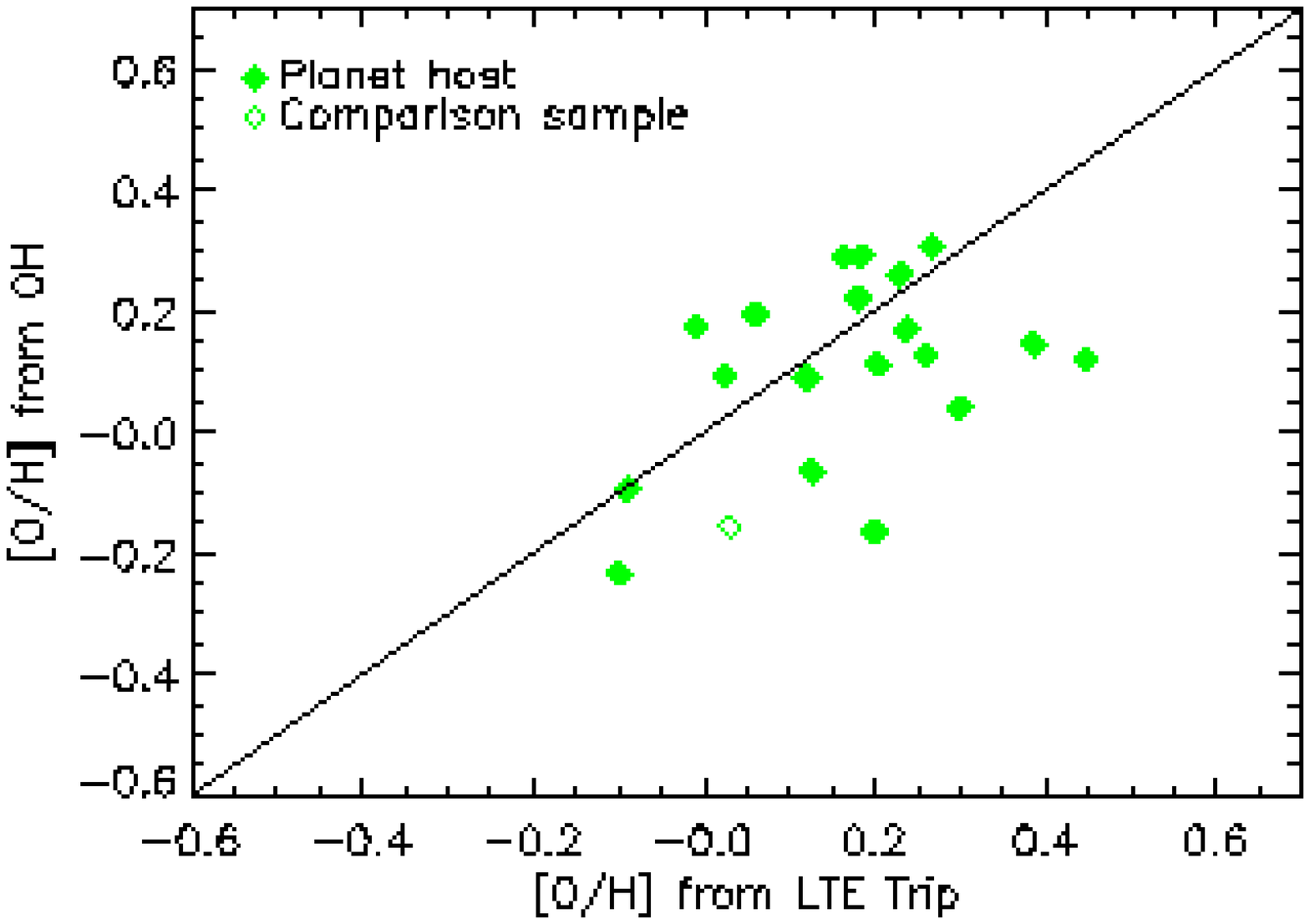}
\caption{Comparisons of the results from [O\,I] and OH lines with those obtained from triplet lines in LTE.}
\label{fig5}
\end{figure}
Abundances from the [O\,I] line at 6300 \AA\ were obtained in 103 dwarfs (see Tables~\ref{longtab4} and~\ref{longtab5}),
 while 77 stars were analysed using near-UV OH lines (see Table~\ref{longtab6} and~\ref{longtab7}). 
LTE abundances were derived from the triplet lines at 7771--5 \AA\ and then corrected for NLTE effects for 
87 stars (see Tables~\ref{longtab8} and~\ref{longtab9}). Altogether we present oxygen abundances derived from 
these three indicators for 96 and 59 stars with and without known planets, respectively.

\subsection{Triplet vs.\ [O\,I] and near-UV OH}
\label{Trip-Forb}
Figure~\ref{fig4} (top panel) shows the comparison between the results obtained from the NLTE study of the 
7771--5 \AA\ triplet and the analysis of the [O\,I] 6300 \AA\ line. NLTE triplet abundances are systematically
lower with respect to the forbidden-line results, with discrepancies of the order of 0.3\,dex on average. 

The values obtained from the synthesis of OH lines present a better consistency with the NLTE triplet 
results (see Figure~\ref{fig4}, middle panel). In this case, the comparison is less meaningful because of 
the limited number of targets in common between the two analyses. Nevertheless, a large portion of targets 
show underabundances of the order of 0.2\,dex in NLTE triplet values with respect to near-UV OH 
results. 

The NLTE corrections applied to the LTE analysis results correspond to the 
maximum effect, since collisions with H atoms are not taken into account (see Sect.~\ref{AnTrip}).This can 
produce an underestimation of the final triplet abundances, and could be the reason for the systematic 
underabundance of the NLTE triplet results.

If we compare results from LTE triplet analysis with those from [O\,I] 6300 \AA\ and OH (see
Figure~\ref{fig5}), the consistency with these indicators improves, with typical
discrepancies of 0.1\,dex. Moreover, a suggestive number of targets shows an overabundance of the order of 
0.2\,dex in LTE triplet values. This means that the NLTE and LTE analyses give lower and upper limits, respectively, for the oxygen abundance.   

\subsection{[O\,I] vs.\ near-UV OH} 
\begin{figure}
\centering 
\includegraphics[width=6.7cm]{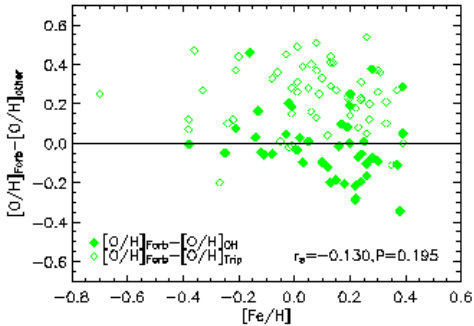}
\caption{Difference between [O/H] ratios derived from [O\,I] and from another indicator, OH (filled symbols) or triplet (open symbols), vs. [Fe/H]. The Spearman rank-order correlation coefficient and its significance value are written at the bottom of the plot.}
\label{Forbcomp}
\end{figure}
Figure~\ref{fig4} (bottom panel) represents the comparison between oxygen abundances obtained from the 
analysis of the [O\,I] 6300 \AA\ line and from the synthesis of near-UV OH lines. In most cases the two 
indicators agree quite well, with discrepancies of the order of 0.1\,dex. Abundances obtained  from OH lines
are generally lower than those from the 6300 \AA\ [O\,I] line. 

The cause of this behaviour could be an underestimation of the Ni\,I component blended with the 6300 \AA\ 
[O\,I] line. The atomic parameters, especially the $\log{gf}$, for this Ni\,I line are uncertain (e.g.\ 
Allende Prieto, Lambert \& Asplund \cite{All01}), and this can introduce uncertainities in the estimation of 
the Ni\,I line contribution to the total spectral feature at 6300.3 \AA. Moreover, dwarfs have very weak 
[O\,I] lines (e.g.\ $EW\sim$5\,m\AA\ in the Sun). Thus, measurement uncertainties may be very large, depending on the resolution and S/N of the data. 

If the uncertainties related to Ni\,I were the main responsible of the discrepancies with the other two indicators, a correlation should exist between these discrepancies and metallicity. Figure~\ref{Forbcomp} shows the difference of abundances from [O\,I] and from other indicators as a function of [Fe/H], as well as the correlation coefficient and its significance value. Since no correlation with metallicity exists, we can discard the possibility that the Ni\,I blend introduces significant errors into our [O\,I] results.      

Another source of uncertainty could be the strong dependence of the OH lines on temperature, and therefore
to surface inhomogeneities. However, previous works (e.g.\ Israelian et al.\ \cite{Isr98}; Boesgaard et al.\ \cite{Boe99}; Ecuvillon et al.\ 
\cite{Ecu04a}) obtained a good consistence between abundances based on molecular and atomic features 
using classical 1D atmosphere models. This fact, added to the agreement we have found between [O\,I] and
OH measures, makes us confident about the reliability of our OH results. We found that 
discrepancies between OH, O\,I triplet and [O\,I] barely exceed 0.2\,dex.

\section{Comparison between planet host and comparison sample stars}
Several studies have been published about abundances of metals other than iron in planet host stars
(Santos et al.\ \cite{San00}; Gonzalez et al.\ \cite{Gon01}; Takeda et al.\ \cite{Tak01}; Sadakane et al.\ 
\cite{Sad02}; Bodaghee et al.\ \cite{Bod03}; Ecuvillon et al.\ \cite{Ecu04a}, \cite{Ecu04b}). Oxygen 
abundances have been analysed by some of them (Santos et al.\ \cite{San00}; Gonzalez et al.\ \cite{Gon01}; 
Takeda et al.\ \cite{Tak01}; Sadakane et al.\ \cite{Sad02}; Takeda \& Honda \cite{Tak05}). However the limited 
number of planet host stars considered, and the comparison realized by some authors between targets with 
planets and field stars extracted from  the literature, have prevented definitive conclusions from being reached.

We have carried out a homogeneous study of oxygen abundances in an almost complete set of 
96 stars with extrasolar giant planets, as well as in a large volume-limited sample of 59 stars with no 
known planetary-mass companion, all belonging to the CORALIE planet search survey (see Udry et al.\
\cite{Udr00}). The volume-limited comparison sample consists of stars from the CORALIE southern planet search sample 
without any known planets, with distances below 20\,pc, as derived from {\itshape Hipparcos} parallaxes (ESA 1997). 
All these stars and their parameters come from Santos et al.\ (\cite{San01}, \cite{San03b}, \cite{San04a}, \cite{San05}).  
Three different indicators were used in order to obtain more reliable and solid results.

\subsection{[O/H] distributions}
\begin{figure*}
\centering 
\includegraphics[width=6.7cm]{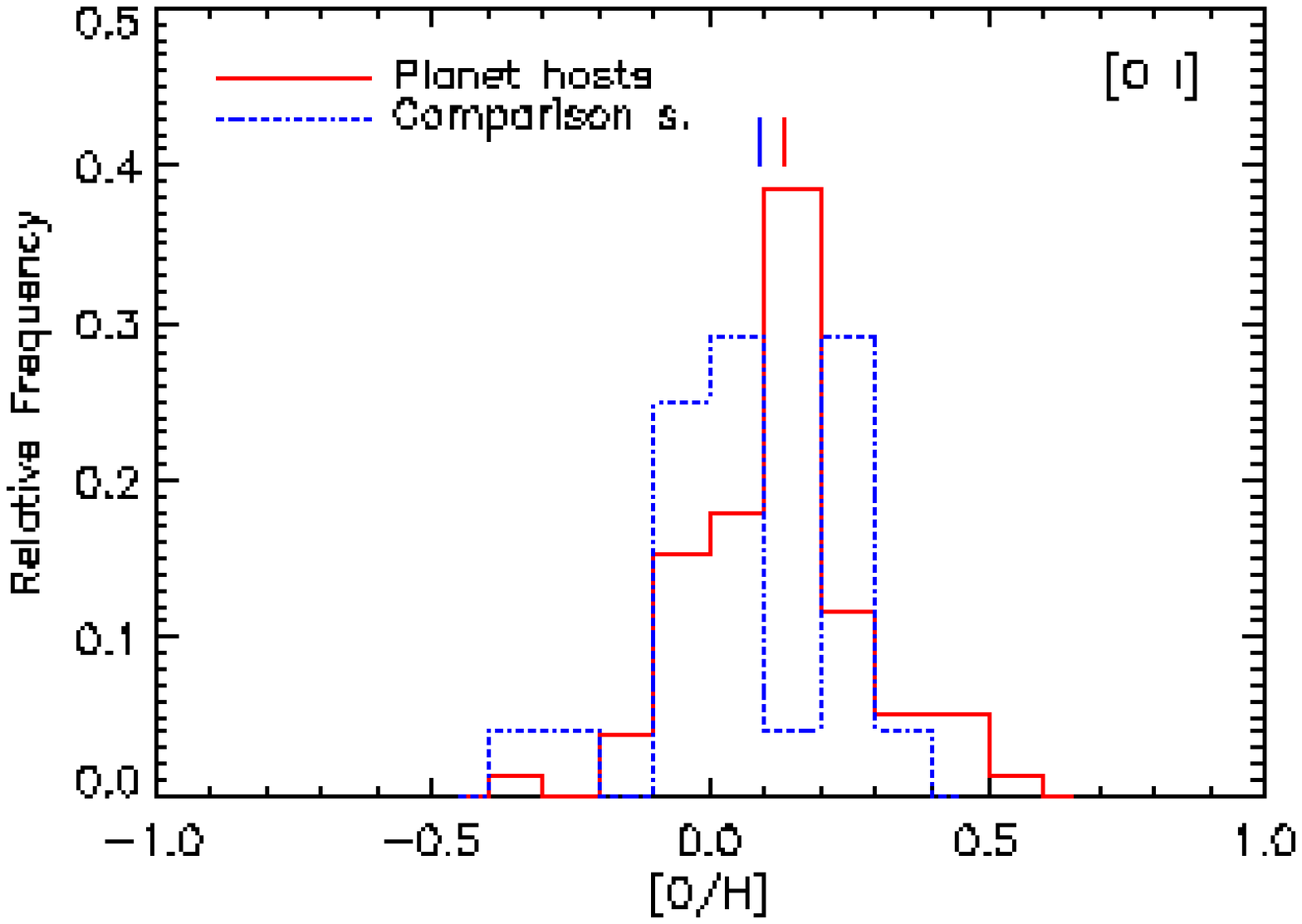}
\includegraphics[width=6.7cm]{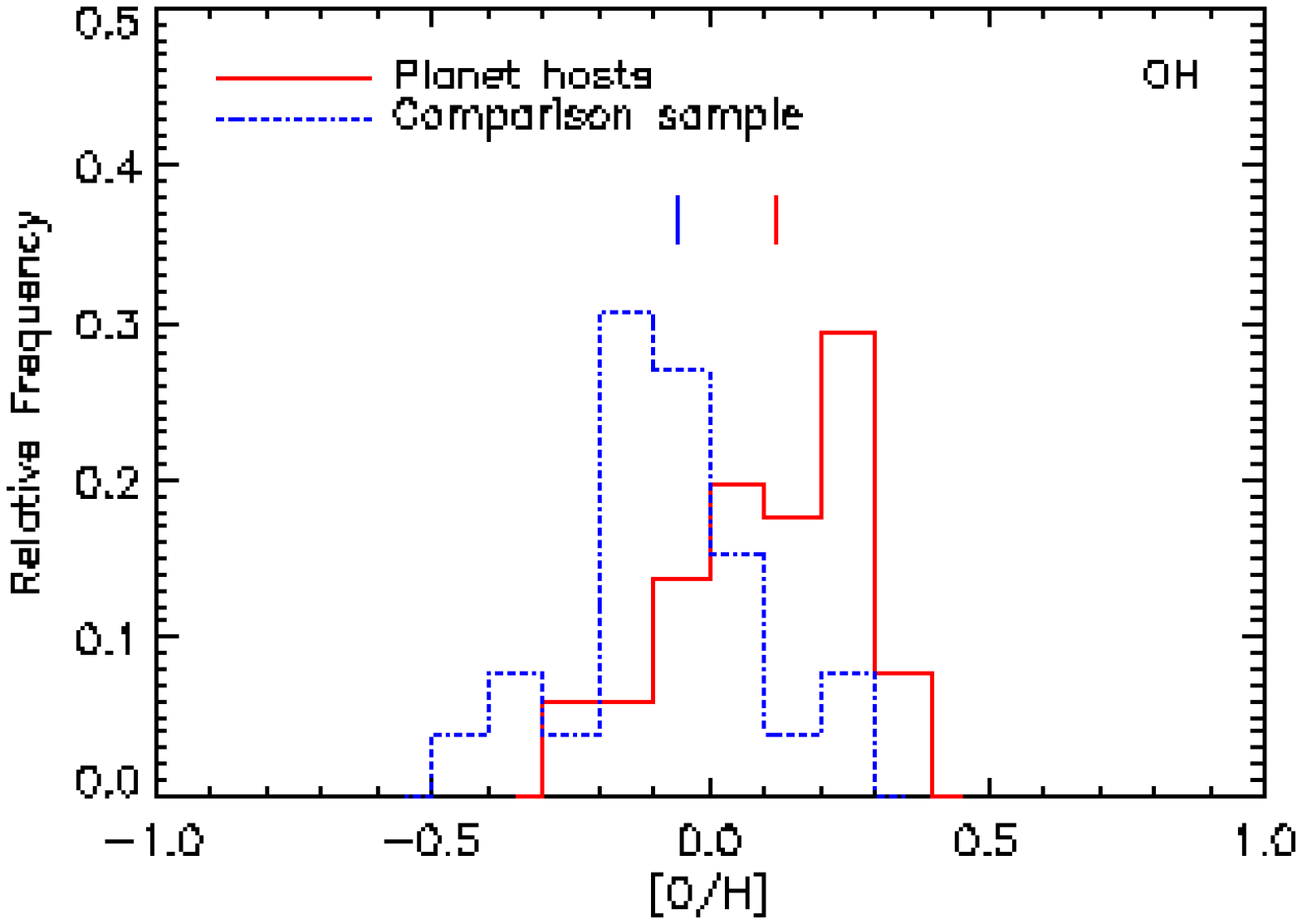}
\includegraphics[width=6.7cm]{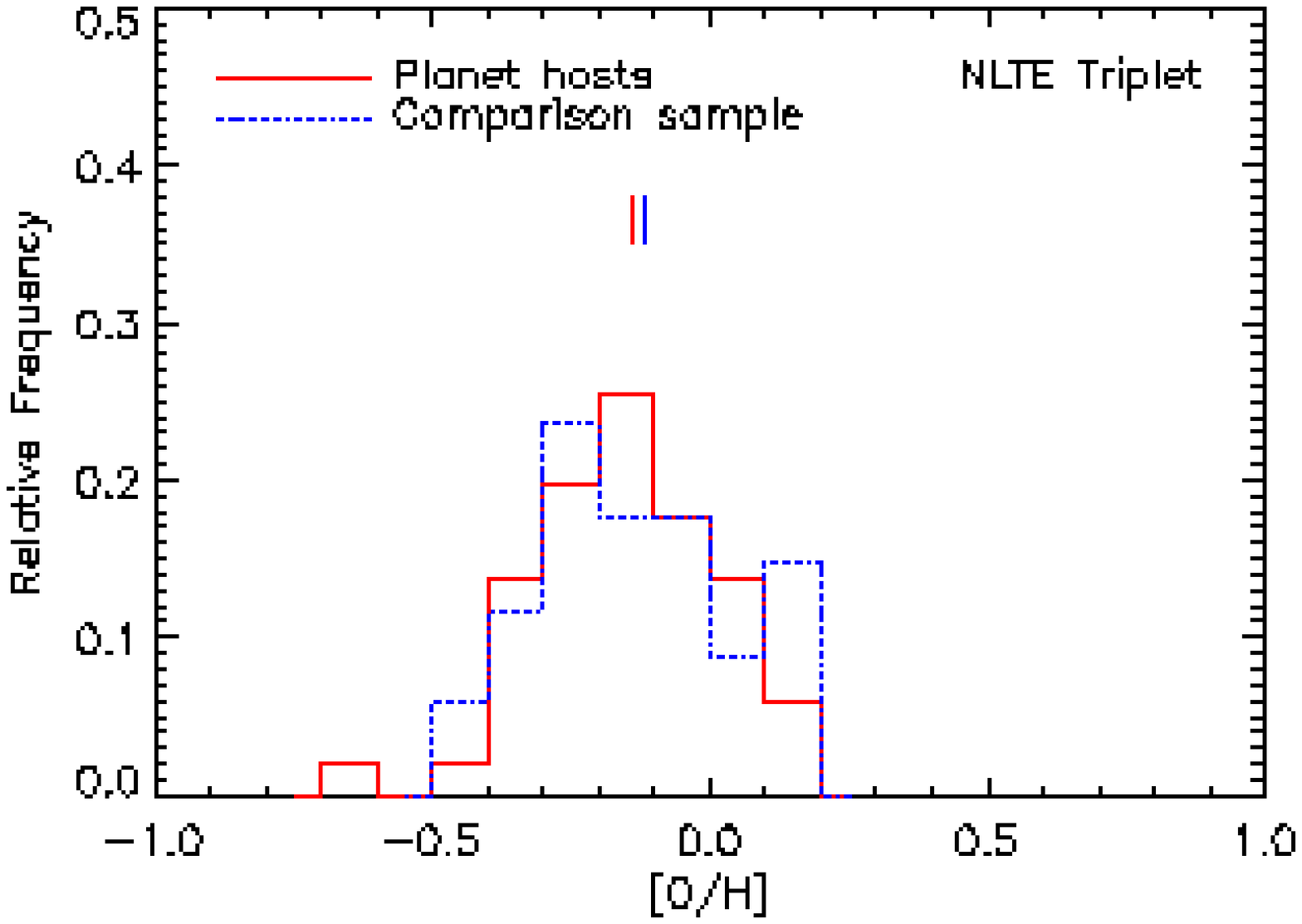}
\includegraphics[width=6.7cm]{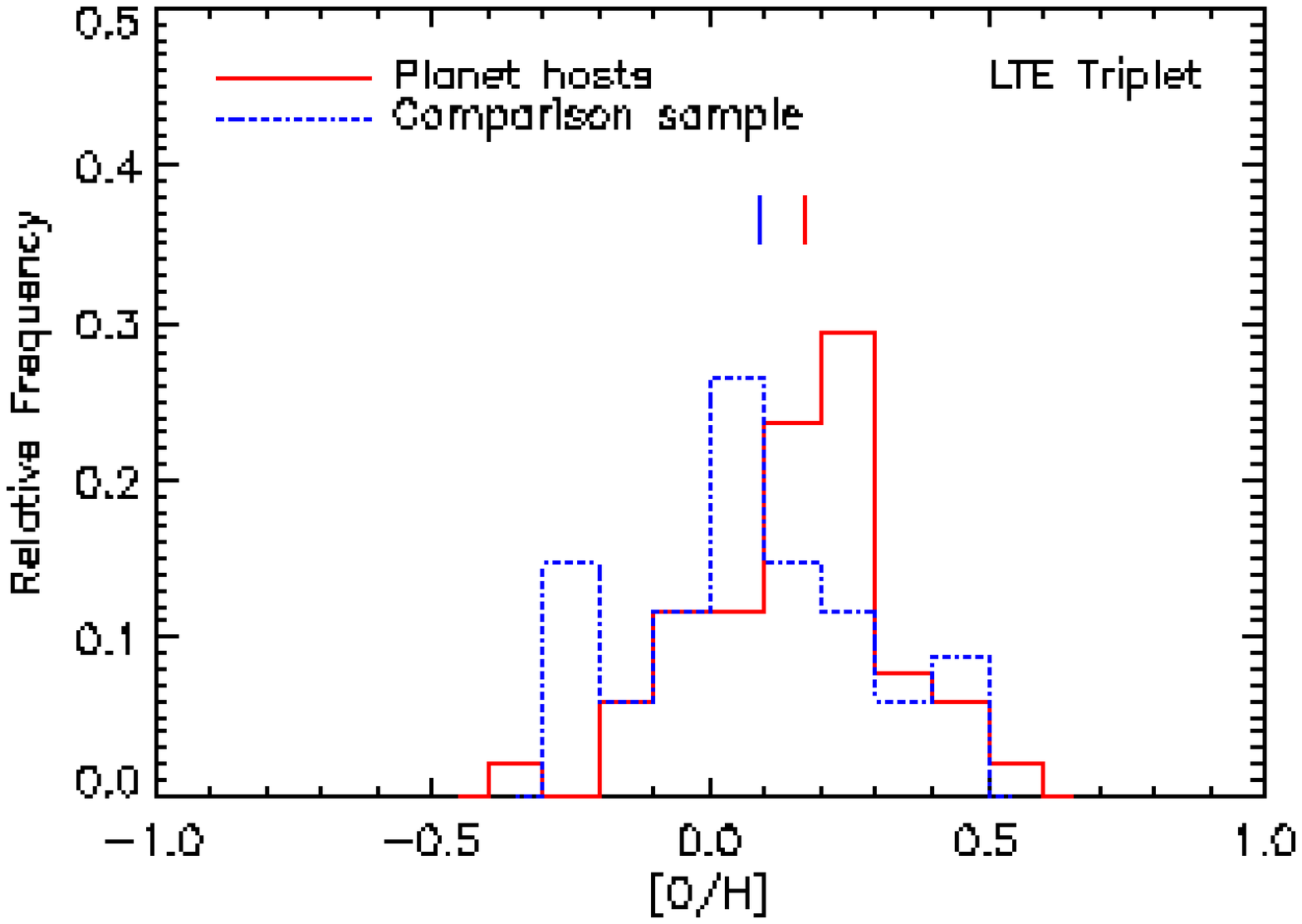}
\caption{[O/H] distributions from different indicators. The solid and dotted lines represent planet host and
comparison sample stars, respectively. The vertical lines represent the average abundance values of the two
samples, stars with and without planets.}
\label{fig6}
\end{figure*}

\begin{figure*}
\centering 
\includegraphics[width=6.7cm]{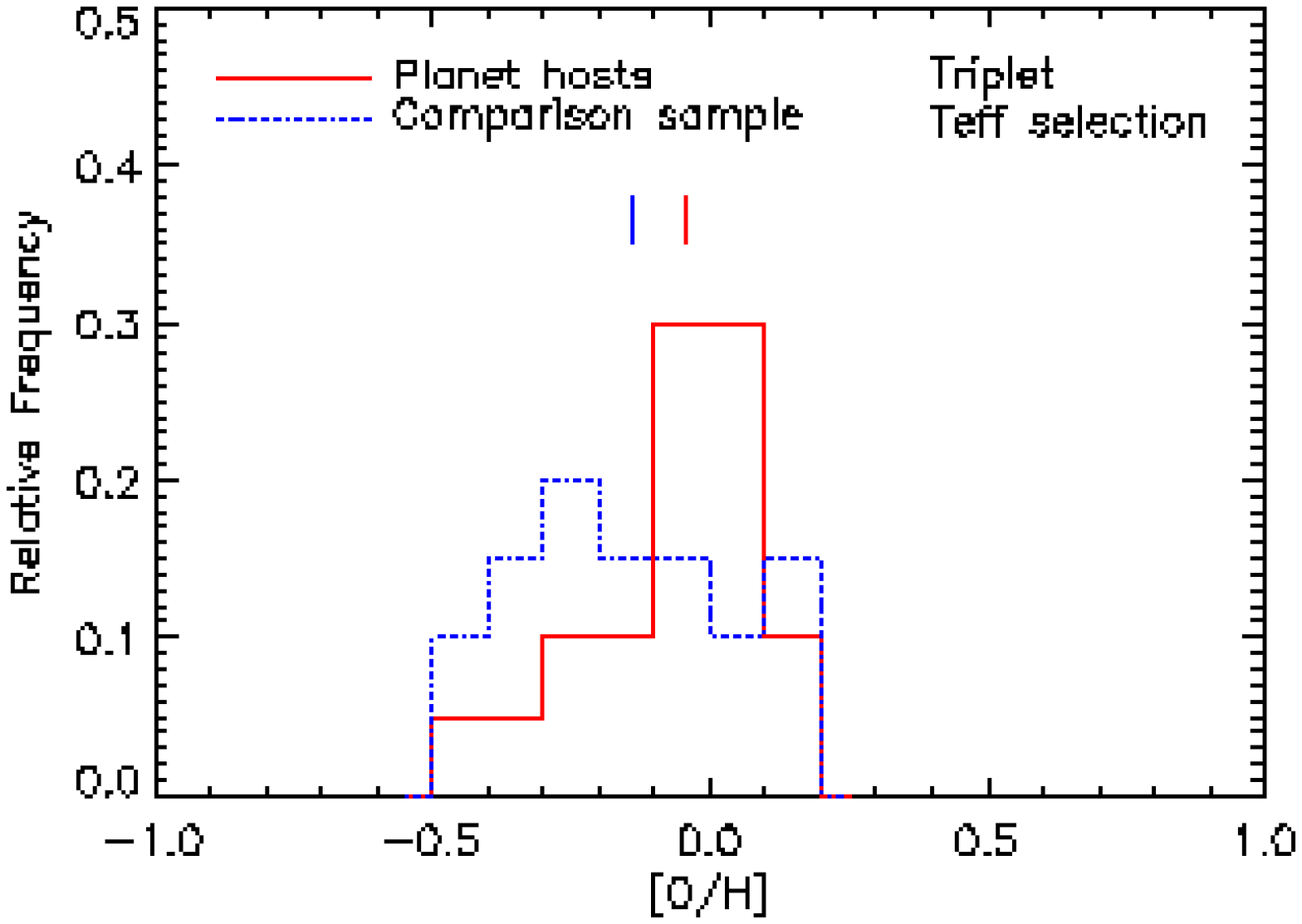}
\includegraphics[width=6.7cm]{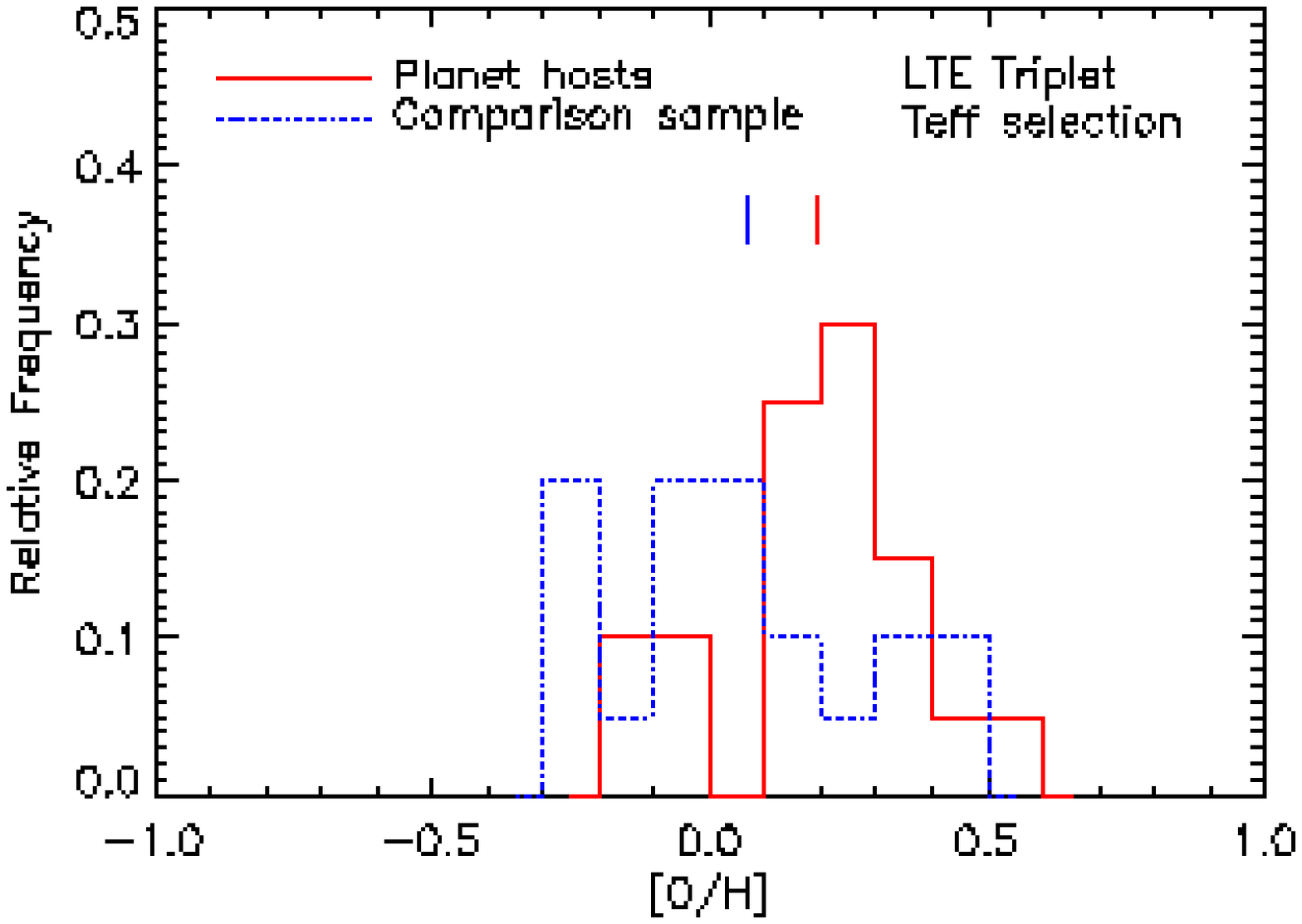}
\caption{[O/H] distributions for NLTE (left panel) and LTE (right panel) triplet results for the two subsamples of planet host (solid line) and comparison sample (dotted line) stars with the {\it same} $T_{\rm eff}$ distributions. The vertical lines represent the average abundance values of the two subsamples, stars with and without planets.}
\label{OHsel}
\end{figure*}

\begin{figure*}
\centering 
\includegraphics[width=6.7cm]{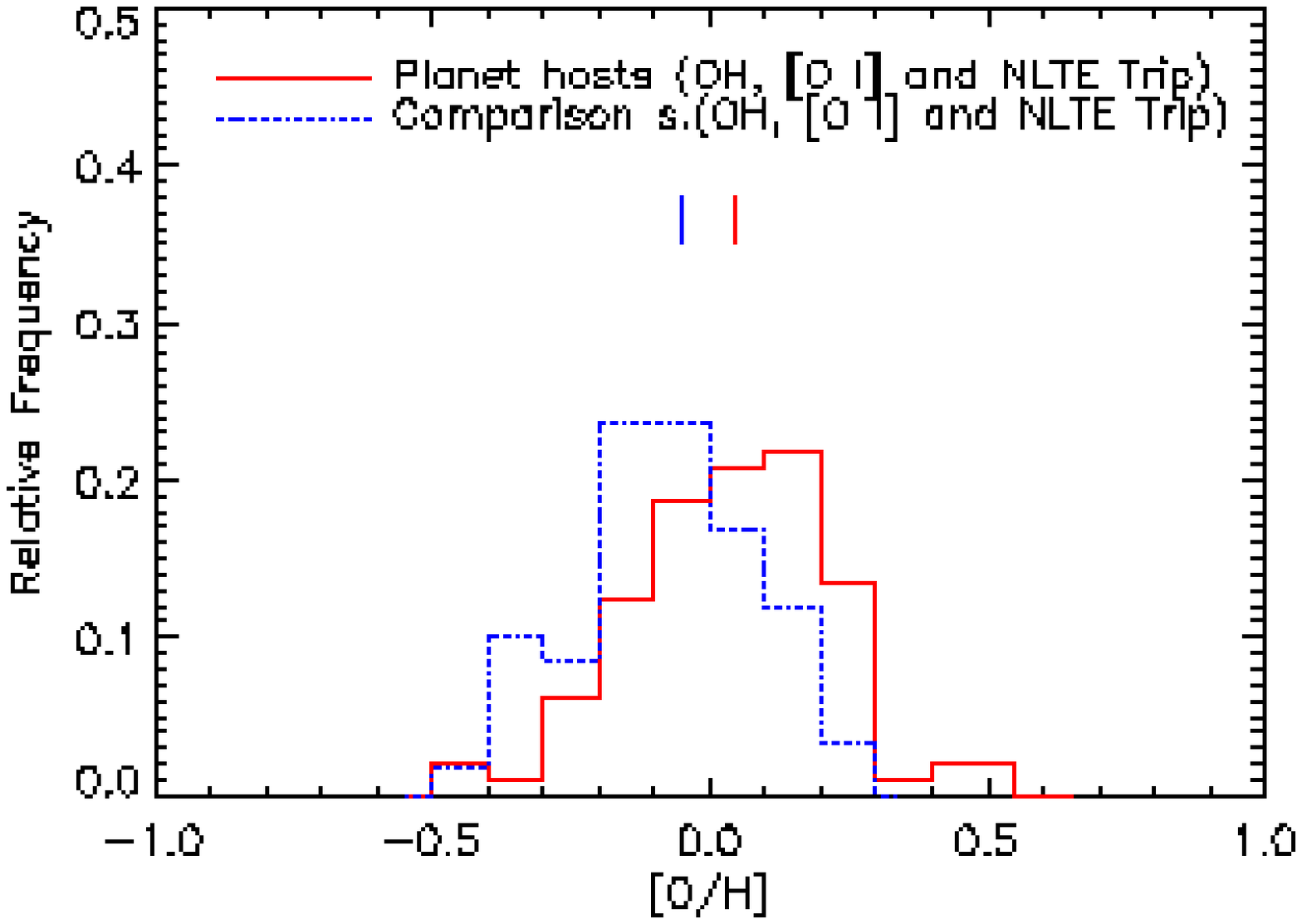}
\includegraphics[width=6.7cm]{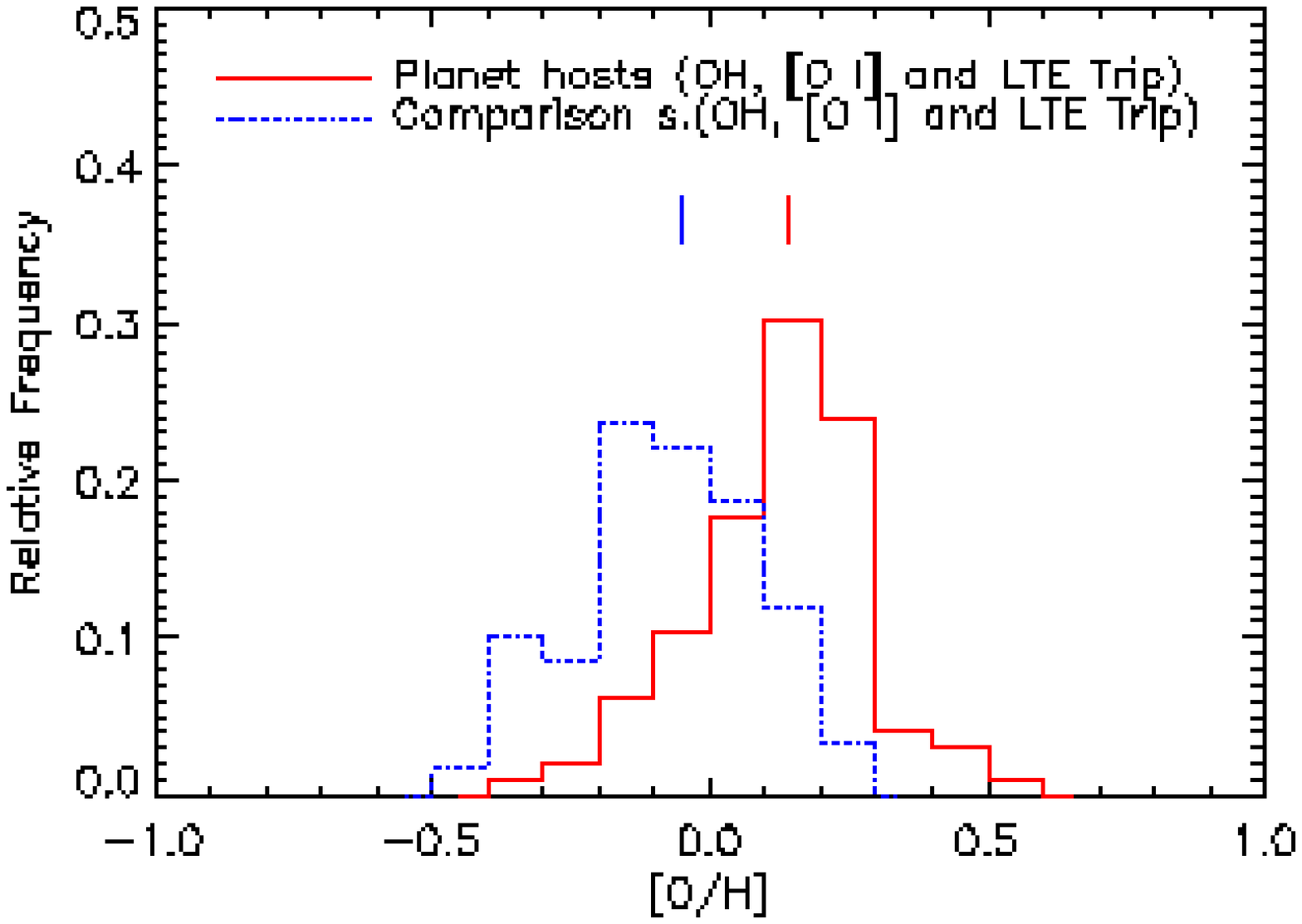}
\caption{Left panel: [O/H] distributions using average results from OH lines, [O\,I] line and O\,I triplet in
NLTE. Right panel: [O/H] distributions using average results from OH lines, [O\,I] line and O\,I triplet in
LTE. The solid and dotted lines represent planet host and comparison sample stars, respectively. The 
vertical lines represent the average abundance values of the two samples, stars with and without planets.}
\label{fig7}
\end{figure*}

\begin{table}[!]
\caption[]{Average [O/H] values from the different indicators, with the corresponding dispersions, for the 
set of planet host stars and the comparison sample}
\begin{center}
\begin{tabular}{cccr}
\hline
\noalign{\smallskip}
Indicator & $<$[O/H]$>\pm$ rms & $<$[O/H]$>\pm$ rms & Difference \\
 &  (comp. sample) & (planet hosts) & \\
\hline 
\noalign{\smallskip}
OH         & $-0.07\pm0.16$ & $ 0.10\pm0.16$ & 0.17 \\
$[$O\,I$]$ & $ 0.07\pm0.17$ & $ 0.12\pm0.16$ & 0.05 \\
NLTE O\,I  & $-0.14\pm0.18$ & $-0.16\pm0.17$ &-0.02 \\
LTE O\,I   & $ 0.08\pm0.20$ & $ 0.15\pm0.17$ & 0.07 \\
mean(NLTE) & $-0.07\pm0.16$ & $ 0.03\pm0.18$ & 0.10 \\
mean(LTE)  & $-0.07\pm0.16$ & $ 0.12\pm0.16$ & 0.19 \\
\noalign{\smallskip}
\hline	      
\end{tabular} 
\end{center}
\label{tab3}  
\end{table}

Figure~\ref{fig6} presents [O/H] distributions of the two samples, stars with and without planets, for the
different indicators: [O\,I] 6300 \AA\ (left top panel), near-UV OH (top right panel) and triplet 7771--5 
\AA\ with NLTE (bottom left panel) and LTE (bottom right panel) treatments. 
Forbidden line results for the comparison sample present a bimodal shape, characterized by a steep 
descent from the peak around [O/H] $\sim$ 0.0 towards negative values. The distribution of planet host stars
obtained with the same indicator exhibits a symmetrical shape, with a steep descent from the peak around 
[O/H] $\sim$ 0.2 towards [O/H] $>$ 0.2. 
An asymmetric distribution is obtained from the OH results for planet host stars, while the 
comparison sample has a symmetric shape (see Figure~\ref{fig6}, top right panel). 
The NLTE and LTE triplet distributions are quite symmetrical for the two samples of stars with and without
planets (see Figure~\ref{fig6}, bottom panels).  

The average values of [O/H] for the samples with and without planets for each indicator, and for all the
indicators together, as well as the rms dispersions and the differences between the mean [O/H] values, are
listed in Table~\ref{tab3}. In general, the mean [O/H] value corresponding to the comparison sample is lower 
than the mean abundance value obtained for the set of planet host stars. From the [O\,I] analysis we obtain 
a difference of the order of 0.05\,dex between the mean abundances of the two 
samples, while the OH synthesis leads to a larger difference of 0.17\,dex. 
 
The NLTE triplet values present a particular characteristic: the mean abundance value is lower in planet host stars than in the comparison sample. This could be because the 
comparison stars with available triplet measurements are generally cooler than planet host stars with 
triplet determinations, and NLTE corrections are much less important for lower $T_{\rm eff}$. The 
abundance underestimation related to our NLTE treatment (see Sect.~\ref{Trip-Forb}) is therefore on average 
less significant in the comparison sample than in the planet host stars. In fact, if we select two subsamples of stars with and without planets having the same $T_{\rm eff}$ distributions, and compare their triplet [O/H] distributions, this effect disappears (see Figure~\ref{OHsel}). The planet host subsample shows an average oxygen overabundance of the order of 0.1\,dex with respect to the comparison subsample, in both NLTE (Figure~\ref{OHsel}, left panel) and LTE (Figure~\ref{OHsel}, right panel) analyses. 

Figure~\ref{fig7} presents the [O/H] distributions of the two samples, stars with and without planets, 
obtained by averaging for each target the abundances obtained from different indicators, adopting NLTE 
(left panel) or LTE (right panel) triplet values. The distributions of the comparison sample obtained in 
both cases are very similar: both have slightly asymmetric shapes, with mean values of -0.10\,dex. 
Concerning the planet host sample, both distributions issued from including NLTE and LTE triplet values, respectively, show 
 asymmetric shapes. 
The latter present a mean value ($<$[O/H]$>=$ 0.12) much larger than the former ($<$ [O/H] $>=$ 0.03 -- see 
Table~\ref{tab3}). 
As our NLTE treatement considers maximum NLTE corrections, we may consider  the abundances obtained 
from including the NLTE and LTE triplet values to correspond to lower and upper limits, respectively 
(see Sect.~\ref{Trip-Forb}). Therefore we propose that planet host stars present an average oxygen 
overabundance between 0.1 and 0.2\,dex with respect to the comparison sample.

\subsection{[O/H] and [O/Fe] vs.\ [Fe/H]}
\begin{figure*}
\centering 
\includegraphics[width=6.7cm]{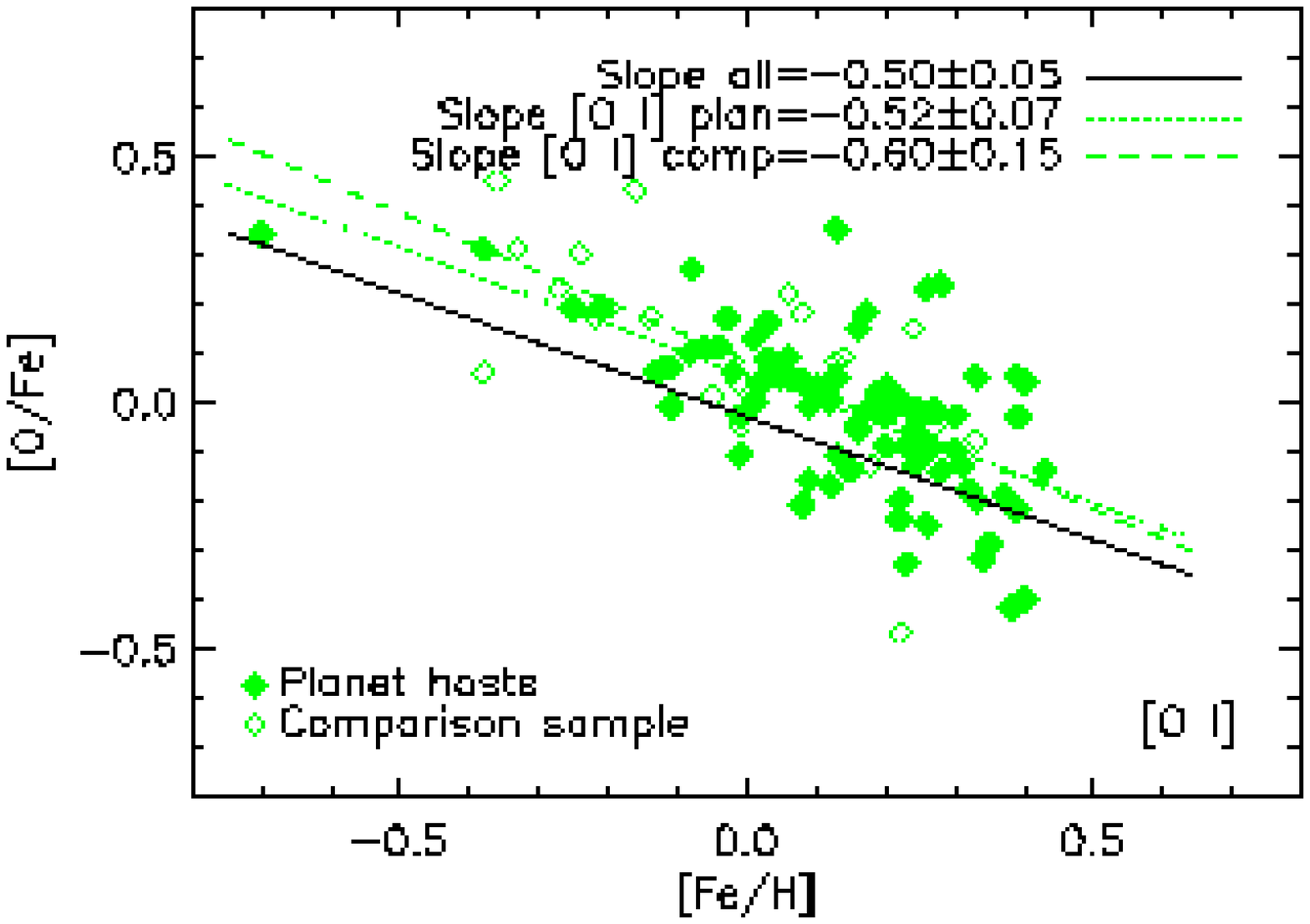}
\includegraphics[width=6.7cm]{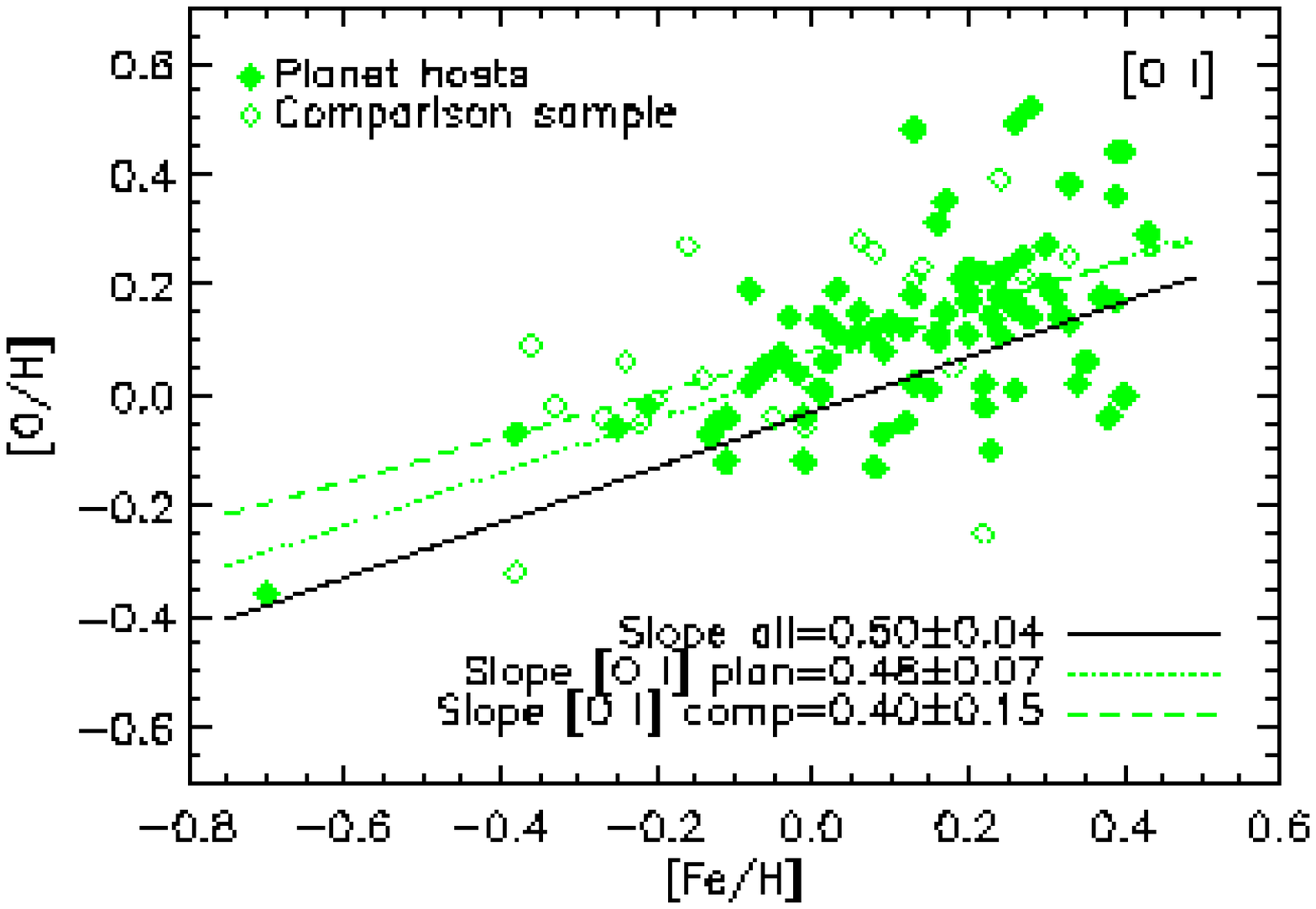}
\includegraphics[width=6.7cm]{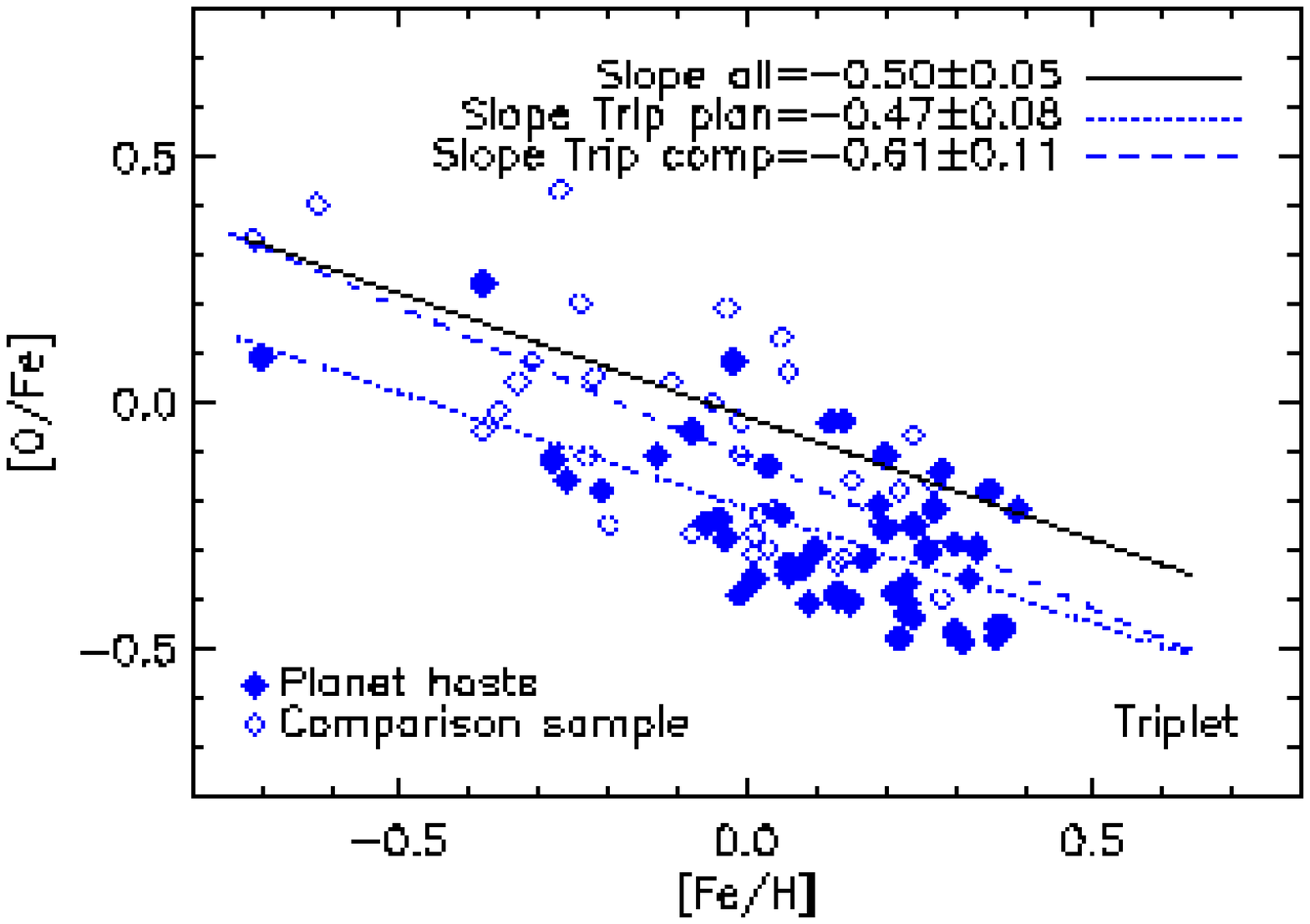}
\includegraphics[width=6.7cm]{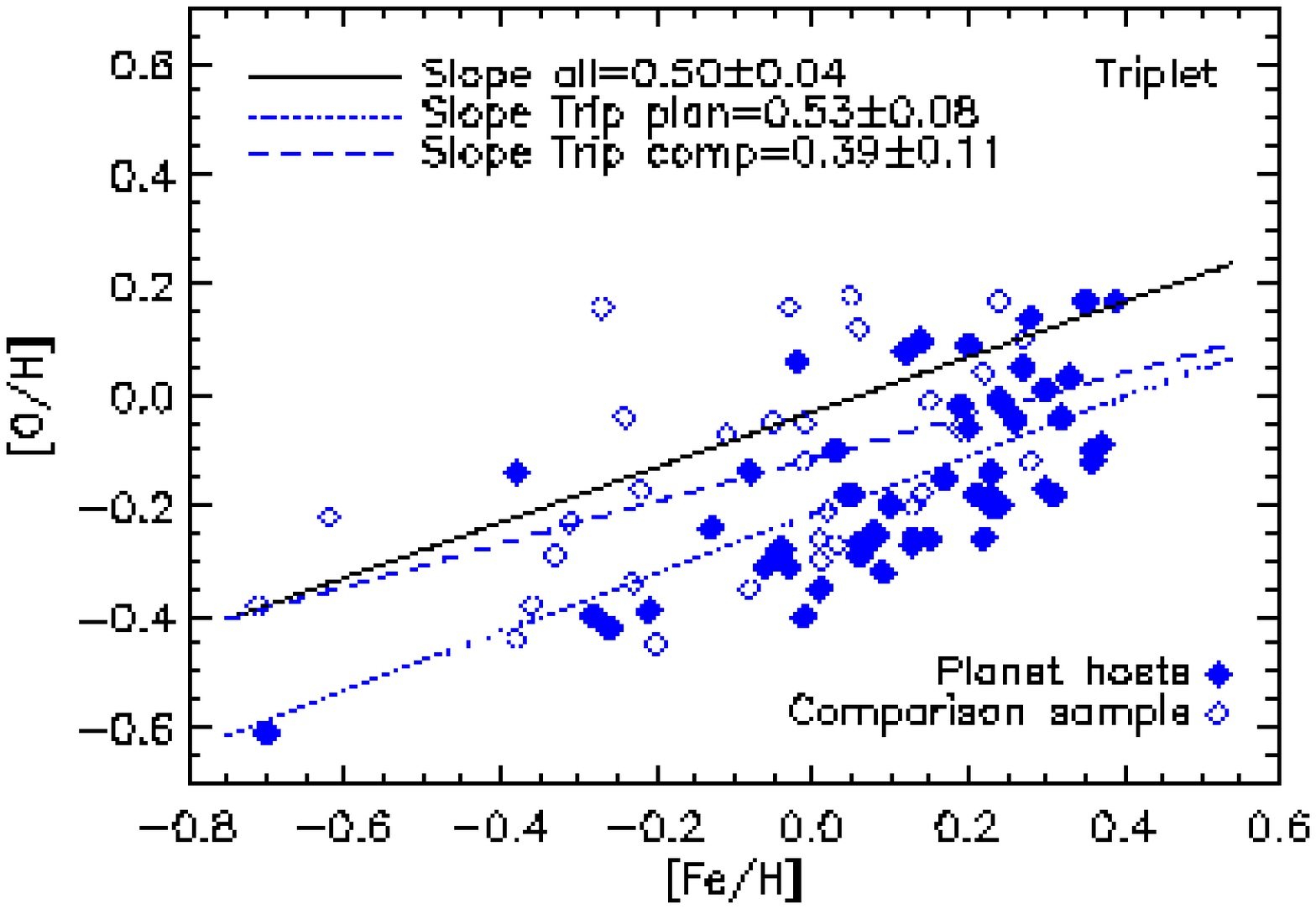}
\includegraphics[width=6.7cm]{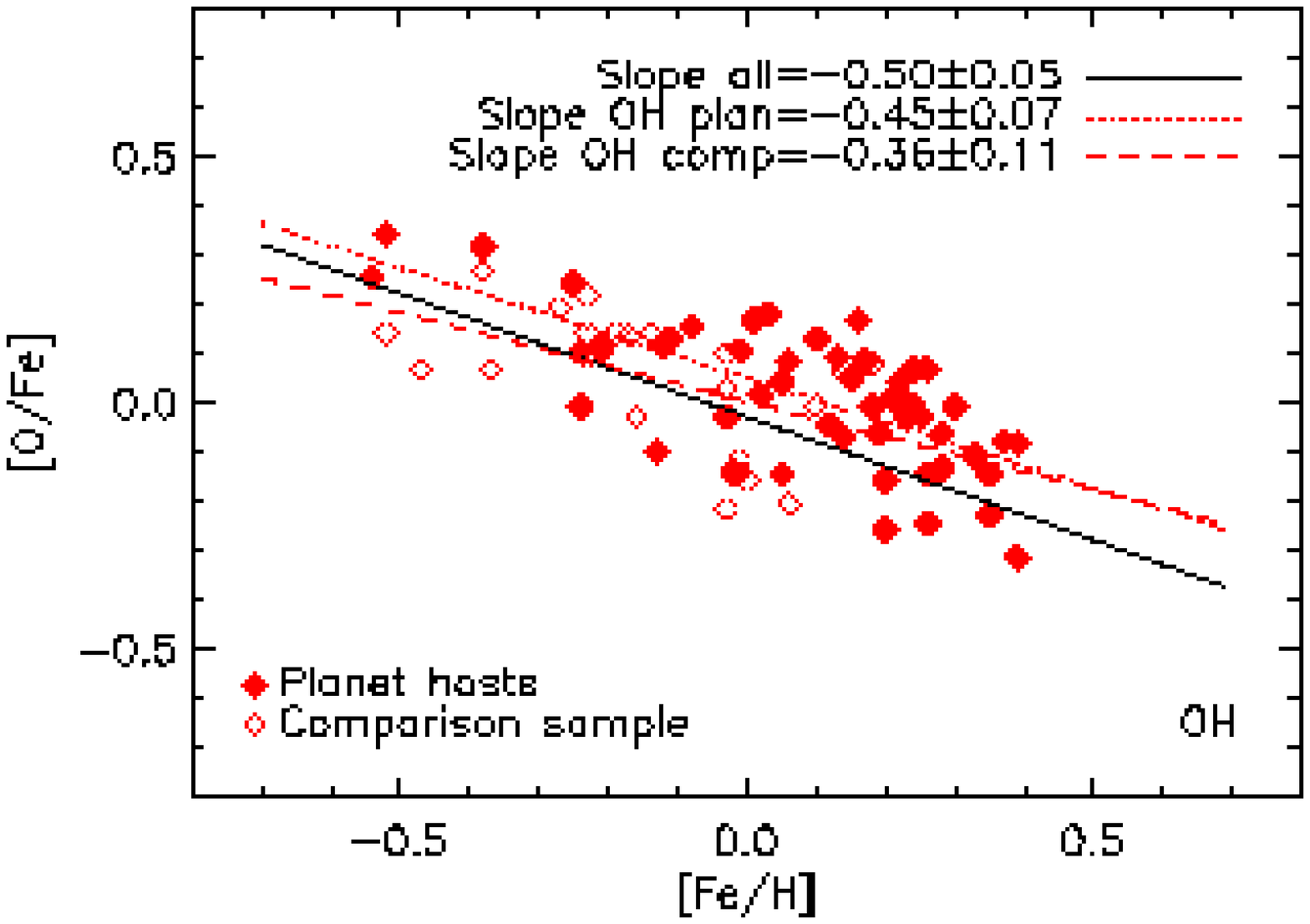}
\includegraphics[width=6.7cm]{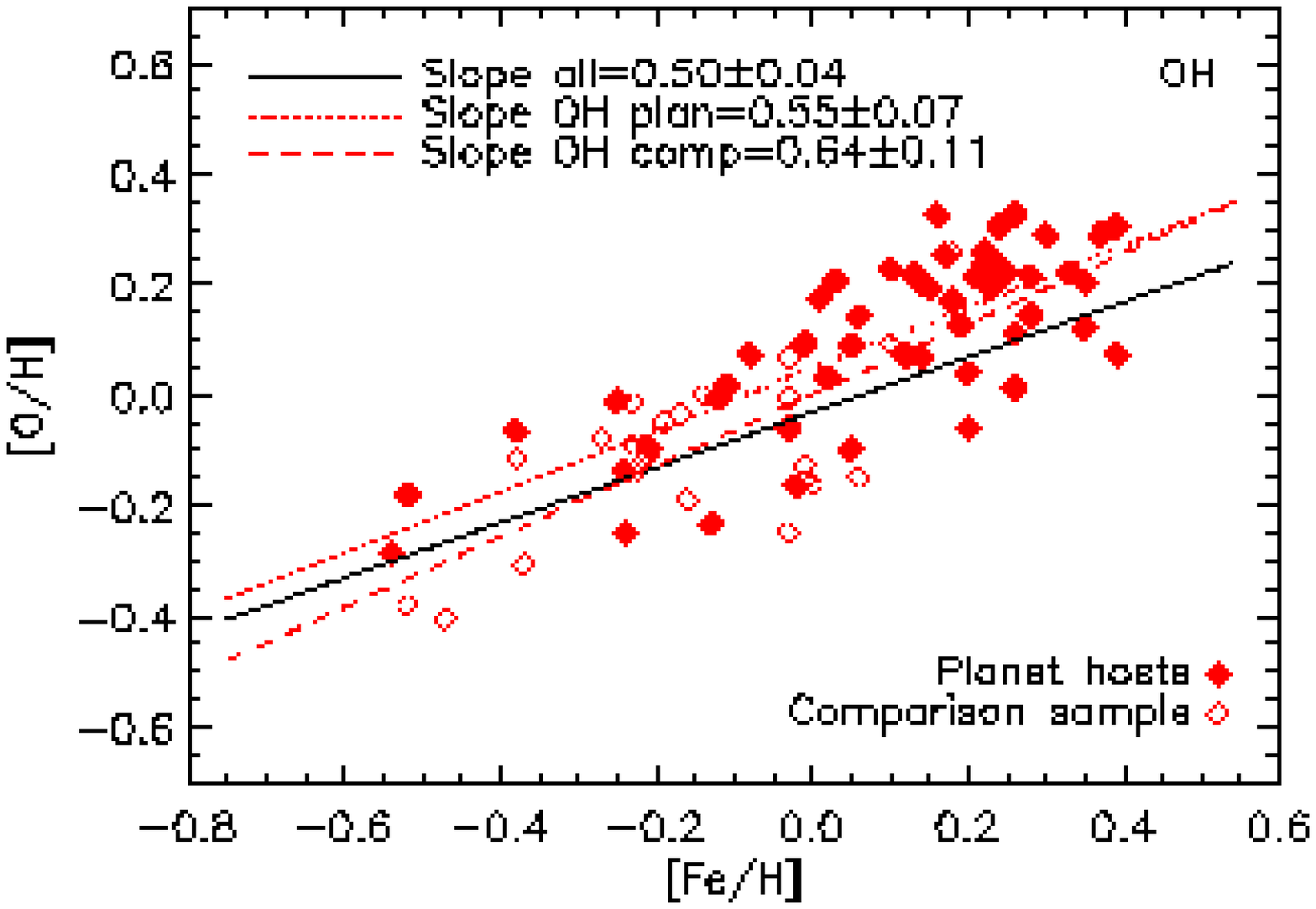}
\caption{[O/Fe] and [O/H] vs. [Fe/H] plots for the three indicators. Filled and open symbols represent planet 
host and comparison sample stars, respectively. Linear least-squares fits to both samples, stars with and
without planets, for each indicator and for the three indicators together are represented and slope values 
are indicated at the bottom of each plot.}
\label{fig8}
\end{figure*}

In Figure~\ref{fig8}, [O/Fe] and [O/H] ratios as functions of [Fe/H] for [O\,I] 6300 \AA\ (top panel), 
O\,I 7771--5 \AA\ (middle panel), and near-UV OH (bottom panel) are presented. No clear differences 
appear between the behaviours of the two samples, stars with and without planets. There seem to be no  
anomalies in oxygen abundances related to the presence of planets. The average trends that planet host 
stars mark are similar to those traced by the comparison sample, although discrepancies between the two trends are slightly larger for the triplet than for the other indicators. Since targets with planets are on average more metal rich than comparison sample stars, their abundance distributions correspond to the extensions of the comparison sample trends at high [Fe/H]. 

The average trends resulting from different indicators present slight discrepancies, but similar behaviours. 
The abundances obtained from OH line synthesis present less dispersion than those derived from 
the other two indicators. The [O/Fe] vs.\ [Fe/H] plots for all the indicators show that, on average, [O/Fe] 
clearly decreases with [Fe/H] in the metallicity range $-0.8 <$ [Fe/H] $< 0.5$, with significantly negative 
slopes in all the linear least-squares fits. The linear least-squares fit for the [O/Fe] values averaged 
from the three indicators gives a slope of $-0.50\pm 0.04$.

\begin{figure}
\centering 
\includegraphics[width=6.7cm]{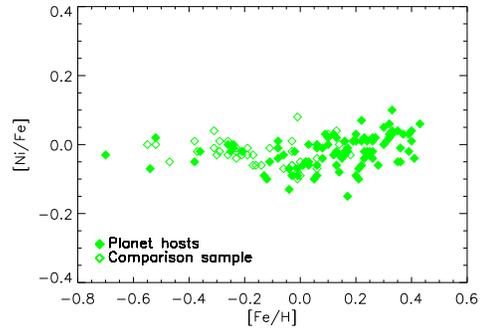}
\caption{[Ni/Fe] vs. [Fe/H] plot. Filled and open symbols represent planet 
host and comparison sample stars, respectively.}
\label{niq}
\end{figure}

Bensby, Feltzing \& Lundstr\"om (\cite{Ben04}) obtained a similar trend of decreasing [O/Fe] with increasing [Fe/H] by analyzing [O\,I] in a large set of disk dwarfs. This behaviour could be caused by the steep rise these authors found in [Ni/Fe] vs.\ [Fe/H] for [Fe/H] $>$ 0. Nevertheless, since our [Ni/Fe] vs.\ [Fe/H] plot does not show such a obvious increase (see Fig.\ref{niq}), it is very unlikely that our [O\,I] results are affected by this phenomenon. Moreover, the three indicators reproduce similar steep descents, which additionally supports that the [O/Fe] decrease with increasing [Fe/H] found from [O\,I] analysis is ``real''. 

\section{Galactic trends or effects related with planets?}
The abundances of volatile elements are a key factor in searching for chemical anomalies associated with the
presence of planets. If the accretion of large amounts of planetary material were the dominant source of the 
metallicity excess observed in planet host stars, a relative overabundance of refractory elements with 
respect to volatiles, or at least some anomaly related to the presence of planets, would be expected in the 
majority of these kinds of targets (e.g.\ Smith et al.\ \cite{Smi01}). Thus knowing how the abundances of 
volatile and refractory elements behave in stars with and without planets can give valuable clues to the 
relative importance of the differential 
accretion. 

Our results show that the oxygen abundances do not present clear anomalies in planet host stars with respect 
to comparison sample dwarfs. The trends traced by the two samples, stars with and without planets, are nearly indistinguishable.
This supports the ``primordial'' hypothesis suggested by Santos et al.\ (\cite{San00}, \cite{San01}), which
proposes the high metal content of the protoplanetary cloud the system planets-star has formed out of 
as an explanation for the observed iron overabundance in planet-harbouring stars. 
Likewise, although the occurrence of accretion is not excluded, the possibility that pollution is
the principal source of the observed metallicity enhancement is unlikely. Therefore the observed trends 
would simply be a product of the chemical evolution of the Galaxy.

Previous studies have already led to results supporting a ``primordial'' origin of the iron excess in planet
host stars (Pinsonneault, DePoye \& Coffee \cite{Pin01}; Santos et al.\ \cite{San01}, \cite{San03b}, 
\cite{San04a}, \cite{San05}). Takeda et al.\ (\cite{Tak01}) and Sadakane et al.\ (\cite{Sad02}) found no 
differences between the abundances of some refractory and volatile elements for a set of planet host stars 
and some field dwarfs from the literature. Recently, Ecuvillon et al.\ (\cite{Ecu04a}, \cite{Ecu04b}) have 
found that the volatiles N, C, S, and Zn behave identically in a large set of planet host and comparison sample 
stars analysed homogeneously. Similar results have been found by Takeda \& Honda (\cite{Tak05}) for CNO
abundances in a set of 27 planet host stars.

We have also obtained a clear monotonic decrease of [O/Fe] with [Fe/H] in the metallicity range 
$-0.8 <$ [Fe/H] $< 0.5$ for all the spectroscopic indicators. The corresponding linear least-squares fits have 
significantly negative slopes, with values around $-$0.5. Some previous studies of oxygen abundances in 
Galactic disc stars (Nissen \& Edvardsson \cite{Nis92}; Edvardsson et al.\ \cite{Edv93}; Nissen et al.\
\cite{Nis02}) found [O/Fe] ratios flattening in the metallicity range $-0.3<$ [Fe/H] $<$ 0.3. Nevertheless,
further studies have revealed that oxygen continues to decline with increasing [Fe/H] (e.g. Feltzing \& 
Gustafsson \cite{Fel98}; Takeda \cite{Tak03}). A recent analysis of a large number of F and G 
disk dwarfs by Bensby, Feltzing \& Lundstr\"om (\cite{Ben04}) has obtained a monotonic decrease of 
[O/Fe] ratios in the metallicity range $-0.9<$ [Fe/H] $< 0.4$, which is in concordance with the predictions of 
chemical evolution models of the Milky Way (e.g.\ Chiappini et al.\ \cite{Chi03}). This implies that oxygen is 
 produced only in SNe\,II, with no SN\,Ia signature  contribution, which would produce a levelling out of 
[O/Fe] at [Fe/H] = 0, as  has been observed in other $\alpha$-elements (Bensby, Feltzing \& Lundstr\"om 
\cite{Ben03}, \cite{Ben04}). Takeda \& Honda (\cite{Tak05}) have found [O/Fe] increasing with
decreasing [Fe/H] with a slope of $\sim$0.4--0.5 for a large sample of 160 dwarfs with metallicities
$-1<$ [Fe/H] $< +0.4$. Our study has obtained a similar monotonic decrease and thus confirms this issue.

An average oxygen overabundance of between 0.1 and 0.2\,dex in the planet host stars with respect to 
the comparison sample has been obtained. It is not clear if this difference is due to the presence of planets.
In order to check this possibility, models of Galactic chemical evolution in this metallicity range must be studied. Unfortunately, such models are not available, and our conclusions can only be based on the best guess. For example, assuming that the main parameters which govern the trends of chemical elements in the galactic disk do not vary in the metallicity range -1.$<$[Fe/H]$<$0.5, we would expect that the elements of the same nuclesynthetic origin present similar behaviours at super-solar metallicities. However, this is not what is observed. For instance, the [X/Fe] ratios of the $\alpha$-elements Si, Ti and Mg decrease monotonically in the metallicity range -1$<$[Fe/H]$<$0 and become constant in the super-solar regime (e.g. Bodaghee et al.\ \cite{Bod03}; Gilli et al.\ \cite{Gil05}), while, as we have seen above, the [O/Fe] ratio continously decreasing in the whole metallicity range. Since at these metallicities the fraction of planet-harbouring stars is important, we cannot exclude the possibility of global effects on abundance trends linked to planets. More works needs to be done before we can tackle this problem.

\section{Concluding remarks}
We presented the first detailed and uniform study of oxygen abundances in an almost complete
set of planet-harbouring stars and in a unbiased volume-limited comparison sample of solar-type dwarfs with
no known planetary-mass companions. An homogeneous set of atmospheric parameters, spectroscopically
determined, was adopted, and three independent analyses from different indicators were performed. We also
provide the first accurate and homogeneous comparison of near-UV OH, 6300 \AA\ [O\,I] and 7771--5 \AA\ 
O\,I triplet in a large set of solar-type stars. 

Oxygen is one of the most controversial elements due to its problematic indicators. We found a good
agreement between the results of [O\,I] and OH analyses. The NLTE treatement for the triplet 
led to an underestimation of the oxygen abundances, while the LTE values may be considered as 
abundance upper limits. The OH results show a much smaller dispersion in the [O/Fe] and [O/H] vs.\ [Fe/H]
plots with respect to the others values, while the [O\,I] analysis generally reveals a better agreement 
with other indicators.   

We found that [O/Fe] and [O/H] trends as function of metallicity show the same behaviour in planet 
host and comparison sample stars. No anomalies associated with the presence of planets have appeared in those
representations. For all the indicators, [O/Fe] ratios decrease monotonically with [Fe/H], with significantly
negative slopes of the order of $-0.5$. 
Planet host stars present on average an oxygen overabundance between 0.1 and 0.2\,dex with respect to the
comparison sample.

We have discussed whether these characteristics are effects related to the presence of planets or products 
of  Galactic chemical evolution. Further investigations on the refractory/volatile abundance ratios
in stars with and without planets, as well as detailed comparisons with theoretical models at super-solar 
metallicities, are required to provide more conclusive evidence for the debate.

\begin{acknowledgements}
The authors acknowledge the data analysis facilities provided by the Starlink Project which is run by CCLRC 
on behalf of PPARC. We thank the referee Dr.~P.~E.~Nissen for many useful suggestions and comments. Support from Funda\c{c}\~ao para a Ci\^encia e a Tecnologia (Portugal) to N.C.S. in the form of a scholarship (reference  SFRH/BPD/8116/2002) and a grant (reference POCI/CTE-AST/56453/2004) is gratefully
acknowledged.
\end{acknowledgements}

\onecolumn
\begin{longtable}{lccccccr}
\caption[]{Oxygen abundances from [O\,I] line in a set of planet host stars.}\\
\hline
Star & $T_\mathrm{eff}$ & $\log {g}$ & $\xi_t$ & [Fe/H] & EW$_{[\ion{O}{i}]}$ & [O/H]$_{[\ion{O}{i}]}$ & Instr.\\
 & (K) & (cm\,s$^{-2}$) & (km\,s$^{-1}$) & & (m\AA) & & \\
\hline 
\hline
\endfirsthead
\caption[]{Continued.}\\
\hline
Star & $T_\mathrm{eff}$ & $\log {g}$ & $\xi_t$ & [Fe/H] & EW$_{[\ion{O}{i}]}$ & [O/H]$_{[\ion{O}{i}]}$ & Instr.\\
 & (K) & (cm\,s$^{-2}$) & (km\,s$^{-1}$) & & (m\AA) & & \\
\hline 
\hline
\endhead
\noalign{\smallskip}
\hline
\noalign{\footnotesize{The instruments used to obtain the spectra were: [1]UVES; [2]UES; [3]FEROS; [4]SARG; [5]CORALIE.}}
\endfoot
\noalign{\smallskip}
\hline
\noalign{\footnotesize{The instruments used to obtain the spectra were: [1]UVES; [2]UES; [3]FEROS; [4]SARG; [5]CORALIE.}}
\noalign{\footnotesize{$^1$ This value was excluded from the figures since it is inconsistent with the triplet result.}}
\endlastfoot
\noalign{\smallskip}
\object{HD\,1237}  & $5536\pm50 $ & $4.56\pm0.12$& $1.33\pm0.06$ & $ 0.12\pm0.06$ & $ 2.9\pm0.5$ & $-0.05\pm0.10$ & [1]\\
\object{HD\,23079} & $5959\pm46 $ & $4.35\pm0.12$& $1.20\pm0.10$ & $-0.11\pm0.06$ & $ 3.5\pm0.5$ & $-0.12\pm0.09$ & [1]\\
\object{HD\,28185} & $5656\pm44 $ & $4.45\pm0.08$& $1.01\pm0.06$ & $ 0.22\pm0.05$ & $ 5.4\pm0.8$ & $ 0.22\pm0.09$ & [1]\\
\object{HD\,30177} & $5591\pm50 $ & $4.35\pm0.12$& $1.03\pm0.06$ & $ 0.39\pm0.06$ & $ 8.3\pm0.5$ & $ 0.44\pm0.07$ & [1]\\
\object{HD\,33636 }& $6046\pm49 $ & $4.71\pm0.09$& $1.79\pm0.19$ & $-0.08\pm0.06$ & $ 3.2\pm0.7$ & $ 0.02\pm0.11$ & [1]\\
\object{HD\,37124 }& $5546\pm30 $ & $4.50\pm0.03$& $0.80\pm0.07$ & $-0.38\pm0.04$ & $ 4.5\pm0.3$ & $-0.07\pm0.04$ & [1]\\
\object{HD\,39091 }& $5991\pm27 $ & $4.42\pm0.10$& $1.24\pm0.04$ & $ 0.10\pm0.04$ & $ 4.9\pm0.5$ & $ 0.13\pm0.07$ & [1]\\
\object{HD\,50554 }& $6026\pm30 $ & $4.41\pm0.13$& $1.11\pm0.06$ & $ 0.01\pm0.04$ & $ 4.0\pm0.4$ & $ 0.01\pm0.07$ & [1]\\
\object{HD\,65216 }& $5666\pm31 $ & $4.53\pm0.09$& $1.06\pm0.05$ & $-0.12\pm0.04$ & $ 3.7\pm0.5$ & $-0.05\pm0.08$ & [1]\\
\object{HD\,72659 }& $5995\pm45 $ & $4.30\pm0.07$& $1.42\pm0.09$ & $ 0.03\pm0.06$ & $ 5.5\pm0.5$ & $ 0.11\pm0.05$ & [1]\\
\object{HD\,73256 }& $5518\pm49 $ & $4.42\pm0.12$& $1.22\pm0.06$ & $ 0.26\pm0.06$ & $ 4.8\pm1.0$ & $ 0.17\pm0.12$ & [1]\\
\object{HD\,74156 }& $6112\pm39 $ & $4.34\pm0.10$& $1.38\pm0.07$ & $ 0.16\pm0.05$ & $ 4.6\pm0.3$ & $ 0.11\pm0.06$ & [1]\\
\object{HD\,106252}& $5899\pm35 $ & $4.34\pm0.07$& $1.08\pm0.06$ & $-0.01\pm0.05$ & $ 4.0\pm0.8$ & $-0.04\pm0.11$ & [1]\\
\object{HD\,114729}& $5886\pm36 $ & $4.28\pm0.13$& $1.25\pm0.09$ & $-0.25\pm0.05$ & $ 4.9\pm0.5$ & $-0.06\pm0.07$ & [1]\\
\object{HD\,213240}& $5984\pm33 $ & $4.25\pm0.10$& $1.25\pm0.05$ & $ 0.17\pm0.05$ & $ 5.7\pm0.4$ & $ 0.15\pm0.06$ & [1]\\
\object{HD\,216435}& $5938\pm42 $ & $4.12\pm0.05$& $1.28\pm0.06$ & $ 0.24\pm0.05$ & $ 6.7\pm0.7$ & $ 0.18\pm0.06$ & [1]\\
\object{HD\,216437}& $5887\pm32 $ & $4.30\pm0.07$& $1.31\pm0.04$ & $ 0.25\pm0.04$ & $ 6.2\pm0.5$ & $ 0.23\pm0.05$ & [1]\\
\object{BD\,103166}&$ 5325\pm45  $&$ 4.36\pm0.07 $&$ 0.95\pm0.05 $&$  0.35\pm0.05$ &$ 3.9\pm1.0 $&$ 0.06\pm0.13$&[3]\\
\object{HD\,2039}  & $5976\pm51 $ & $4.45\pm0.10$ & $1.26\pm0.07$ &$ 0.32\pm0.06 $&$ 4.2\pm1.0  $&$ 0.14\pm0.12$&[5]\\
\object{HD\,3651}  & $5173\pm35 $ & $4.37\pm0.12$ & $0.74\pm0.05$ &$ 0.12\pm0.04 $&$ 5.1\pm0.5  $&$ 0.12\pm0.08$&[4]\\
\object{HD\,4203}  & $5636\pm40 $ & $4.23\pm0.14$ & $1.12\pm0.05$ &$ 0.40\pm0.05 $&$ 3.8\pm1.5  $&$ 0.00\pm0.21$&[3]\\
\object{HD\,9826}  & $6212\pm64 $ & $4.26\pm0.13$ & $1.69\pm0.16$ &$ 0.13\pm0.08 $&$ 4.0\pm0.7  $&$ 0.02\pm0.10$&[2]\\
\object{HD\,16141} & $5801\pm30 $ & $4.22\pm0.12$ & $1.34\pm0.04$ &$ 0.15\pm0.04 $&$4.7\pm1.0   $&$ 0.01\pm0.11$&[3]\\
\object{HD\,17051} & $6252\pm53 $ & $4.61\pm0.16$ & $1.18\pm0.10$ &$ 0.26\pm0.06 $&$ 3.6\pm0.7  $&$ 0.16\pm0.11$&[1]\\
\object{HD\,19994} & $6190\pm00 $ & $4.19\pm0.00$ & $1.54\pm0.00$ &$ 0.24\pm0.00 $&$ 5.0\pm1.0  $&$ 0.11\pm0.09$&[1]\\
\object{HD\,22049} & $5073\pm42 $ & $4.43\pm0.08$ & $1.05\pm0.06$ &$-0.13\pm0.05 $&$ 3.8\pm0.8  $&$-0.07\pm0.10$&[2]\\
\object{HD\,23079} & $5959\pm46 $ & $4.35\pm0.12$ & $1.20\pm0.10$ &$-0.11\pm0.06 $&$ 4.2\pm0.2  $&$-0.04\pm0.06$&[3]\\
\object{HD\,23596} & $6108\pm36 $ & $4.25\pm0.10$ & $1.30\pm0.05$ &$ 0.31\pm0.05 $&$ 5.4\pm0.3  $&$ 0.18\pm0.05$&[2]\\
\object{HD\,27442} & $4825\pm107$ & $3.55\pm0.32$ & $1.18\pm0.12$ &$ 0.39\pm0.13 $&$15.6\pm1.5  $&$ 0.36\pm0.16$&[3]\\
\object{HD\,28185} & $5656\pm44 $ & $4.45\pm0.08$ & $1.01\pm0.06$ &$ 0.22\pm0.05 $&$ 3.6\pm1.0  $&$ 0.02\pm0.15$&[5]\\
\object{HD\,30177} & $5587\pm00 $ & $4.29\pm0.00$ & $1.08\pm0.00$ &$ 0.38\pm0.00 $&$ 3.3\pm1.0  $&$-0.04\pm0.15$&[5]\\
\object{HD\,38529} & $5674\pm40 $ & $3.94\pm0.12$ & $1.38\pm0.05$ &$ 0.40\pm0.06 $&$13.3\pm0.7  $&$ 0.44\pm0.06$&[3]\\
\object{HD\,46375} & $5268\pm55 $ & $4.41\pm0.16$ & $0.97\pm0.06$ &$ 0.20\pm0.06 $&$ 5.9\pm0.5  $&$ 0.23\pm0.09$&[2]\\
\object{HD\,50554} & $6026\pm30 $ & $4.41\pm0.13$ & $1.11\pm0.06$ &$ 0.01\pm0.04 $&$ 5.3\pm0.9  $&$ 0.14\pm0.09$&[2]\\
\object{HD\,52265} & $6105\pm00 $ & $4.28\pm0.00$ & $1.36\pm0.00$ &$ 0.23\pm0.00 $&$ 5.1\pm0.7  $&$ 0.14\pm0.05$&[3]\\
\object{HD\,59686 }& $4871\pm135$ & $3.15\pm0.24$ & $1.85\pm0.12$ &$  0.28\pm0.18$ &$33.8\pm1.0 $&$ 0.52\pm0.13$&[3]\\
\object{HD\,70642 }&$ 5671\pm46  $&$ 4.39\pm0.05 $&$ 1.01\pm0.06 $&$  0.20\pm0.06$ &$ 5.5\pm0.5 $&$ 0.17\pm0.05$&[3]\\
\object{HD\,73256} & $5518\pm49 $ & $4.42\pm0.12$ & $1.22\pm0.06$ &$ 0.26\pm0.06 $&$ 3.5\pm0.9  $&$ 0.01\pm0.14$&[3]\\
\object{HD\,74156} & $6112\pm39 $ & $4.34\pm0.10$ & $1.38\pm0.07$ &$ 0.16\pm0.05 $&$ 7.3\pm0.7  $&$ 0.31\pm0.06$& [3]\\
\object{HD\,75289} & $6143\pm53 $ & $4.42\pm0.13$ & $1.53\pm0.09$ &$ 0.28\pm0.07 $&$ 5.6\pm0.4  $&$ 0.14\pm0.08$& [1]\\
\object{HD\,75732} & $5279\pm62 $ & $4.37\pm0.18$ & $0.98\pm0.07$ &$ 0.33\pm0.07 $&$ 4.5\pm0.7  $&$ 0.13\pm0.11$& [2]\\
\object{HD\,82943} & $6015\pm00 $ & $4.46\pm0.00$ & $1.13\pm0.00$ &$ 0.30\pm0.00 $&$ 4.8\pm0.7  $&$ 0.20\pm0.06$& [1]\\
\object{HD\,83443 }&$ 5501\pm63  $&$ 4.46\pm0.09 $&$ 1.07\pm0.08 $&$  0.39\pm0.07$ &$ 4.4\pm1.0 $&$ 0.17\pm0.13$&[3]\\
\object{HD\,92788 }&$ 5758\pm37  $&$ 4.3 \pm0.04 $&$ 1.1 \pm0.04 $&$  0.34\pm0.05$ &$ 3.8\pm1.0 $&$ 0.02\pm0.13$&[3]\\
\object{HD\,95128} & $5954\pm25 $ & $4.44\pm0.10$ & $1.30\pm0.04$ &$ 0.06\pm0.03 $&$ 5.2\pm0.6  $&$ 0.15\pm0.06$&[4]\\
\object{HD\,106252}&$ 5834\pm37  $&$ 4.22\pm0.02 $&$ 1.06\pm0.06 $&$ -0.03\pm0.05$ &$ 7.0\pm1.0 $&$ 0.14\pm0.07$&[3]\\
\object{HD\,108147}& $6248\pm42 $ & $4.49\pm0.16$ & $1.35\pm0.08$ &$ 0.20\pm0.05 $&$ 3.7\pm0.4  $&$ 0.11\pm0.09$&[1]\\
\object{HD\,108874}& $5596\pm42 $ & $4.37\pm0.12$ & $0.89\pm0.05$ &$ 0.23\pm0.05 $&$ 3.0\pm0.7  $&$-0.10\pm0.12$&[2]\\
\object{HD\,114386}&$ 4865\pm93  $&$ 4.3 \pm0.17 $&$ 0.86\pm0.12 $&$ -0.04\pm0.07$ &$ 5.1\pm1.0 $&$ 0.07\pm0.15$&[3]\\
\object{HD\,114762}& $5884\pm34 $ & $4.22\pm0.02$ & $1.31\pm0.17$ &$-0.70\pm0.04 $&$ 3.2\pm0.1  $&$-0.36\pm0.02$&[1]\\
\object{HD\,114783}& $5098\pm36 $ & $4.45\pm0.11$ & $0.74\pm0.05$ &$ 0.09\pm0.04 $&$ 4.4\pm1.5  $&$ 0.08\pm0.18$&[4]\\
\object{HD\,117176}& $5560\pm34 $ & $4.07\pm0.05$ & $1.18\pm0.05$ &$-0.06\pm0.05 $&$ 7.1\pm0.5  $&$ 0.05\pm0.04$&[4]\\
\object{HD\,121504}& $6075\pm40 $ & $4.64\pm0.12$ & $1.31\pm0.07$ &$ 0.16\pm0.05 $&$ 3.5\pm0.2  $&$ 0.10\pm0.06$&[1]\\
\object{HD\,128311}& $4835\pm72 $ & $4.44\pm0.21$ & $0.89\pm0.11$ &$ 0.03\pm0.07 $&$ 6.0\pm1.5  $&$ 0.19\pm0.16$&[2]\\
\object{HD\,134987}& $5776\pm29$ & $4.36\pm0.07$ & $1.09\pm0.04$ &$ 0.30\pm0.04 $&$ 6.5\pm0.5   $&$ 0.27\pm0.05$&[4]\\
\object{HD\,137759}& $4775\pm113$& $3.09\pm0.40$ & $1.78\pm0.11$ &$ 0.13\pm0.14 $&$34.7\pm1.0   $&$ 0.48\pm0.19$&[4]\\
\object{HD\,141937}& $5909\pm39$ & $4.51\pm0.08$ & $1.13\pm0.06$ &$ 0.10\pm0.05 $&$ 4.6\pm0.7   $&$ 0.13\pm0.07$&[2]\\
\object{HD\,143761}& $5853\pm25$ & $4.41\pm0.15$ & $1.35\pm0.07$ &$-0.21\pm0.04 $&$ 4.5\pm0.8   $&$-0.02\pm0.11$&[4]\\
\object{HD\,145675}& $5311\pm87$ & $4.42\pm0.18$ & $0.92\pm0.10$ &$ 0.43\pm0.08 $&$ 5.7\pm1.0   $&$ 0.29\pm0.12$&[4]\\
\object{HD\,147513}&$ 5894\pm31  $&$ 4.43\pm0.02 $&$ 1.26\pm0.05 $&$  0.08\pm0.04$ &$ 5.0\pm1.5 $&$ 0.12\pm0.16$&[3]\\
\object{HD\,150706}& $5961\pm27$ & $4.50\pm0.10$ & $1.11\pm0.06$ &$-0.01\pm0.04 $&$ 2.8\pm0.5   $&$-0.12\pm0.09$&[2]\\
\object{HD\,168443}& $5617\pm35$ & $4.22\pm0.05$ & $1.21\pm0.05$ &$ 0.06\pm0.05 $&$ 6.3\pm1.0   $&$ 0.11\pm0.08$&[4]\\
\object{HD\,168746}& $5601\pm33$ & $4.41\pm0.12$ & $0.99\pm0.05$ &$-0.08\pm0.05 $&$ 6.8\pm0.5   $&$ 0.19\pm0.07$&[4]\\
\object{HD\,177830}& $4804\pm77$ & $3.57\pm0.17$ & $1.14\pm0.09$ &$ 0.33\pm0.09 $&$16.0\pm0.7   $&$ 0.38\pm0.09$&[4]\\
\object{HD\,178911B}&$5600\pm42$ & $4.44\pm0.08$ & $0.95\pm0.05$ &$ 0.27\pm0.05 $&$ 5.9\pm1.0   $&$ 0.25\pm0.08$&[4]\\
\object{HD\,179949}& $6260\pm43$ & $4.43\pm0.05$ & $1.41\pm0.09$ &$ 0.22\pm0.05 $&$ 2.9\pm1.0   $&$-0.02\pm0.18$&[4]\\
\object{HD\,186427}& $5772\pm25$ & $4.40\pm0.07$ & $1.07\pm0.04$ &$ 0.08\pm0.04 $&$ 3.0\pm0.3   $&$-0.13\pm0.05$&[4]\\
\object{HD\,187123}& $5845\pm22$ & $4.42\pm0.07$ & $1.10\pm0.03$ &$ 0.13\pm0.03 $&$ 5.6\pm0.5   $&$ 0.18\pm0.05$&[4]\\
\object{HD\,190360}& $5584\pm36$ & $4.37\pm0.06$ & $1.07\pm0.05$ &$ 0.24\pm0.05 $&$ 6.1\pm0.6   $&$ 0.22\pm0.04$&[2]\\
\object{HD\,192263}& $4947\pm58$ & $4.51\pm0.20$ & $0.86\pm0.09$ &$-0.02\pm0.06 $&$ 4.1\pm1.0   $&$ 0.04\pm0.15$&[3]\\
\object{HD\,195019}& $5842\pm31$ & $4.32\pm0.07$ & $1.27\pm0.05$ &$ 0.09\pm0.04 $&$ 3.7\pm0.7   $&$-0.07\pm0.09$&[2]\\
\object{HD\,209458}& $6117\pm26$ & $4.48\pm0.08$ & $1.40\pm0.06$ &$ 0.02\pm0.03 $&$ 4.0\pm0.8   $&$ 0.06\pm0.10$&[1]\\
\object{HD\,210277}& $5532\pm00$ & $4.29\pm0.00$ & $1.03\pm0.00$ &$ 0.19\pm0.00 $&$ 6.6\pm0.5   $&$ 0.21\pm0.04$&[4]\\
\object{HD\,213240}& $5984\pm33$ & $4.25\pm0.10$ & $1.25\pm0.05$ &$ 0.17\pm0.05 $&$ 9.0\pm1.5   $&$ 0.35\pm0.09$&[5]\\
\object{HD\,216770}& $5423\pm41$ & $4.40\pm0.13$ & $1.01\pm0.05$ &$ 0.26\pm0.04 $&$10.2\pm2.0   $&$ 0.49\pm0.12$&[4]\\
\object{HD\,217014}& $5804\pm36$ & $4.42\pm0.07$ & $1.20\pm0.05$ &$ 0.20\pm0.05 $&$ 5.4\pm1.0   $&$ 0.19\pm0.11$&[3]\\
\object{HD\,217107}& $5645\pm00$ & $4.31\pm0.00$ & $1.06\pm0.00$ &$ 0.37\pm0.00 $&$ 5.4\pm0.2   $&$ 0.18\pm0.01$&[2]\\
\object{HD\,222582}& $5843\pm38$ & $4.45\pm0.07$ & $1.03\pm0.06$ &$ 0.05\pm0.05 $&$ 4.8\pm1.0   $&$ 0.10\pm0.11$&[2]\\
\object{HD\,104985}& $4773\pm62 $ & $2.76\pm0.14$ & $1.71\pm0.07$ &$-0.28\pm0.09 $ &$42.7\pm1.5 $&$ 0.30\pm0.07^1$&[4]\\
\label{longtab4}
\end{longtable}

\begin{longtable}{lccccccr}
\caption[]{Oxygen abundances from [O\,I] line in a set of comparison stars.}\\
\hline
Star & $T_\mathrm{eff}$ & $\log {g}$ & $\xi_t$ & [Fe/H] & EW$_{[\ion{O}{i}]}$ & [O/H]$_{[\ion{O}{i}]}$ & Instr.\\
 & (K) & (cm\,s$^{-2}$) & (km\,s$^{-1}$) & & (m\AA) & & \\
\hline 
\hline
\endfirsthead
\caption[]{Continued.}\\
\hline
Star & $T_\mathrm{eff}$ & $\log {g}$ & $\xi_t$ & [Fe/H] & EW$_{[\ion{O}{i}]}$ & [O/H]$_{[\ion{O}{i}]}$ & Instr.\\
 & (K) & (cm\,s$^{-2}$) & (km\,s$^{-1}$) & & (m\AA) & & \\
\hline 
\hline
\endhead
\noalign{\smallskip}
\hline
\noalign{\footnotesize{The instruments used to obtain the spectra were: [1]UVES; [2]UES; [3]FEROS; [4]SARG; [5]CORALIE.}}
\endfoot
\noalign{\smallskip}
\hline
\noalign{\footnotesize{The instruments used to obtain the spectra were: [1]UVES; [2]UES; [3]FEROS; [4]SARG; [5]CORALIE.}}
\noalign{\footnotesize{$^1$ This value was excluded from the figures since it is inconsistent with the triplet result.}}
\endlastfoot
\noalign{\smallskip}
\object{HD\,1581}  & $5956\pm44$ & $4.39\pm0.13$ & $1.07\pm0.09 $&$-0.14\pm0.05 $&$ 4.8\pm0.2   $&$ 0.03\pm0.07$& [3]\\
\object{HD\,7570}  & $6140\pm41$ & $4.39\pm0.16$ & $1.50\pm0.08 $&$ 0.18\pm0.05 $&$ 3.8\pm0.3   $&$ 0.05\pm0.09$& [1]\\
\object{HD\,52919 }&$ 4740\pm49  $&$ 4.25\pm0.1  $&$ 1.03\pm0.09 $&$ -0.05\pm0.06$ &$ 4.8\pm1.0 $&$-0.04\pm0.11$&[3]\\
\object{HD\,53143 }&$ 5462\pm54  $&$ 4.47\pm0.07 $&$ 1.08\pm0.07 $&$  0.22\pm0.06$ &$ 1.9\pm0.5 $&$-0.25\pm0.14$&[3]\\
\object{HD\,67199 }&$ 5136\pm56  $&$ 4.54\pm0.08 $&$ 0.84\pm0.08 $&$  0.06\pm0.06$ &$ 6.3\pm2.0 $&$ 0.28\pm0.18$&[3]\\
\object{HD\,100623}&$ 5246\pm37  $&$ 4.54\pm0.05 $&$ 1   \pm0.06 $&$ -0.38\pm0.05$ &$ 2.3\pm0.5 $&$-0.32\pm0.10$&[3]\\
\object{HD\,102365}&$ 5667\pm27  $&$ 4.59\pm0.02 $&$ 1.05\pm0.06 $&$ -0.27\pm0.04$ &$ 3.8\pm0.5 $&$-0.04\pm0.07$&[3]\\
\object{HD\,104304}&$ 5562\pm50  $&$ 4.37\pm0.06 $&$ 1.1 \pm0.06 $&$  0.27\pm0.06$ &$ 6.0\pm0.5 $&$ 0.22\pm0.05$&[3]\\
\object{HD\,109200}&$ 5103\pm46  $&$ 4.47\pm0.07 $&$ 0.75\pm0.07 $&$ -0.24\pm0.05$ &$ 5.2\pm0.4 $&$ 0.06\pm0.05$&[3]\\
\object{HD\,115617}&$ 5577\pm33  $&$ 4.34\pm0.03 $&$ 1.07\pm0.04 $&$  0.01\pm0.05$ &$ 4.6\pm1.0 $&$ 0.00\pm0.11$&[3]\\
\object{HD\,118972}&$ 5241\pm66  $&$ 4.43\pm0.1  $&$ 1.24\pm0.08 $&$ -0.01\pm0.08$ &$ 3.6\pm0.9 $&$-0.06\pm0.19$&[3]\\
\object{HD\,125072}&$ 5001\pm115 $&$ 4.39\pm0.21 $&$ 1.21\pm0.15 $&$  0.24\pm0.11$ &$ 8.3\pm1.0 $&$ 0.39\pm0.12$&[3]\\
\object{HD\,140901}&$ 5645\pm37  $&$ 4.4 \pm0.04 $&$ 1.14\pm0.05 $&$  0.13\pm0.05$ &$ 6.2\pm1.5 $&$ 0.21\pm0.12$&[3]\\
\object{HD\,144628}&$ 5071\pm43  $&$ 4.41\pm0.06 $&$ 0.82\pm0.07 $&$ -0.36\pm0.06$ &$ 6.5\pm0.8 $&$ 0.09\pm0.07$&[3]\\
\object{HD\,146233}&$ 5786\pm35  $&$ 4.31\pm0.03 $&$ 1.18\pm0.05 $&$  0.08\pm0.05$ &$ 7.7\pm1.0 $&$ 0.26\pm0.07$&[3]\\
\object{HD\,152391}&$ 5521\pm43  $&$ 4.54\pm0.05 $&$ 1.29\pm0.06 $&$  0.03\pm0.05$ &$ 4.7\pm0.7 $&$ 0.12\pm0.08$&[3]\\
\object{HD\,154088}&$ 5414\pm60  $&$ 4.28\pm0.08 $&$ 1.14\pm0.07 $&$  0.33\pm0.07$ &$ 6.6\pm0.7 $&$ 0.25\pm0.07$&[3]\\
\object{HD\,156274}&$ 5300\pm32  $&$ 4.41\pm0.04 $&$ 1   \pm0.05 $&$ -0.33\pm0.05$ &$ 5.0\pm0.9 $&$-0.02\pm0.09$&[3]\\
\object{HD\,165499}&$ 5950\pm45  $&$ 4.31\pm0.02 $&$ 1.14\pm0.08 $&$  0.01\pm0.06$ &$ 4.5\pm1.0 $&$ 0.01\pm0.10$&[3]\\
\object{HD\,170657}&$ 5115\pm52  $&$ 4.48\pm0.08 $&$ 1.13\pm0.07 $&$ -0.22\pm0.06$ &$ 4.0\pm0.5 $&$-0.05\pm0.07$&[3]\\
\object{HD\,172051}&$ 5634\pm30  $&$ 4.43\pm0.03 $&$ 1.06\pm0.05 $&$ -0.20\pm0.04$ &$ 4.7\pm0.5 $&$-0.01\pm0.05$&[3]\\
\object{HD\,177565}&$ 5664\pm28  $&$ 4.43\pm0.02 $&$ 1.02\pm0.04 $&$  0.14\pm0.04$ &$ 6.2\pm0.5 $&$ 0.23\pm0.04$&[3]\\
\object{HD\,192310}&$ 5069\pm49  $&$ 4.38\pm0.19 $&$ 0.79\pm0.07 $&$ -0.01\pm0.05$ &$ 4.5\pm0.5 $&$ 0.03\pm0.11$&[4]\\
\object{HD\,222335}&$5260\pm41 $ & $ 4.45\pm0.11$ &$ 0.92\pm0.06 $&$ -0.16\pm0.05 $&$ 8.0\pm1.0 $&$ 0.27\pm0.07$&[3]\\
\label{longtab5}
\end{longtable}

\begin{longtable}{lccccccccr}
\caption[]{Oxygen abundances from OH band synthesis for a set of stars with planets and brown dwarf 
companions.}\\
\hline
Star & $T_\mathrm{eff}$ & $\log\mathrm{g}$ & $\xi_t$ & $[Fe/H]$ & $[O/H]_1$
&$[O/H]_2$ & $[O/H]_3$ & $[O/H]_4$ & $[O/H]_{avg}$ \\
 & (K) & (cm\,s$^{-2}$) & (km\,s$^{-1}$) & & & & & & \\
\hline 
\hline
\endfirsthead
\caption[]{Continued.}\\
\hline
Star & $T_\mathrm{eff}$ & $\log\mathrm{g}$ & $\xi_t$ & $[Fe/H]$ & $[O/H]_1$
&$[O/H]_2$ & $[O/H]_3$ & $[O/H]_4$ & $[O/H]_{avg}$ \\
 & (K) & (cm\,s$^{-2}$) & (km\,s$^{-1}$) & & & & & & \\
\hline 
\hline
\endhead
\noalign{\smallskip}
\hline
\endfoot
\noalign{\smallskip}
\hline
\endlastfoot
\noalign{\smallskip}
\object{HD\,142	  }& $6302\pm56 $ & $4.34\pm0.13$  &   $1.86\pm0.17$ &  $ 0.14\pm0.07$ &  -0.02 &  0.03 &  0.08 &  0.18  &  $ 0.07\pm0.13$\\
\object{HD\,1237  }& $5536\pm50 $ & $4.56\pm0.12$  &   $1.33\pm0.06$ &  $ 0.12\pm0.06$ &   0.06 &  0.01 &  0.06 &  0.16  &  $ 0.07\pm0.10$\\
\object{HD\,4208  }& $5626\pm32 $ & $4.49\pm0.10$  &   $0.95\pm0.06$ &  $-0.24\pm0.04$ &  -0.15 & -0.10 & -0.20 & -0.10  &  $-0.14\pm0.08$\\
\object{HD\,23079 }& $5959\pm46 $ & $4.35\pm0.12$  &   $1.20\pm0.10$ &  $-0.11\pm0.06$ &  -0.02 &  0.03 &  0.03 &  0.03  &  $ 0.02\pm0.09$\\
\object{HD\,28185 }& $5656\pm44 $ & $4.45\pm0.08$  &   $1.01\pm0.06$ &  $ 0.22\pm0.05$ &   0.16 &  0.26 &  0.21 &  0.31  &  $ 0.23\pm0.10$\\
\object{HD\,30177 }& $5591\pm50 $ & $4.35\pm0.12$  &   $1.03\pm0.06$ &  $ 0.39\pm0.06$ &   0.23 &  0.23 &  0.38 &  0.38  &  $ 0.31\pm0.12$\\
\object{HD\,33636 }& $6046\pm49 $ & $4.71\pm0.09$  &   $1.79\pm0.19$ &  $-0.08\pm0.06$ &   0.01 &  0.06 &  0.11 &  0.11  &  $ 0.07\pm0.10$\\
\object{HD\,37124 }& $5546\pm30 $ & $4.50\pm0.03$  &   $0.80\pm0.07$ &  $-0.38\pm0.04$ &  -0.14 & -0.04 & -0.04 & -0.04  &  $-0.06\pm0.08$\\
\object{HD\,39091 }& $5991\pm27 $ & $4.42\pm0.10$  &   $1.24\pm0.04$ &  $ 0.10\pm0.04$ &   0.19 &  0.19 &  0.24 &  0.29  &  $ 0.23\pm0.08$\\
\object{HD\,47536 }& $4554\pm85 $ & $2.48\pm0.23$  &   $1.82\pm0.08$ &  $-0.54\pm0.12$ &  -0.30 & -0.30 & -0.20 & -0.35  &  $-0.29\pm0.12$\\
\object{HD\,50554 }& $6026\pm30 $ & $4.41\pm0.13$  &   $1.11\pm0.06$ &  $ 0.01\pm0.04$ &   0.15 &  0.15 &  0.15 &  0.25  &  $ 0.17\pm0.09$\\
\object{HD\,59686 }& $4871\pm135$ & $3.15\pm0.41$  &   $1.85\pm0.12$ &  $ 0.28\pm0.18$ &   0.12 &  0.12 &  0.22 &  0.12  &  $ 0.14\pm0.16$\\
\object{HD\,65216 }& $5666\pm31 $ & $4.53\pm0.09$  &   $1.06\pm0.05$ &  $-0.12\pm0.04$ &  -0.08 &  0.02 &  0.02 &  0.02  &  $-0.00\pm0.08$\\
\object{HD\,70642 }& $5693\pm26 $ & $4.41\pm0.09$  &   $1.01\pm0.04$ &  $ 0.18\pm0.04$ &   0.12 &  0.12 &  0.22 &  0.22  &  $ 0.17\pm0.09$\\
\object{HD\,72659 }& $5995\pm45 $ & $4.30\pm0.07$  &   $1.42\pm0.09$ &  $ 0.03\pm0.06$ &   0.17 &  0.17 &  0.22 &  0.27  &  $ 0.21\pm0.09$\\
\object{HD\,73256 }& $5518\pm49 $ & $4.42\pm0.12$  &   $1.22\pm0.06$ &  $ 0.26\pm0.06$ &   0.00 &  0.10 &  0.15 &  0.20  &  $ 0.11\pm0.11$\\
\object{HD\,74156 }& $6112\pm39 $ & $4.34\pm0.10$  &   $1.38\pm0.07$ &  $ 0.16\pm0.05$ &   0.30 &  0.30 &  0.35 &  0.35  &  $ 0.32\pm0.08$\\
\object{HD\,88133 }& $5438\pm34 $ & $3.94\pm0.11$  &   $1.16\pm0.03$ &  $ 0.33\pm0.05$ &   0.17 &  0.17 &  0.27 &  0.27  &  $ 0.22\pm0.09$\\
\object{HD\,99492 }& $4810\pm72 $ & $4.21\pm0.21$  &   $0.72\pm0.13$ &  $ 0.26\pm0.07$ &   0.00 &  0.00 &  0.05 &  0.00  &  $ 0.01\pm0.08$\\
\object{HD\,106252}& $5899\pm35 $ & $4.34\pm0.07$  &   $1.08\pm0.06$ &  $-0.01\pm0.05$ &   0.08 &  0.08 &  0.13 &  0.08  &  $ 0.09\pm0.07$\\
\object{HD\,114729}& $5886\pm36 $ & $4.28\pm0.13$  &   $1.25\pm0.09$ &  $-0.25\pm0.05$ &  -0.01 & -0.01 & -0.01 & -0.01  &  $-0.01\pm0.07$\\
\object{HD\,117207}& $5654\pm33 $ & $4.32\pm0.05$  &   $1.13\pm0.04$ &  $ 0.23\pm0.05$ &   0.12 &  0.17 &  0.22 &  0.27  &  $ 0.19\pm0.09$\\
\object{HD\,117618}& $6013\pm41 $ & $4.39\pm0.07$  &   $1.73\pm0.09$ &  $ 0.06\pm0.06$ &   0.12 &  0.10 &  0.15 &  0.20  &  $ 0.14\pm0.09$\\
\object{HD\,213240}& $5984\pm33 $ & $4.25\pm0.10$  &   $1.25\pm0.05$ &  $ 0.17\pm0.05$ &   0.23 &  0.21 &  0.26 &  0.31  &  $ 0.25\pm0.08$\\
\object{HD\,216435}& $5938\pm42 $ & $4.12\pm0.05$  &   $1.28\pm0.06$ &  $ 0.24\pm0.05$ &   0.20 &  0.13 &  0.28 &  0.33  &  $ 0.23\pm0.12$\\
\object{HD\,216437}& $5887\pm32 $ & $4.30\pm0.07$  &   $1.31\pm0.04$ &  $ 0.25\pm0.04$ &   0.20 &  0.13 &  0.28 &  0.28  &  $ 0.22\pm0.10$\\
\object{HD\,219449}& $4757\pm102$ & $2.71\pm0.25$  &   $1.71\pm0.09$ &  $ 0.05\pm0.14$ &  -0.16 & -0.11 & -0.01 & -0.11  &  $-0.10\pm0.13$\\
\object{HD\,6434}  & $5835\pm50$& $4.60\pm0.15$& $1.53\pm0.10$& $-0.52\pm0.05$ &-0.26 &-0.13 &-0.20 &-0.13 &$ -0.18\pm0.10$\\ 
\object{HD\,9826}  & $6212\pm64$& $4.26\pm0.13$& $1.69\pm0.16$& $ 0.13\pm0.08$ & 0.17 & 0.27 & 0.17 & 0.27 &$  0.22\pm0.12$\\
\object{HD\,10647} & $6143\pm31$& $4.48\pm0.08$& $1.40\pm0.08$& $-0.03\pm0.04$ &-0.01 &-0.11 &-0.01 &-0.11 &$ -0.06\pm0.09$\\
\object{HD\,13445} & $5163\pm37$& $4.52\pm0.13$& $0.72\pm0.06$& $-0.24\pm0.05$ &-0.40 &-0.20 &-0.20 &-0.20 &$ -0.25\pm0.12$\\
\object{HD\,16141} & $5801\pm30$& $4.22\pm0.12$& $1.34\pm0.04$& $ 0.15\pm0.04$ & 0.09 & 0.24 & 0.25 & 0.20 &$  0.19\pm0.10$\\
\object{HD\,17051} & $6252\pm53$& $4.61\pm0.16$& $1.18\pm0.10$& $ 0.26\pm0.06$ & 0.29 & 0.34 & 0.39 & 0.29 &$  0.33\pm0.11$\\
\object{HD\,19994} & $6190\pm57$& $4.19\pm0.13$& $1.54\pm0.13$& $ 0.24\pm0.07$ & 0.28 & 0.33 & 0.38 & 0.23 &$  0.31\pm0.12$\\
\object{HD\,22049} & $5073\pm42$& $4.43\pm0.08$& $1.05\pm0.06$& $-0.13\pm0.04$ &-0.34 &-0.18 &-0.22 &-0.19 &$ -0.23\pm0.10$\\
\object{HD\,27442} &$4825\pm107$& $3.55\pm0.32$& $1.18\pm0.12$& $ 0.39\pm0.13$ &-0.02 & 0.16 & 0.08 & -    &$  0.07\pm0.14$\\
\object{HD\,46375} & $5268\pm55$& $4.41\pm0.16$& $0.97\pm0.06$& $ 0.20\pm0.06$ &-0.06 & 0.09 & 0.09 & 0.04 &$  0.04\pm0.10$\\
\object{HD\,52265} & $6103\pm52$& $4.28\pm0.15$& $1.36\pm0.09$& $ 0.23\pm0.07$ & 0.19 & 0.19 & 0.27 & 0.19 &$  0.21\pm0.10$\\
\object{HD\,75289} & $6143\pm53$& $4.42\pm0.13$& $1.53\pm0.09$& $ 0.28\pm0.07$ & 0.19 & 0.19 & 0.29 & 0.19 &$  0.22\pm0.11$\\
\object{HD\,82943} & $6016\pm30$& $4.46\pm0.08$& $1.13\pm0.04$& $ 0.30\pm0.04$ & 0.24 & 0.29 & 0.39 & 0.24 &$  0.29\pm0.10$\\
\object{HD\,83443} & $5454\pm61$& $4.33\pm0.17$& $1.08\pm0.08$& $ 0.35\pm0.08$ &-0.01 & 0.19 & 0.19 & -    &$  0.12\pm0.13$\\
\object{HD\,143761} & $5853\pm25$& $4.41\pm0.15$& $1.35\pm0.07$& $-0.21\pm0.04$&-0.17 &-0.07 &-0.07 &-0.07 &$ -0.09\pm0.08$\\
\object{HD\,169830} & $6299\pm41$& $4.10\pm0.02$& $1.42\pm0.09$& $ 0.21\pm0.05$& 0.19 & 0.29 & 0.29 & 0.09 &$  0.22\pm0.12$\\
\object{HD\,179949} & $6260\pm43$& $4.43\pm0.05$& $1.41\pm0.09$& $ 0.22\pm0.05$& 0.22 & 0.34 & 0.29 & 0.19 &$  0.26\pm0.11$\\
\object{HD\,192263} & $4947\pm58$& $4.51\pm0.20$& $0.86\pm0.09$& $-0.02\pm0.06$&-0.21 &-0.11 &-0.18 &-0.16 &$ -0.17\pm0.09$\\ 
\object{HD\,202206} & $5752\pm53$& $4.50\pm0.09$& $1.01\pm0.06$& $ 0.35\pm0.06$& 0.14 & 0.24 & 0.24 & 0.19 &$  0.20\pm0.09$\\
\object{HD\,209458} & $6117\pm26$& $4.48\pm0.08$& $1.40\pm0.06$& $ 0.02\pm0.03$&-0.01 & 0.06 & 0.06 & 0.02 &$  0.03\pm0.07$\\
\object{HD\,210277} & $5532\pm28$& $4.29\pm0.09$& $1.04\pm0.03$& $ 0.19\pm0.04$&-0.01 & 0.19 & 0.22 & 0.11 &$  0.13\pm0.12$\\
\object{HD\,217014} & $5804\pm36$& $4.42\pm0.07$& $1.20\pm0.05$& $ 0.20\pm0.05$& 0.04 &-0.11 &-0.01 &-0.16 &$ -0.06\pm0.11$\\
\object{HD\,217107} & $5646\pm34$& $4.31\pm0.10$& $1.06\pm0.04$& $ 0.37\pm0.05$& 0.14 & 0.34 & 0.39 & 0.29 &$  0.29\pm0.13$\\
\object{HD\,222582 }& $5843\pm38$& $4.45\pm0.07$& $1.03\pm0.06$& $ 0.05\pm0.05$& 0.04 & 0.14 & 0.09 & 0.09 &$  0.09\pm0.08$\\
\label{longtab6}
\end{longtable}

\begin{longtable}{lccccccccr}
\caption[]{Oxygen abundances from OH band synthesis for a set of comparison stars (stars without giant
planets).}\\
\hline
Star & $T_\mathrm{eff}$ & $\log\mathrm{g}$ & $\xi_t$ & $[Fe/H]$ & $[O/H]_1$
&$[O/H]_2$ & $[O/H]_3$ & $[O/H]_4$ & $[O/H]_{avg}$ \\
 & (K) & (cm\,s$^{-2}$) & (km\,s$^{-1}$) & & & & & & \\
\hline 
\hline
\endfirsthead
\caption[]{Continued.}\\
\hline
Star & $T_\mathrm{eff}$ & $\log\mathrm{g}$ & $\xi_t$ & $[Fe/H]$ & $[O/H]_1$
&$[O/H]_2$ & $[O/H]_3$ & $[O/H]_4$ & $[O/H]_{avg}$ \\
 & (K) & (cm\,s$^{-2}$) & (km\,s$^{-1}$) & & & & & & \\
\hline 
\hline
\endhead
\noalign{\smallskip}
\hline
\endfoot
\noalign{\smallskip}
\hline
\endlastfoot
\noalign{\smallskip}
\object{HD\,1461}  & $5785\pm50$ & $4.47\pm0.15$ & $1.00\pm0.10$ & $ 0.18\pm0.05$ & 0.09 & 0.19 & 0.24 & 0.19 &$  0.18\pm0.10$\\ 
\object{HD\,1581}  & $5956\pm44$ & $4.39\pm0.13$ & $1.07\pm0.09$ & $-0.14\pm0.05$ &-0.05 & 0.00 & 0.05 & 0.00 &$  0.00\pm0.09$\\
\object{HD\,3823}  & $5950\pm50$ & $4.12\pm0.15$ & $1.00\pm0.10$ & $-0.27\pm0.05$ &-0.13 &-0.05 &-0.05 &-0.08 &$ -0.08\pm0.09$\\
\object{HD\,4391}  & $5878\pm53$ & $4.74\pm0.15$ & $1.13\pm0.10$ & $-0.03\pm0.06$ &-0.09 & 0.04 & 0.03 & 0.00 &$ -0.00\pm0.10$\\
\object{HD\,7570}  & $6140\pm41$ & $4.39\pm0.16$ & $1.50\pm0.08$ & $ 0.18\pm0.05$ & 0.22 & 0.27 & 0.32 & 0.22 &$  0.26\pm0.09$\\
\object{HD\,10700} & $5344\pm29$ & $4.57\pm0.09$ & $0.91\pm0.06$ & $-0.52\pm0.04$ &-0.48 &-0.28 &-0.38 &-0.38 &$ -0.38\pm0.10$\\
\object{HD\,14412} & $5368\pm24$ & $4.55\pm0.05$ & $0.88\pm0.05$ & $-0.47\pm0.03$ &-0.53 &-0.33 &-0.43 &-0.33 &$ -0.41\pm0.11$\\
\object{HD\,20010} & $6275\pm57$ & $4.40\pm0.37$ & $2.41\pm0.41$ & $-0.19\pm0.06$ &-0.08 &-0.07 & 0.00 &-0.05 &$ -0.05\pm0.12$\\
\object{HD\,20766} & $5733\pm31$ & $4.55\pm0.10$ & $1.09\pm0.06$ & $-0.21\pm0.04$ &-0.17 &-0.07 &-0.12 &-0.07 &$ -0.11\pm0.08$\\
\object{HD\,20794} & $5444\pm31$ & $4.47\pm0.07$ & $0.98\pm0.06$ & $-0.38\pm0.04$ &-0.24 &-0.04 &-0.14 &-0.04 &$ -0.12\pm0.11$\\
\object{HD\,20807} & $5843\pm26$ & $4.47\pm0.10$ & $1.17\pm0.06$ & $-0.23\pm0.04$ &-0.12 &-0.06 &-0.09 &-0.09 &$ -0.09\pm0.07$\\
\object{HD\,23484} & $5176\pm45$ & $4.41\pm0.17$ & $1.03\pm0.06$ &  $0.06\pm0.05$ &-0.27 &-0.09 &-0.10 &-0.13 &$ -0.15\pm0.11$\\
\object{HD\,30495} & $5868\pm30$ & $4.55\pm0.10$ & $1.24\pm0.05$ &  $0.02\pm0.04$ &-0.04 & 0.11 & 0.06 & 0.01 &$  0.04\pm0.09$\\
\object{HD\,36435} & $5479\pm37$ & $4.61\pm0.07$ & $1.12\pm0.05$ & $-0.00\pm0.05$ &-0.26 &-0.11 &-0.11 &-0.16 &$ -0.16\pm0.10$\\
\object{HD\,38858} & $5752\pm32$ & $4.53\pm0.07$ & $1.26\pm0.07$ & $-0.23\pm0.05$ &-0.19 &-0.09 &-0.14 &-0.14 &$ -0.14\pm0.08$\\
\object{HD\,43162} & $5633\pm35$ & $4.48\pm0.07$ & $1.24\pm0.05$ & $-0.01\pm0.04$ &-0.20 &-0.07 &-0.07 &-0.17 &$ -0.13\pm0.09$\\
\object{HD\,43834} & $5594\pm36$ & $4.41\pm0.09$ & $1.05\pm0.04$ &  $0.10\pm0.05$ & 0.00 & 0.14 & 0.14 & 0.09 &$  0.09\pm0.10$\\
\object{HD\,69830} & $5410\pm26$ & $4.38\pm0.07$ & $0.89\pm0.03$ & $-0.03\pm0.04$ &-0.09 & 0.06 & 0.01 & 0.01 &$ -0.00\pm0.09$\\
\object{HD\,72673} & $5242\pm28$ & $4.50\pm0.09$ & $0.69\pm0.05$ & $-0.37\pm0.04$ &-0.43 &-0.23 &-0.33 &-0.23 &$ -0.31\pm0.11$\\
\object{HD\,74576} & $5000\pm55$ & $4.55\pm0.13$ & $1.07\pm0.08$ & $-0.03\pm0.06$ &-0.34 &-0.19 &-0.22 &-0.24 &$ -0.25\pm0.10$\\  
\object{HD\,76151} & $5803\pm29$ & $4.50\pm0.08$ & $1.02\pm0.04$ &  $0.14\pm0.04$ & 0.13 & 0.26 & 0.23 & 0.18 &$  0.20\pm0.09$\\
\object{HD\,84117} & $6167\pm37$ & $4.35\pm0.10$ & $1.42\pm0.09$ & $-0.03\pm0.05$ & 0.04 & 0.11 & 0.11 & 0.01 &$  0.07\pm0.09$\\
\object{HD\,189567}& $5765\pm24$ & $4.52\pm0.05$ & $1.22\pm0.05$ & $-0.23\pm0.04$ &-0.09 & 0.06 &-0.04 & 0.01 &$ -0.01\pm0.09$\\
\object{HD\,192310}& $5069\pm49$ & $4.38\pm0.19$ & $0.79\pm0.07$ & $-0.01\pm0.05$ &-0.27 &-0.07 &-0.17 &-0.12 &$ -0.16\pm0.11$\\
\object{HD\,211415}& $5890\pm30$ & $4.51\pm0.07$ & $1.12\pm0.07$ & $-0.17\pm0.04$ &-0.03 & 0.00 &-0.08 &-0.03 &$ -0.04\pm0.07$\\
\object{HD\,222335}& $5260\pm41$ & $4.45\pm0.11$ & $0.92\pm0.06$ & $-0.16\pm0.05$ &-0.32 &-0.17 &-0.10 &-0.17 &$ -0.19\pm0.12$\\
\label{longtab7}
\end{longtable}

\begin{landscape}
\begin{scriptsize}
\begin{longtable}{lccccccccccccccr}
\caption[]{Oxygen abundances from Triplet lines in a set of planet host stars.}\\
\hline
Star & $T_\mathrm{eff}$ & $\log {g}$ & $\xi_t$ & [Fe/H] & EW$_{1}$ & [O/H]$_{1}$ & [O/H]$_{1}^{NLTE}$ &
 EW$_{2}$ & [O/H]$_{2}$& [O/H]$_{2}^{NLTE}$ & EW$_{3}$ & [O/H]$_{3}$ & [O/H]$_{3}^{NLTE}$ &[O/H]$_{avg}^{NLTE}$ & Instr. \\
 & (K) & (cm\,s$^{-2}$) & (km\,s$^{-1}$) & & (m\AA) & 7775 \AA & 7775 \AA & (m\AA) & 7774 \AA  & 7774  \AA & (m\AA) & 7772 \AA & 7772 \AA & & \\
\hline 
\hline
\endfirsthead
\caption[]{Continued.}\\
\hline
Star & $T_\mathrm{eff}$ & $\log {g}$ & $\xi_t$ & [Fe/H] & EW$_{1}$ & [O/H]$_{1}$ & [O/H]$_{1}^{NLTE}$ &
 EW$_{2}$ & [O/H]$_{2}$& [O/H]$_{2}^{NLTE}$ & EW$_{3}$ & [O/H]$_{3}$ & [O/H]$_{3}^{NLTE}$ &[O/H]$_{avg}^{NLTE}$ & Instr. \\
 & (K) & (cm\,s$^{-2}$) & (km\,s$^{-1}$) & & (m\AA) & 7775 \AA & 7775 \AA & (m\AA) & 7774 \AA  & 7774  \AA & (m\AA) & 7772 \AA & 7772 \AA & & \\
\hline 
\hline
\endhead
\noalign{\smallskip}
\hline
\noalign{\footnotesize{The instruments used to obtain the spectra were: [1]UVES; [2]UES; [3]FEROS; [4]SARG; [5]CORALIE.}}
\noalign{\footnotesize{$^1$ This value was excluded from the figures since it is inconsistent with the [O I] result.}}
\endfoot
\noalign{\smallskip}
\hline
\noalign{\footnotesize{The instruments used to obtain the spectra were: [1]UVES; [2]UES; [3]FEROS; [4]SARG; [5]CORALIE.}}
\endlastfoot
\noalign{\smallskip}
\object{HD\,3651}  & $5173\pm35 $ & $4.37\pm0.12$ & $0.74\pm0.05$ &$ 0.12\pm0.04 $ & 44.7 & 0.21 & 0.00 & 42.8 & 0.31 & 0.11 & 33.4 & 0.31 & 0.14 &$ 0.08\pm0.10$&[4]\\
\object{HD\,8574}  & $6151\pm57 $ & $4.51\pm0.10$ & $1.45\pm0.15$ &$ 0.06\pm0.07 $ &121.0 & 0.22 &-0.26 &100.2 & 0.15 &-0.26 & 80.5 & 0.12 &-0.29 &$-0.27\pm0.08$&[4]\\
\object{HD\,9826}  & $6212\pm64 $ & $4.26\pm0.13$ & $1.69\pm0.16$ &$ 0.13\pm0.08 $ &125.9 & 0.18 &-0.33 &112.0 & 0.18 &-0.28 & 94.0 & 0.18 &-0.21 &$-0.27\pm0.09$&[2]\\
\object{HD\,10697} & $5641\pm28 $ & $4.05\pm0.05$ & $1.13\pm0.03$ &$ 0.14\pm0.04 $ &118.7 & 0.61 & 0.13 & 90.2 & 0.43 & 0.04 & 81.1 & 0.52 & 0.13 &$ 0.10\pm0.08$&[4]\\
\object{HD\,12661} & $5702\pm36 $ & $4.33\pm0.08$ & $1.05\pm0.04$ &$ 0.36\pm0.05 $ & 82.3 & 0.20 &-0.13 & 70.5 & 0.18 &-0.10 & 56.6 & 0.16 &-0.07 &$-0.10\pm0.07$&[2]\\
\object{HD\,16141} & $5801\pm30 $ & $4.22\pm0.12$ & $1.34\pm0.04$ &$ 0.15\pm0.04 $ & 83.0 & 0.04 &-0.32 & 74.3 & 0.07 &-0.26 & 60.8 & 0.07 &-0.20 &$-0.26\pm0.08$&[2]\\
\object{HD\,19994} & $6190\pm00 $ & $4.19\pm0.00$ & $1.54\pm0.00$ &$ 0.24\pm0.00 $ &133.5 & 0.29 &-0.24 &123.4 & 0.34 &-0.15 & 92.0 & 0.17 &-0.20 &$-0.20\pm0.07$&[2]\\
\object{HD\,22049} & $5073\pm42 $ & $4.43\pm0.08$ & $1.05\pm0.06$ &$-0.13\pm0.05 $ & 26.5 &-0.12 &-0.28 & 24.2 &-0.04 &-0.19 & 15.6 &-0.14 &-0.26 &$-0.24\pm0.10$&[2]\\
\object{HD\,23596} & $6108\pm36 $ & $4.25\pm0.10$ & $1.30\pm0.05$ &$ 0.31\pm0.05 $ &117.8 & 0.23 &-0.23 &105.1 & 0.23 &-0.18 & 85.2 & 0.19 &-0.14 &$-0.18\pm0.08$&[2]\\
\object{HD\,37124} & $5546\pm30 $ & $4.50\pm0.03$ & $0.80\pm0.07$ &$-0.38\pm0.04 $ & 69.2 & 0.14 &-0.16 & 58.2 & 0.13 &-0.14 & 43.6 & 0.11 &-0.11 &$-0.14\pm0.07$&[2]\\
\object{HD\,40979} & $6145\pm42 $ & $4.31\pm0.15$ & $1.29\pm0.09$ &$ 0.21\pm0.05 $ &122.5 & 0.25 &-0.23 &111.4 & 0.28 &-0.16 & 88.0 & 0.21 &-0.14 &$-0.18\pm0.08$&[4]\\
\object{HD\,46375} & $5268\pm55 $ & $4.41\pm0.16$ & $0.97\pm0.06$ &$ 0.20\pm0.06 $ & 51.2 & 0.22 &-0.01 & 48.3 & 0.31 & 0.09 & 40.5 & 0.37 & 0.18 &$ 0.09\pm0.14$&[2]\\
\object{HD\,50554} & $6026\pm30 $ & $4.41\pm0.13$ & $1.11\pm0.06$ &$ 0.01\pm0.04 $ & 88.6 &-0.04 &-0.42 & 81.1 & 0.01 &-0.34 & 64.6 & 0.00 &-0.29 &$-0.35\pm0.10$&[2]\\
\object{HD\, 59686}&$4871\pm135 $&$ 3.15\pm0.24 $&$ 1.85\pm0.12 $&$  0.28\pm0.18 $ & 47.3 & 0.27 & 0.01 & 48.0 & 0.43 & 0.17 & 40.2 & 0.46 & 0.23 &$ 0.14\pm0.30$&[3] \\
\object{HD\,68988} & $5988\pm52 $ & $4.45\pm0.15$ & $1.25\pm0.08$ &$ 0.36\pm0.06 $ &105.8 & 0.23 &-0.16 & 92.5 & 0.22 &-0.12 & 75.5 & 0.21 &-0.07 &$-0.12\pm0.10$&[4]\\
\object{HD\, 70642}&$5671\pm 46 $&$ 4.39\pm0.05 $&$ 1.01\pm0.06 $&$  0.20\pm0.06 $ & 80.3 & 0.20 &-0.13 & 68.9 & 0.18 &-0.11 & 64.0 & 0.33 & 0.06 &$-0.06\pm0.13$&[3] \\
\object{HD\,73256} & $5518\pm49 $ & $4.42\pm0.12$ & $1.22\pm0.06$ &$ 0.26\pm0.06 $ & 68.3 & 0.22 &-0.06 & 59.2 & 0.22 &-0.03 & 44.6 & 0.17 &-0.03 &$-0.04\pm0.08$&[3]\\
\object{HD\,75732} & $5279\pm62 $ & $4.37\pm0.18$ & $0.98\pm0.07$ &$ 0.33\pm0.07 $ & 51.7 & 0.22 & 0.00 & 47.1 & 0.28 & 0.07 & 33.7 & 0.19 & 0.03 &$ 0.03\pm0.11$&[2]\\
\object{HD\,80606} & $5574\pm72 $ & $4.46\pm0.20$ & $1.14\pm0.09$ &$ 0.32\pm0.09 $ & 71.9 & 0.20 &-0.09 & 61.8 & 0.19 &-0.06 & 51.1 & 0.23 & 0.02 &$-0.04\pm0.13$&[2]\\
\object{HD\,82943} & $6015\pm00 $ & $4.46\pm0.00$ & $1.13\pm0.00$ &$ 0.30\pm0.00 $ &103.4 & 0.18 &-0.21 & 87.1 & 0.13 &-0.20 & 74.3 & 0.18 &-0.10 &$-0.17\pm0.08$&[2]\\
\object{HD\, 83443}&$5501\pm 63 $&$ 4.46\pm0.09 $&$ 1.07\pm0.08 $&$  0.39\pm0.07 $ & 85.9 & 0.49 & 0.17 & 73.8 & 0.48 & 0.20 & 53.5 & 0.37 & 0.16 &$ 0.17\pm0.09$&[3] \\
\object{HD\,89744} & $6234\pm45 $ & $3.98\pm0.05$ & $1.62\pm0.08$ &$ 0.22\pm0.05 $ &140.2 & 0.30 &-0.29 &124.8 & 0.27 &-0.26 &101.0 & 0.20 &-0.23 &$-0.26\pm0.07$&[2]\\
\object{HD\, 92788}&$5758\pm 37 $&$ 4.30\pm0.04 $&$ 1.10\pm0.04 $&$  0.34\pm0.05 $&  95.3 & 0.99 & 0.62 & 72.7 & 0.93 & 0.63 & 67.3 & 1.03 & 0.76 &$ 0.67\pm0.09^1 $& [3] \\
\object{HD\,104985}& $4773\pm62 $ & $2.76\pm0.14$ & $1.71\pm0.07$ &$-0.28\pm0.09 $ & 54.7 &-0.19 &-0.48 & 50.3 &-0.12 &-0.38 & 40.1 &-0.11 &-0.33 &$-0.40\pm0.20$&[4]\\
\object{HD\,106252}&$5834\pm 37 $&$ 4.22\pm0.02 $&$ 1.06\pm0.06 $&$ -0.03\pm0.05 $&  84.7 & 0.04 &-0.34 & 72.9 & 0.02 &-0.32 & 57.8 & 0.01 &-0.27 &$-0.31\pm0.07$&[3] \\
\object{HD\,108874}& $5596\pm42 $ & $4.37\pm0.12$ & $0.89\pm0.05$ &$ 0.23\pm0.05 $ & 64.2 & 0.05 &-0.23 & 58.5 & 0.10 &-0.16 & 44.1 & 0.05 &-0.15 &$-0.18\pm0.08$&[2]\\
\object{HD\,114386}&$4865\pm 93 $&$ 4.30\pm0.17 $&$ 0.86\pm0.12 $&$ -0.04\pm0.07 $&  37.8 & 1.27 & 1.10 & 24.9 & 1.20 & 1.06 &-     &-     &-     &$ 1.15\pm0.18^1 $& [3] \\
\object{HD\,114762}& $5884\pm34 $ & $4.22\pm0.02$ & $1.31\pm0.17$ &$-0.70\pm0.04 $ & 65.1 &-0.31 &-0.64 & 53.2 &-0.34 &-0.63 & 42.0 &-0.31 &-0.55 &$-0.61\pm0.08$&[2]\\
\object{HD\,117176}& $5560\pm34 $ & $4.07\pm0.05$ & $1.18\pm0.05$ &$-0.06\pm0.05 $ & 70.0 & 0.06 &-0.28 & 56.3 &-0.02 &-0.31 & 40.2 &-0.11 &-0.33 &$-0.31\pm0.07$&[4]\\
\object{HD\,120136}& $6339\pm73 $ & $4.19\pm0.10$ & $1.70\pm0.16$ &$ 0.23\pm0.07 $ &166.2 & 0.49 &-0.16 &140.8 & 0.40 &-0.16 &118.1 & 0.37 &-0.10 &$-0.14\pm0.08$&[2]\\
\object{HD\,130322}& $5392\pm36$ & $4.48\pm0.06$ & $0.85\pm0.05$ &$ 0.03\pm0.04 $  & 51.5 & 0.06 &-0.18 & 47.6 & 0.14 &-0.09 & 37.4 & 0.15 &-0.04 &$-0.10\pm0.10$&[4]\\
\object{HD\,134987}& $5776\pm29$ & $4.36\pm0.07$ & $1.09\pm0.04$ &$ 0.30\pm0.04 $  &101.2 & 0.36 &-0.03 & 85.4 & 0.31 &-0.03 & 75.0 & 0.39 & 0.10 &$ 0.01\pm0.10$&[4]\\
\object{HD\,136118}& $6222\pm39$ & $4.27\pm0.15$ & $1.79\pm0.12$ &$-0.04\pm0.05 $  &145.0 & 0.34 &-0.27 &121.5 & 0.25 &-0.28 & 90.5 & 0.11 &-0.29 &$-0.28\pm0.06$&[4]\\
\object{HD\,141937}& $5909\pm39$ & $4.51\pm0.08$ & $1.13\pm0.06$ &$ 0.10\pm0.05 $  & 90.3 & 0.10 &-0.27 & 83.6 & 0.17 &-0.17 & 63.6 & 0.10 &-0.17 &$-0.20\pm0.09$&[2]\\
\object{HD\,143761}& $5853\pm25$ & $4.41\pm0.15$ & $1.35\pm0.07$ &$-0.21\pm0.04 $  & 74.0 &-0.12 &-0.46 & 66.2 &-0.08 &-0.39 & 52.8 &-0.07 &-0.33 &$-0.39\pm0.10$&[4]\\
\object{HD\,147513}&$5894\pm 31 $&$ 4.43\pm0.02 $&$ 1.26\pm0.05 $&$ 0.08\pm0.04 $ &  90.6 & 0.09 &-0.28 & 71.9 & 0.01 &-0.30 & 64.8 & 0.11 &-0.17 &$-0.25\pm0.09$& [3] \\
\object{HD\,150706}& $5961\pm27$ & $4.50\pm0.10$ & $1.11\pm0.06$ &$-0.01\pm0.04 $  & 80.3 &-0.08 &-0.43 & 70.1 &-0.07 &-0.38 & 50.1 &-0.16 &-0.39 &$-0.40\pm0.07$&[2]\\
\object{HD\,168443}& $5617\pm35$ & $4.22\pm0.05$ & $1.21\pm0.05$ &$ 0.06\pm0.05 $ & 68.5 & 0.02 &-0.30 & 54.8 &-0.06 &-0.33 & 45.3 &-0.02 &-0.25 &$-0.29\pm0.07$&[4]\\
\object{HD\,168746}&$5601\pm 33 $&$ 4.41\pm0.12 $&$ 0.99\pm0.05 $&$ -0.08\pm0.05 $& 73.2 & 0.14 &-0.19 & 64.6 & 0.17 &-0.13 & 49.8 & 0.14 &-0.10 &$-0.14\pm0.08$&[4]\\
\object{HD\,178911B}&$5600\pm42$ & $4.44\pm0.08$ & $0.95\pm0.05$ &$ 0.27\pm0.05 $ & 79.2 & 0.28 &-0.03 & 72.0 & 0.33 & 0.04 & 62.6 & 0.40 & 0.15 &$ 0.05\pm0.12$&[4]\\
\object{HD\,179949}& $6260\pm43$ & $4.43\pm0.05$ & $1.41\pm0.09$ &$ 0.23\pm0.05 $ &130.9 & 0.26 &-0.23 &115.6 & 0.25 &-0.19 & 91.3 & 0.18 &-0.17 &$-0.20\pm0.07$&[4]\\
\object{HD\,186427}& $5772\pm25$ & $4.40\pm0.07$ & $1.07\pm0.04$ &$ 0.08\pm0.04 $ & 73.7 &-0.01 &-0.33 & 67.4 & 0.05 &-0.25 & 53.2 & 0.04 &-0.20 &$-0.26\pm0.09$&[4]\\
\object{HD\,187123}& $5845\pm22$ & $4.42\pm0.07$ & $1.10\pm0.03$ &$ 0.13\pm0.03 $ & 82.6 & 0.05 &-0.29 & 70.0 & 0.02 &-0.28 & 56.1 & 0.02 &-0.22 &$-0.26\pm0.07$&[4]\\
\object{HD\,190228}& $5325\pm00$ & $3.90\pm0.00$ & $1.11\pm0.00$ &$-0.26\pm0.00 $ & 43.9 &-0.19 &-0.44 & 34.4 &-0.25 &-0.47 & 30.4 &-0.14 &-0.34 &$-0.42\pm0.09$&[2]\\
\object{HD\,190360}& $5584\pm36$ & $4.37\pm0.06$ & $1.07\pm0.05$ &$ 0.24\pm0.05 $ & 78.6 & 0.26 &-0.06 & 67.6 & 0.25 &-0.03 & 57.0 & 0.30 & 0.06 &$-0.01\pm0.09$&[2]\\
\object{HD\,192263}& $4947\pm58$ & $4.51\pm0.20$ & $0.86\pm0.09$ &$-0.02\pm0.06 $ & 30.2 & 0.21 & 0.05 & 23.0 & 0.14 & 0.01 & 20.0 & 0.25 & 0.13 &$ 0.06\pm0.12$&[3]\\
\object{HD\,195019}& $5859\pm31$ & $4.32\pm0.07$ & $1.27\pm0.05$ &$ 0.09\pm0.04 $ & 83.0 & 0.01 &-0.35 & 72.0 & 0.00 &-0.32 & 55.3 &-0.04 &-0.29 &$-0.32\pm0.07$&[2]\\
\object{HD\,210277}& $5532\pm00$ & $4.29\pm0.00$ & $1.03\pm0.00$ &$ 0.19\pm0.00 $ & 76.6 & 0.27 &-0.05 & 65.2 & 0.25 &-0.03 & 53.1 & 0.26 & 0.02 &$-0.02\pm0.06$&[2]\\
\object{HD\,216770}& $5423\pm41$ & $4.40\pm0.13$ & $1.01\pm0.05$ &$ 0.26\pm0.04 $ & 57.0 & 0.13 &-0.12 & 53.5 & 0.21 &-0.02 & 40.1 & 0.16 &-0.02 &$-0.05\pm0.10$&[4]\\
\object{HD\,217107}& $5645\pm00$ & $4.31\pm0.00$ & $1.06\pm0.00$ &$ 0.37\pm0.00 $ & 73.8 & 0.13 &-0.17 & 69.7 & 0.22 &-0.06 & 55.3 & 0.20 &-0.03 &$-0.09\pm0.09$&[2]\\
\object{HD\,219542B}&$5732\pm31$ & $4.40\pm0.05$ & $0.99\pm0.04$ &$ 0.17\pm0.04 $ & 85.3 & 0.20 &-0.15 & 73.5 & 0.19 &-0.12 & 52.0 & 0.06 &-0.17 &$-0.15\pm0.07$&[4]\\
\object{HD\,222582}& $5843\pm38$ & $4.45\pm0.07$ & $1.03\pm0.06$ &$ 0.05\pm0.05 $ & 79.2 & 0.12 &-0.22 & 70.1 & 0.14 &-0.17 & 53.9 & 0.10 &-0.14 &$-0.18\pm0.08$&[2]\\
\object{BD\,103166}&$5325\pm 45 $&$ 4.36\pm0.07 $&$ 0.95\pm0.05 $&$  0.35\pm0.05 $& 66.2 & 0.41 & 0.14 & 61.3 & 0.48 & 0.23 & 41.0 & 0.31 & 0.13 &$ 0.17\pm0.09$&[3]\\
\label{longtab8}
\end{longtable}

\clearpage
\begin{longtable}{lccccccccccccccr}
\caption[]{Oxygen abundances from Triplet lines in a set of comparison stars.}\\
\hline
Star & $T_\mathrm{eff}$ & $\log {g}$ & $\xi_t$ & [Fe/H] & EW$_{1}$ & [O/H]$_{1}$ & [O/H]$_{1}^{NLTE}$ &
 EW$_{2}$ & [O/H]$_{2}$& [O/H]$_{2}^{NLTE}$ & EW$_{3}$ & [O/H]$_{3}$ & [O/H]$_{3}^{NLTE}$ &[O/H]$_{avg}^{NLTE}$ & Instr. \\
 & (K) & (cm\,s$^{-2}$) & (km\,s$^{-1}$) & & (m\AA) & 7775 \AA & 7775 \AA & (m\AA) & 7774 \AA  & 7774  \AA & (m\AA) & 7772 \AA & 7772 \AA & & \\
\hline 
\hline
\endfirsthead
\caption[]{Continued.}\\
\hline
Star & $T_\mathrm{eff}$ & $\log {g}$ & $\xi_t$ & [Fe/H] & EW$_{1}$ & [O/H]$_{1}$ & [O/H]$_{1}^{NLTE}$ &
 EW$_{2}$ & [O/H]$_{2}$& [O/H]$_{2}^{NLTE}$ & EW$_{3}$ & [O/H]$_{3}$ & [O/H]$_{3}^{NLTE}$ &[O/H]$_{avg}^{NLTE}$ & Instr. \\
 & (K) & (cm\,s$^{-2}$) & (km\,s$^{-1}$) & & (m\AA) & 7775 \AA & 7775 \AA & (m\AA) & 7774 \AA  & 7774  \AA & (m\AA) & 7772 \AA & 7772 \AA & & \\
\hline 
\hline
\endhead
\noalign{\smallskip}
\hline
\noalign{\footnotesize{The instruments used to obtain the spectra were: [1]UVES; [2]UES; [3]FEROS; [4]SARG; [5]CORALIE.}}
\endfoot
\noalign{\smallskip}
\hline
\noalign{\footnotesize{The instruments used to obtain the spectra were: [1]UVES; [2]UES; [3]FEROS; [4]SARG; [5]CORALIE.}}
\endlastfoot
\noalign{\smallskip}
\object{HD\, 52919}&$4740\pm 49 $&$ 4.25\pm0.10 $&$ 1.03\pm0.09 $&$ -0.05\pm0.06 $&   18.1 & 0.06 &-0.07 & 14.7 & 0.06 &-0.05 & 10.8 & 0.06 &-0.03 &$-0.05\pm0.10 $& [3] \\
\object{HD\, 53143}&$5462\pm 54 $&$ 4.47\pm0.07 $&$ 1.08\pm0.07 $&$  0.22\pm0.06 $&   66.7 & 0.24 &-0.04 & 60.1 & 0.29 & 0.04 & 42.6 & 0.30 & 0.11 &$ 0.04\pm0.11 $& [3] \\
\object{HD\, 57095}&$4945\pm 54 $&$ 4.45\pm0.10 $&$ 1.09\pm0.08 $&$ -0.03\pm0.06 $&   32.8 & 0.25 & 0.08 &-     &-     &-     & 24.0 & 0.37 & 0.24 &$ 0.16\pm0.13 $& [3] \\
\object{HD\, 64606}&$5351\pm 68 $&$ 4.62\pm0.07 $&$ 0.92\pm0.19 $&$ -0.71\pm0.09 $&   38.0 &-0.14 &-0.33 & 27.9 &-0.21 &-0.38 & 17.7 &-0.29 &-0.43 &$-0.38\pm0.11 $& [3] \\
\object{HD\, 67199}&$5136\pm 56 $&$ 4.54\pm0.08 $&$ 0.84\pm0.08 $&$  0.06\pm0.06 $&   44.7 & 0.28 & 0.08 & 40.0 & 0.33 & 0.14 & 28.9 & 0.28 & 0.13 &$ 0.12\pm0.10 $& [3] \\
\object{HD\,100623}&$5246\pm 37 $&$ 4.54\pm0.05 $&$ 1.00\pm0.06 $&$ -0.38\pm0.05 $&   27.2 &-0.29 &-0.47 & 23.6 &-0.24 &-0.40 & 14.8 &-0.33 &-0.46 &$-0.44\pm0.08 $& [3] \\
\object{HD\,102365}&$5667\pm 27 $&$ 4.59\pm0.02 $&$ 1.05\pm0.06 $&$ -0.27\pm0.04 $&   63.4 & 0.39 & 0.10 & 52.0 & 0.37 & 0.11 & 48.1 & 0.49 & 0.26 &$ 0.16\pm0.11 $& [3] \\
\object{HD\,102438}&$5639\pm 51 $&$ 4.60\pm0.04 $&$ 1.12\pm0.10 $&$ -0.23\pm0.06 $&   54.1 &-0.16 &-0.42 & 47.4 &-0.12 &-0.36 & 40.0 &-0.04 &-0.25 &$-0.34\pm0.12 $& [3] \\
\object{HD\,104304}&$5562\pm 50 $&$ 4.37\pm0.06 $&$ 1.10\pm0.06 $&$  0.27\pm0.06 $&   89.0 & 0.42 & 0.07 & 73.1 & 0.36 & 0.06 & 63.3 & 0.43 & 0.17 &$ 0.10\pm0.10 $& [3] \\
\object{HD\,109200}&$5103\pm 46 $&$ 4.47\pm0.07 $&$ 0.75\pm0.07 $&$ -0.24\pm0.05 $&   37.0 & 0.12 &-0.07 & 30.2 & 0.10 &-0.07 & 24.6 & 0.17 & 0.03 &$-0.04\pm0.10 $& [3] \\
\object{HD\,115617}&$5577\pm 33 $&$ 4.34\pm0.03 $&$ 1.07\pm0.04 $&$  0.01\pm0.05 $&   61.9 &-0.01 &-0.30 & 56.0 & 0.04 &-0.23 & 40.1 &-0.04 &-0.25 &$-0.26\pm0.07 $& [3] \\
\object{HD\,118972}&$5241\pm 66 $&$ 4.43\pm0.10 $&$ 1.24\pm0.08 $&$ -0.01\pm0.08 $&   49.0 & 0.17 &-0.06 & 40.0 & 0.14 &-0.06 & 31.0 & 0.15 &-0.02 &$-0.05\pm0.11 $& [3] \\
\object{HD\,125072}&$5001\pm115 $&$ 4.39\pm0.21 $&$ 1.21\pm0.15 $&$  0.24\pm0.11 $&-       &-     &-     & 33.0 & 0.34 & 0.18 & 24.0 & 0.29 & 0.16 &$ 0.17\pm0.18 $& [3] \\
\object{HD\,128620}&$5844\pm 42 $&$ 4.30\pm0.04 $&$ 1.18\pm0.05 $&$  0.28\pm0.06 $&   98.0 & 0.24 &-0.15 & 81.2 & 0.17 &-0.16 & 70.7 & 0.23 &-0.06 &$-0.12\pm0.09 $& [3] \\
\object{HD\,128621}&$5199\pm 80 $&$ 4.37\pm0.12 $&$ 1.05\pm0.10 $&$  0.19\pm0.09 $&   43.6 & 0.14 &-0.06 & 35.2 & 0.10 &-0.08 & 27.2 & 0.10 &-0.04 &$-0.06\pm0.12 $& [3] \\
\object{HD\,135204}&$5332\pm 37 $&$ 4.31\pm0.04 $&$ 0.84\pm0.05 $&$ -0.11\pm0.05 $&   56.8 & 0.18 &-0.09 & 50.2 & 0.21 &-0.04 & 35.1 & 0.12 &-0.07 &$-0.07\pm0.07 $& [3] \\
\object{HD\,136352}&$5667\pm 43 $&$ 4.39\pm0.03 $&$ 1.08\pm0.09 $&$ -0.31\pm0.06 $&   75.0 & 0.07 &-0.27 & 68.0 & 0.13 &-0.18 & 46.6 & 0.00 &-0.24 &$-0.23\pm0.08 $& [3] \\
\object{HD\,140901}&$5645\pm 37 $&$ 4.40\pm0.04 $&$ 1.14\pm0.05 $&$  0.13\pm0.05 $&   70.6 & 0.07 &-0.24 & 60.5 & 0.06 &-0.21 & 47.5 & 0.05 &-0.17 &$-0.20\pm0.07 $& [3] \\
\object{HD\,144628}&$5071\pm 43 $&$ 4.41\pm0.06 $&$ 0.82\pm0.07 $&$ -0.36\pm0.06 $&   22.6 &-0.24 &-0.40 & 17.4 &-0.28 &-0.42 & 13.8 &-0.21 &-0.33 &$-0.38\pm0.09 $& [3] \\
\object{HD\,146233}&$5786\pm 35 $&$ 4.31\pm0.03 $&$ 1.18\pm0.05 $&$  0.08\pm0.05 $&   82.2 & 0.07 &-0.28 & 67.6 & 0.01 &-0.29 & 58.5 & 0.08 &-0.18 &$-0.25\pm0.09 $& [3] \\
\object{HD\,149661}&$5290\pm 52 $&$ 4.39\pm0.07 $&$ 1.11\pm0.07 $&$  0.04\pm0.06 $&   46.7 & 0.07 &-0.16 & 36.2 &-0.01 &-0.20 & 27.2 &-0.02 &-0.17 &$-0.18\pm0.09 $& [3] \\
\object{HD\,150689}&$4867\pm 87 $&$ 4.43\pm0.16 $&$ 1.12\pm0.13 $&$ -0.08\pm0.08 $&   14.0 &-0.28 &-0.39 & 11.0 &-0.29 &-0.39 &  9.0 &-0.20 &-0.28 &$-0.35\pm0.16 $& [3] \\
\object{HD\,152391}&$5521\pm 43 $&$ 4.54\pm0.05 $&$ 1.29\pm0.06 $&$  0.03\pm0.05 $&   55.0 &-0.03 &-0.28 & 44.2 &-0.07 &-0.29 & 33.4 &-0.08 &-0.25 &$-0.27\pm0.08 $& [3] \\
\object{HD\,154088}&$5414\pm 60 $&$ 4.28\pm0.08 $&$ 1.14\pm0.07 $&$  0.33\pm0.07 $&   67.4 & 0.28 & 0.00 & 59.0 & 0.29 & 0.04 & 47.3 & 0.29 & 0.08 &$ 0.04\pm0.10 $& [3] \\
\object{HD\,154577}&$4973\pm 55 $&$ 4.73\pm0.08 $&$ 0.94\pm0.12 $&$ -0.62\pm0.07 $&   20.0 &-0.10 &-0.23 & 14.4 &-0.16 &-0.27 &  9.1 &-0.08 &-0.17 &$-0.22\pm0.12 $& [3] \\
\object{HD\,156274}&$5300\pm 32 $&$ 4.41\pm0.04 $&$ 1.00\pm0.05 $&$ -0.33\pm0.05 $&   43.0 &-0.03 &-0.26 & 27.9 &-0.23 &-0.41 & 27.0 &-0.03 &-0.20 &$-0.29\pm0.13 $& [3] \\
\object{HD\,165185}&$5942\pm 85 $&$ 4.53\pm0.05 $&$ 1.39\pm0.16 $&$  0.02\pm0.10 $&   96.0 & 0.12 &-0.27 & 82.1 & 0.10 &-0.24 & 70.9 & 0.17 &-0.13 &$-0.21\pm0.12 $& [3] \\
\object{HD\,165499}&$5950\pm 45 $&$ 4.31\pm0.02 $&$ 1.14\pm0.08 $&$  0.01\pm0.06 $&   95.7 & 0.09 &-0.32 & 79.4 & 0.03 &-0.32 & 64.4 & 0.03 &-0.26 &$-0.30\pm0.08 $& [3] \\
\object{HD\,170493}&$4854\pm120 $&$ 4.49\pm0.23 $&$ 1.36\pm0.20 $&$  0.15\pm0.13 $&   22.0 & 0.10 &-0.03 & 19.1 & 0.14 & 0.03 & 12.4 & 0.05 &-0.04 &$-0.01\pm0.21 $& [3] \\
\object{HD\,170657}&$5115\pm 52 $&$ 4.48\pm0.08 $&$ 1.13\pm0.07 $&$ -0.22\pm0.06 $&   33.1 & 0.00 &-0.18 & 24.7 &-0.07 &-0.23 & 20.7 & 0.02 &-0.11 &$-0.17\pm0.11 $& [3] \\
\object{HD\,172051}&$5634\pm 30 $&$ 4.43\pm0.03 $&$ 1.06\pm0.05 $&$ -0.20\pm0.04 $&   52.5 &-0.22 &-0.49 & 42.3 &-0.25 &-0.48 & 35.2 &-0.18 &-0.38 &$-0.45\pm0.08 $& [3] \\
\object{HD\,177565}&$5664\pm 28 $&$ 4.43\pm0.02 $&$ 1.02\pm0.04 $&$  0.14\pm0.04 $&   68.4 & 0.03 &-0.26 & 63.3 & 0.10 &-0.18 & 50.6 & 0.11 &-0.11 &$-0.18\pm0.10 $& [3] \\
\object{HD\,192310}&$5069\pm 49 $&$ 4.38\pm0.19 $&$ 0.79\pm0.07 $&$ -0.01\pm0.05 $&   29.5 & 0.34 & 0.17 & 25.9 & 0.02 &-0.13 & 21.6 &-0.27 &-0.40 &$-0.12\pm0.31 $& [4] \\
\object{HD\,219449}&$ 4757\pm102 $&$ 2.71\pm0.25 $&$ 1.71\pm0.09 $&$  0.05\pm0.14$ &  52.2 & 0.41 & 0.12 & 47.5 & 0.45 & 0.18 & 40.2 & 0.48 & 0.24 &$ 0.18\pm0.32 $& [4]\\

\label{longtab9}
\end{longtable}
\end{scriptsize}

\end{landscape}

\end{document}